\documentclass[options]{JHEP3}

\usepackage{amssymb}
\usepackage{epsfig} 
\usepackage{amsfonts}
\usepackage{graphicx}
\usepackage{amsmath} 
\newcommand{\beq}{\begin{equation}}
\newcommand{\eeq}{\end{equation}}
\newcommand{\beqs}{\begin{eqnarray}}
\newcommand{\eeqs}{\end{eqnarray}}

\numberwithin{equation}{section}

\newcommand{\ee}{\end{equation}}
\newcommand{\eea}{\end{eqnarray}}
\newcommand{\be}{\begin{equation}}
\newcommand{\bea}{\begin{eqnarray}}

\abstract{
We generate charged black brane solutions in $D-$dimensions in a theory of gravity 
coupled to a dilaton and an antisymmetric form, by using a Harrison-type transformation. 
The seed vacuum solutions that we use correspond to uplifted Kaluza-Klein 
black strings and black holes in $(D-p)$-dimensions. 
A generalization of the Marolf-Mann quasilocal 
formalism to the Kaluza-Klein theory is also presented, 
the global charges of the black objects being computed  in this way.
We argue that the thermodynamics of the charged solutions can be derived from that of the vacuum configurations. 
Our results show that all charged Kaluza-Klein solutions constructed  by means of Harrison transformations are 
thermodynamically unstable in a grand canonical ensemble. 
The general formalism is applied to the case of nonuniform black strings and caged black hole solutions
 in $D=5, 6$ Einstein-Maxwell-dilaton gravity, whose geometrical properties and thermodynamics are discussed.
 We argue that the topology changing transition scenario, which was previously proposed in the vacuum case, also holds in this case.
Spinning generalizations of the charged black strings are constructed in six dimensions
in the slowly rotating limit.
We find that the gyromagnetic ratio of these solutions possesses a nontrivial
dependence on the nonuniformity parameter.
}  
\preprint{hep-th/yymmddd}

\title{ Harrison transformation 
and charged black objects in Kaluza-Klein theory}
\author{{\large Burkhard Kleihaus,}$^{\dagger}$ {\large Jutta Kunz,}$^{\dagger}$
 {\large Eugen Radu}$^{\dagger  }$  and  {\large Cristian Stelea}$^{\ddagger  }$ 
\\ 
\\
$^{\dagger}${\small Institut f\"ur Physik, Universit\"at Oldenburg, Postfach 2503
D-26111 Oldenburg, Germany}
\\ 
$^{\ddagger}${\small Department of Physics and Astronomy, University of British
Columbia, Vancouver, BC V6T 1Z1, Canada}
 
 } 
\begin{document}
%%%%%%%%%%%%%%%
%%%%%%%%%%%%%%%%%%%%%%%%%%%%%%%%%%%%%%%%%%%%%%%%%%%
\section{Introduction}
%%%%%%%%%%%%%%%%%%%%%%%%%%%%%%%%%%%%%%%%%%%%%%%%%%%%%%%%%%%%%%%%%%
Finding  new  solutions of the theory of general relativity by using  solution generating  techniques is 
a subject of long standing interest. One of the main approaches consists in the 
use of hidden symmetries (dualities) arising in dimensionally reduced theories. 
This technique was initially used in four-dimensional general relativity 
coupled to a Maxwell field \cite{Stephani:2003tm} and was generalized in recent 
years to higher dimensional gravity coupled to various matter fields, a great variety of 
important solutions being generated in this way.

This approach is particularly relevant in the case of Kaluza-Klein (KK) solutions with 
a nontrivial dependence on the extra-dimension. The configurations of interest here
are the static nonuniform black strings (NUBS) and the KK localized black 
holes (BHs) (thus we shall not  consider  KK solutions with bubbles  
\cite{Emparan:2001wk}, \cite{Elvang:2002br}, \cite{Elvang:2004iz} 
or multicenter  black objects \cite{Dias:2007hg}). 
The numerical results in the literature reveal in this case the existence 
of a complicated phase structure, the caged BH and the NUBS
 branches merging at a topology changing transition \cite{Kol:2004ww}. 
Most of the results in the literature on KK solutions with a  dependence of the 
extra-dimension concern the case of vacuum configurations. 
 Even  in this case, there are no explicit solutions that one could write 
 down in closed analytical form (both BHs and NUBSs cannot be found by using a Weyl ansatz 
 \cite{Emparan:2001wk} since the number of abelian Killing vectors is too small for them to belong to a generalised Weyl class).  
 However, one can analyse their properties by using a combination of analytical and numerical methods, which is enough for most purposes. 

The more realistic case of localized BHs and NUBS solutions 
with matter fields is relatively scarcely explored. 
Since it is extremely difficult to solve the field equations for such configurations, 
it  would be  very helpful if one  could generate new solutions from known ones on 
the basis of internal symmetries of the field equations. 
The main purpose of this work is to present such a construction for 
 charged  black brane configurations in a theory of gravity 
coupled to a dilaton and a $(p+1)$-form and
formulated in $D$-dimensions. 
These solutions are generated by using a Harrison-type transformation originally 
proposed in \cite{Gal'tsov:1998yu},
applied to the known KK vacuum nonuniform black strings and black holes in $D-p$ dimensions. 
Different from other cases previously discussed in the literature \cite{Harmark:2004ws}, 
this approach 
produces solutions for an arbitrary coupling of the dilaton field to 
the antisymmetric form. 

 To analyze the thermodynamic properties of such non-uniform  brane solutions one has to 
 be able to compute their total action and define the various conserved quantities. 
 For asymptotically flat spacetimes there are several ways to compute the conserved 
 charges and action. One particularly interesting approach is the so-called counterterm 
 method, which was inspired by the AdS/CFT conjecture in string theory. 
 In this approach one supplements the action by including suitable boundary 
 counterterms, which are functionals of curvature invariants of the induced metric on the boundary. 
 By choosing appropriate counterterms that cancel the divergencies in the total action, one 
 further defines a boundary stress-tensor by varying the on-shell action with respect to the boundary metric. 
 Using the boundary stress-tensor one can then construct a conserved charge for 
 any asymptotic Killing vector. However, unlike the AdS  case, here one does not have a 
 universal prescription for defining the boundary counterterms and their form seems to depend heavily on the boundary topology. 
 A novel counterterm for asymptotically flat spaces has been proposed recently by 
 Mann and Marolf \cite{Mann:2005yr}.  In  contrast to the previous counterterms,  by its definition 
 the Mann-Marolf (MM) counterterm assigns zero action to the Minkowski spacetime 
 in any dimension, while the conserved quantities are constructed  essentially 
  using the electric part of the Weyl conformal tensor. 
  In this work, we also propose a generalization of the MM counterterm to extended objects, 
  like the $p$-brane spacetimes. These spaces are not asymptotically flat but `{\it transverse asymptotically flat}'. 
  Leaving a more general analysis for future work, here we show that the total 
  action and conserved charges for $p$-branes can be defined in terms of the electric part of the Weyl 
 tensor and the final results are consistent with those previously derived (by other means) in 
  the literature \cite{Lu:1993vt,Harmark:2004ch,Traschen:2001pb,Townsend:2001rg}. 
  As a check we also compute the action and conserved charges using different 
  choices of counterterms for asymptotically flat spaces and we obtain similar results.
   One should note here that, unlike the case of the MM counterterm, these other 
   counterterm choices are severely limited as in higher than eight dimensions and in 
   certain coordinates they cannot regularize the action even for flat space \cite{Kraus:1999di}.  

 Once one computes the conserved charges and total action, one is ready to perform the standard thermodynamic 
 analysis of the generated solutions. In particular, we proved that the thermodynamic properties of the 
charged solutions generated using the Harrison transformation can be simply derived from those of the   vacuum  seed 
configurations.  The neutral and charged variables are related in a simple way. 
Moreover, we find that the thermodynamics of the system is affected by 
the value of the dilaton coupling constant. 
Our results show that all $D=5,6$ charged KK solutions constructed in this way are 
thermodynamically unstable in a grand canonical ensemble.  
However, when considering instead  a canonical ensemble, one finds stable solutions for large enough
values of the electric charge.

 The outline of the paper is as follows.
  We begin with a presentation of the general ansatz and we give the necessary details
  regarding the Harrison transformation  in the next section. 
  In  Section 3, we consider the  computation of the global charges of black branes, proposing a generalization of the MM 
quasilocal formalism to the case of the KK theory. In Section 4 we review the basic 
properties of the caged BHs and NUBS vacuum configurations of the KK theory that
are used as seed solutions in the Harrison transformation.  Section 5 contains  
a general discussion of the charged black branes in KK theory. 
A more detailed analysis is presented there for $D=5,6$ NUBS and caged BHs 
in Einstein-Maxwell-dilaton theory. The static, charged NUBSs generated in 
this way are used in Section 6 to construct $D=6$ spinning configurations. 
The results there are valid in the limit of slow rotation. 
We give our conclusions and remarks in the final section. 
The Appendix A  contains a brief discussion of some technical aspects involved
in the numerical construction of the caged BHs and NUBS vacuum solutions,
including 
an independent confirmation for the existence of $D=6$ caged BHs.
In Appendix B we present an exact solution
describing slowly rotating, charged uniform black strings in EM theory in $D$ dimensions,
 and
compute the  gyromagnetic ratio of these configurations.

%%%%%%%%%%%%%%%%%%%%%%%%%%%%%%%%%%%%%%%%%%%%%%%%%%%%%%%%%%%%%%%%%%%%%%%%%%%%
\section{General formalism}
%%%%%%%%%%%%%%%%%%%%%%%%%%%%%%%%%%%%%%%%%%%%%%%%%%%%%%%%%%%%%%%%%%%%%%%%%%%%
%%%%%%%%%%%%%%%%%%%%%%%%%%%%%%%%%%%%%%%%%%%%%%%%%%%%%%%%%%%%%%%%%%%%%%%%%%%%
\subsection{The action and field equations}
%%%%%%%%%%%%%%%%%%%%%%%%%%%%%%%%%%%%%%%%%%%%%%%%%%%%%%%%%%%%%%%%%%%%%%%%%%%%
 We consider the following action 
\begin{eqnarray}
\label{actioni}
I&=&-\frac{1}{16 \pi G}\int_M\,d^Dx\sqrt{-g}\left(R-\frac12(\nabla\phi)^2-
\frac{{\rm e}^{-2a\phi}}{2(p+2)!}F^2_{p+2}\right)
-\frac{1}{8\pi G}\int_{\partial M} d^{D-1}x\sqrt{-h}K,\nonumber\\
\end{eqnarray}
with $\phi$ a dilaton field and $F_{ p+2 }$ a $(p+2)$-form
field strength with $F_{ p+2 } = d {\cal A}_{ p+1 }$, where
${\cal A}_{ p+1 }$ is the corresponding $(p+1)$-form gauge field.
The free parameter $a$ governs the strength of the coupling of the dilaton to the 
form field. 
 $h_{ij}$ is the induced metric on the boundary $\partial \mathcal{M}$
and $K_{ij}$ is the extrinsic curvature of this boundary, with $K=K_{ij}h^{ij}$.
For suitable values of 
$D,p,a$, the action \eqref{actioni} corresponds to truncations of various supergravity
models and also of the bosonic part of the low energy
action of type IIA and IIB string theory.

The corresponding  field equations are
\begin{eqnarray}
\nonumber
R_{\mu\nu} -\frac{1}{2}R g_{\mu \nu}&=& \frac{1}{2}~T_{\mu \nu},
\\
\label{eq1}
\partial_\mu(\sqrt{-g}g^{\mu \nu}\partial_\nu\phi)&=&-\frac{a}{ (p+2)!}
{\rm e}^{-2a\phi}F^2_{p+2},
\\
\nonumber
\partial_\mu\left({\rm e}^{-2a\phi}\sqrt{-g}F^{\mu \nu_1\ldots \nu_{p+1}}_{p+2}\right)&=&0,
\end{eqnarray}
where the total stress-energy tensor has contributions from the dilaton and the $(p+2)$-field strength: 
\begin{eqnarray}
\nonumber
&T_{\mu \nu}=\partial_\mu \phi \partial_\nu \phi-\frac{1}{2}g_{\mu \nu}(\nabla\phi)^2  
+e^{-2a \phi}\frac{1}{(p+1)!}(F_{\mu \nu_1\dots \nu_{p+1}}F_\nu^{~\nu_1\dots \nu_{p+1}}
-\frac{1}{2(p+2)}g_{\mu \nu}F^2_{p+2}). 
\end{eqnarray}
Although we shall consider in this paper  only solutions with an `electric'  potential ${\cal A}_{ p+1 }$, 
magnetically charged solutions can be obtained by using the generalized `electric-magnetic duality' by taking:
\begin{eqnarray}
\label{duality}
a \phi \to -a \phi,~~p\to D - p - 4,~~
F_{p+2}\to \tilde F_{D-p-2}= e^{2a\phi}*F_{p+2},
\end{eqnarray}
where  $*$ stands for the Hodge dual, the geometry being not affected by this transformation.

%%%%%%%%%%%%%%%%%%%%%%%%%%%%%%%%%%%%%%%%%%%%%%%%%%%%%%%%%%%%%%%%%%%%%%%%%%%%
\subsection{The Harrison transformation}
%%%%%%%%%%%%%%%%%%%%%%%%%%%%%%%%%%%%%%%%%%%%%%%%%%%%%%%%%%%%%%%%%%%%%%%%%%%%
Following the method outlined in \cite{Gal'tsov:1998yu} (see also \cite{Yazadjiev:2005hr}), we suppose 
that the space-time has (at least) $p+1$ commuting Killing vectors, one of them being timelike. Then the $D-$dimensional line element can be written in the following form:
\begin{eqnarray}
\label{ansa0}
ds^2=g_{ij}^{(0)}(x)dy^i dy^j
+ g_{\alpha \beta}^{(0)}(x)dx^\alpha dx^\beta,
\end{eqnarray}
where $i,~j=1,\dots,p+1$, while $\alpha, ~\beta=1,\dots,D-(p+1)$; also note that $g_0=det(g_{ij}^{(0)}(x))<0$. The above line element is assumed to be a solution of the $vacuum$ Einstein  field equations.

Upon performing the KK reduction with respect to the $y^i-$directions, the  dimensionally reduced action corresponds to
a non-linear $(D-p-1)$-dimensional $\sigma$-model with the target space 
$SL(p+1,R)/SO(p+1) \times SL(2,R)/SO(2) \times R$. These symmetries imply the existence of a Harrison transformation nontrivially acting on the spacetime variables and matter fields. Therefore one may generate nontrivial solutions of the equations \eqref{eq1} starting with known vacuum configurations.
A detailed description of this procedure and the explicit form of the Harrison 
transformation at the $\sigma$-model level is given in ref. \cite{Gal'tsov:1998yu}.

As a result, one finds  a nonvacuum solution of the equations (\ref{eq1}), with the line-element
\begin{eqnarray}
\label{ansa1}
ds^2=\Omega^{-2\alpha(D-p-3)}g_{ij}^{(0)}(x)dy^i dy^j
+ \Omega^{2\alpha (p+1)}g_{\alpha \beta}^{(0)}(x)dx^\alpha dx^\beta,
\end{eqnarray}
the dilaton field and the only nonvanishing component of the $(p+1)$-form being
\begin{eqnarray}
\label{dilaton}
&&\phi=\phi_0-2a \alpha (D-2)\log \Omega,
\\
&& {\cal A}_{ty_1..y_p}=\sqrt{2(D-2)\alpha}e^{a\phi_0}  |g_0|\Omega^{-1} U ,
\end{eqnarray}
where
\begin{eqnarray}
\label{alpha}
\alpha=[2a^2(D-2)+(p+1)(D-p-3)]^{-1},
~~{\rm and}~~~~
\Omega=(1-U^2)^{-1}(1- U^2 |g_0|),
\end{eqnarray}
$U$ being the parameter of the Harrison transformation, with $0\leq U<1$. Here $\phi_0$ is a real constant corresponding to the asymptotic value of the scalar field; in what follows we shall set $\phi_0=0$, without any loss of generality.

In this way, one finds a family of solutions of the Einstein-dilaton-antisymmetric form equations, 
parameterized in terms of $U$. As we shall see later, the parameter $U$ physically 
corresponds to the electrostatic potential difference between the horizon and infinity. 
One should also remark at this point that the new line element (\ref{ansa1}) preserves the symmetries of the seed metric (\ref{ansa0}).

%%%%%%%%%%%%%%%%%%%%%%%%%%%%%%%%%%%%%%%%%%%%%%%%%%%%%%%%%%%%%%%%%%%%%%%%%%%%
\subsection{The metric ansatz and asymptotics}
%%%%%%%%%%%%%%%%%%%%%%%%%%%%%%%%%%%%%%%%%%%%%%%%%%%%%%%%%%%%%%%%%%%%%%%%%%%%
The procedure described above works for any vacuum solution of the Einstein equation that can be written 
in the form (\ref{ansa0}), regardless of its asymptotic expression. However, a $(p+1)-$form gauge potential naturally couples to the world volume of a $p-$brane. We recall that the black $p-$brane solutions are rich dynamical systems, exhibiting new behaviours that have no analogues in the black hole case \cite{Horowitz:1991cd}. 

In this work we consider solutions approaching at infinity the Minkowski spacetime ${\cal M}^{D-p-1}$ 
times a Ricci flat space $T^p\times S^1$. The metric ansatz  that we consider  is
\begin{eqnarray}
\label{an1}
&ds^2=g_{tt}(r,z)dt^2+g_{rr}(r,z)dr^2+g_{zz}(r,z)dz^2+g_{\omega\omega}(r,z)d\omega_{D-p-3}^2+g_{yy}(r,z)\sum_{i=1}^{p}(dY^i)^2,{~~~~~}
\end{eqnarray}
with $r$ the radial coordinate in ${\cal M}^{D-p-1}$  spacetime and  $d\omega^2_{d}$ is the unit metric on $S^{d}$.
Each $Y^i$-direction is compactified on a circle of length 
$L_i$ (we denote ${\cal V}_p=\prod_{i=1}^pL_i)$. 
$z$ is also a periodic coordinate with  period $L$ that parametrizes $S^1$, while $t$ is the time coordinate. For any topology of the event horizon, the  metric functions possess the following asymptotic  behavior, with a
power law decay\footnote{The situation is more complicated in  $D-p=5$ dimensions 
for black objects with an explicit dependence on the $z-$direction. In these cases there are logarithmic terms in the asymptotic expansion of the metric functions. For  $g_{\omega\omega}$, this occurs already at the next to leading order term, $g_{\omega\omega} \simeq r^2(1+u\log r/r$), with 
$u$ a term that depends on $c_t$ and $c_z$. However, the first two terms in the expression of $g_{tt},g_{zz}$ and $g_{yy}$ are still given by (\ref{as2}).  More generally one can have different constants $c_{y}$ 
for each $Y$-direction; however this will not be the case for the solutions considered in this paper.} 
\begin{eqnarray}
\label{as2}
&g_{tt}\simeq -1+\frac{c_t }{r^{D-p-4}},~~g_{yy} \simeq 1+\frac{c_y }{r^{D-p-4}},~~
g_{zz} \simeq 1+\frac{c_z }{r^{D-p-4}},~~g_{\omega\omega} \simeq r^2(1+\mathcal{O}(1/r)),~{~~}
\end{eqnarray} 
$c_t$, $c_y$ and $c_z$ being real constants, which as discussed in the next section, encode 
the global charges of the solutions.
 
 The generic asymptotic expression of the electric $(p+1)$-form is given by
\begin{eqnarray} 
\label{as-V}
 {\cal A}_{ty_1..y_p}=\Phi-\frac{q_e}{r^{D-p-4}}+\dots,
\end{eqnarray}
while the dilaton field decays as
\begin{eqnarray} 
\label{as-d}
\phi=-\frac{q_d}{r^{D-p-4}}+\dots,
\end{eqnarray}
with $q_e$, $q_d$  real constants.
%%%%%%%%%%%%%%%%%%%%%%%%%%%%%%%%%%%%%%%%%%%%%%%%%%%%%%%%%%%%%%%%%%%%%%%%%%%%%%%%%%%%
\section{Conserved charges and the quasilocal formalism}
%%%%%%%%%%%%%%%%%%%%%%%%%%%%%%%%%%%%%%%%%%%%%%%%%%%%%%%%%%%%%%%%%%%%%%%%%%%%%%%%%%%%
\subsection{The Mann-Marolf counterterm}
%%%%%%%%%%%%%%%%%%%%%%%%%%%%%%%%%%%%%%%%%%%%%%%%%%%%%%%%%%%%%%%%%%%%%%%%%%%%%%%%%%%%
In general, the conserved quantities of black $P$-brane spacetimes are encoded in the asymptotic expansions of the metric functions, as given for instance in (\ref{as2}). Their computation is a problem discussed by many authors, starting with \cite{Lu:1993vt} (see also \cite{Harmark:2004ch,Traschen:2001pb,Townsend:2001rg}), mainly by making use of a reference background, whose obvious choice in the case of a general $P$-brane is ${\cal M}^{D-P}\times T^P$, where ${\cal M}^{D-P}$ is the $(D-P)$-dimensional Minkowski spacetime
(to simplify the relations, we denote in this Section $P=p+1$).

An alternative procedure motivated by the AdS/CFT conjecture and referred to as the `counterterm method', which avoids the difficulties encountered in the background subtraction method, makes use of the quasilocal stress-tensor of Brown and York \cite{Brown:1992br}. In this approach one supplements the action by including a suitable boundary counterterm:
\beqs
I_{ct}=\frac{1}{8\pi G}\int_{\partial M}d^{D-1}x\sqrt{-h}\hat{K}.
\eeqs
 As in the better known AdS case, the counterterms $\hat{K}$ are functionals only of curvature invariants of the 
 induced metric on the boundary and so they do not affect the equations of motion.
Conserved quantities can be constructed using a boundary stress-tensor $T_{ij}$ 
obtained via the algorithm of Brown and York, that is by taking the variation 
of the total action (including the counterterms), with respect to the boundary
 metric $h_{ij}$. Provided that the boundary geometry has an isometry generated 
 by a Killing vector $\xi ^{i}$, a conserved charge
\begin{eqnarray}
\label{charge}
{\cal Q}_{\xi }=\oint_{\Sigma }d^{D-2}S^{i}~\xi^{j}T_{ij}
\end{eqnarray}
can be associated with a closed surface $\Sigma $.
Physically, this means that a collection of observers on the boundary with the induced metric $h_{ij}$ measure the same value of ${\cal Q}_{\xi }$. Although the counterterm method  gives results that are equivalent to those obtained using the background subtraction method\footnote{When one can apply the background subtraction method.} \cite{Kraus:1999di,Mann:2005yr,Mann:2006bd}, we employ it here because  it appears to be a more general technique than background subtraction, and moreover, it is interesting to explore the range of problems to which it applies, in particular for configurations with a dependence on the compact extra-dimensions.

Recently, a particularly interesting counterterm for asymptotically locally flat spacetimes 
has been put forward by Mann and Marolf \cite{Mann:2005yr} (see also \cite{Mann:2006bd,Mann:2008ay,Astefanesei:2006zd,Visser:2008gx,Mann:2009id}). This counterterm is taken to be the trace $\hat{K}$ of a symmetric tensor 
$\hat{K}_{ij}$, which in general dimensions\footnote{In four dimensions its explicit form has been recently derived by Visser \cite{Visser:2008gx}.} is defined implicitly in terms of the Ricci tensor ${\cal R}_{ij}$ 
of the induced metric on the boundary via the relation:
\beqs
{\cal R}_{ij}&=&\hat{K}_{ij}\hat{K}-\hat{K}_i^k\hat{K}_{kj}.
\eeqs
In solving this equation one picks a solution $\hat{K}_{ij}$ that asymptotes to the extrinsic curvature of the boundary of the Minkowski space as the manifold boundary $\partial M$ is pushed to infinity. One should note here that the boundary data $(\partial {\cal M}, h)$ is not uniquely defined by the bulk spacetime $({\cal M}, g)$ but it also depends on the limiting procedure used to define the Gibbons-Hawking term. As shown in  \cite{Astefanesei:2006zd}, if one employes the so-called `cylindrical cut-off',
 then the renormalized action leads to a boundary stress-tensor whose leading-order 
 expression in $D>4$ dimensions involves the electric part of the Weyl tensor:
\beqs
T_{ij}&=&-\frac{1}{8\pi G}\frac{rE_{ij}}{D-3},
\eeqs
where $E_{ij}$ is the pull-back to the boundary $\partial {\cal M}$ of the contraction of the bulk Weyl tensor $C_{\mu\nu\rho\tau}$ with the induced metric $h_{\mu\nu}$. More precisely, introducing the unit normal vector $n^{\mu}$ to the boundary  $\partial {\cal M}$ then the electric part of the bulk Weyl tensor is defined by:
\beqs
E_{\mu\nu}&=&C_{\mu\rho\nu\tau}n^{\rho}n^{\tau}=-C_{\mu\rho\nu\tau}h^{\rho\tau},
\eeqs
while $E_{ij}$ is simply the pull-back to the boundary of the above tensor. In the `cylindrical cut-off', which is more suitable for practical computations, the asymptotic form of the metric at spatial infinity takes the form:
\begin{eqnarray}
ds^{2} &=&-\left(1+\mathcal{O}(r ^{-(D-3)})\right) dt^{2}+\left( 1+%
\mathcal{O}(r ^{-(D-3)})\right) dr^{2}+r^{2}\left( \omega _{IJ}+\mathcal{O}(r ^{-(D-3)})\right) d\theta ^{I}d\theta ^{J}  \notag \\
&&+\mathcal{O}(r^{-(D-4)})d\theta ^{I}dt,  \label{cycutold1}
\end{eqnarray}%
where $\omega _{IJ}$ and $\theta ^{I}$ are the metric and coordinates on the unit $(D-2)$-sphere. 

Our aim here is to generalize these results and compute the action 
and conserved charges for the general $P$-branes
considered in the previous Section. 
These spacetimes are not asymptotically flat and they belong to the class of the so-called 
`{\it transverse asymptotically flat}' spacetimes. 
In general, one can compute the conserved mass, angular momenta and brane tensions 
for these spaces by using background subtraction procedures if one takes the 
reference background of the form ${\cal M}^{D-P}\times T^P$ 
\cite{Harmark:2004ch,Traschen:2001pb,Townsend:2001rg}. 
Since the background spacetime is flat, one expects to be able to extend the 
construction of the Mann-Marolf counterterm to apply for this case as well and, 
following the steps in \cite{Astefanesei:2006zd}, we shall proceed to do so. 

Generalizing the ansatz in (\ref{cycutold}) to apply for $P$-branes in $D$-dimensions, 
one takes the asymptotic form of the metric at spatial infinity of the following form:
\beqs
ds^{2} &=&-\left(1+\mathcal{O}(r ^{-(D-P-3)})\right)dt^{2}+\left( 1+
\mathcal{O}(r ^{-(D-P-3)})\right) dr^{2}+ \sum_{a=1}^P
\left( 1+\mathcal{O}(r ^{-(D-P-3)})\right) (dy^a)^2
\notag 
\\
&&
+\sum_{a\neq b}\mathcal{O}(r ^{-(D-P-4)})dy^ady^b+r^{2}
\left( \omega _{IJ}+\mathcal{O}(r ^{-(D-P-3)})\right) 
d\theta ^{I}d\theta ^{J} +\mathcal{O}(r^{-(D-P-4)})d\theta ^{I}dt,  \notag\\\label{cycutold}
\end{eqnarray}
where $\omega _{IJ}$ and $\theta ^{I}$ are the metric and coordinates on the unit $(D-P-2)$-sphere. Here $r$ plays the role of a `radial' coordinate such that spacelike infinity is reached in the $r\rightarrow\infty$ limit and the symbols $\mathcal{O}(r^{-(D-P-3)})$ refer to terms that fall-off at least as fast as $r^{-(D-P-3)}$ in the $r\rightarrow\infty$ limit.

Note that the extrinsic curvature $K_{ij}$ satisfies the Gauss-Codazzi relation:
\beqs
%{\cal R}_{ij}&=&R_{ikjl}h^{kl}+K_{ik}K-K_i^mK_{mk},
{\cal R}_{ij}&=&R_{ikjl}h^{kl}+K_{ij}K-K^m_{~i}K_{jm},
\eeqs
where $R_{ikjl}$ is the pull-back to the boundary $\partial {\cal M}$ of the bulk Riemann tensor. The leading terms of $K_{ij}$ and $\hat{K}_{ij}$ are given by the extrinsic curvature of the standard cylinder of radius $r$ in the reference background ${\cal M}^{D-P}\times T^P$:
\beqs
K_{ij}&\sim&\hat{K}_{ij}=r\omega_{ij}+\mathcal{O}(r^{-(D-P-4)}),
\eeqs
where $\omega_{ij}$ is the pull-back to $\partial{\cal M}$ of the round metric $\omega_{IJ}$ on the unit sphere $S^{D-P-2}$.

As shown in \cite{Mann:2005yr,Mann:2006bd}, in order to compute the difference $\Delta_{ij}\equiv K_{ij}-\hat{K}_{ij}$ to first order one has to solve the linearized Gauss-Codazzi relation:
\begin{equation}
R_{ikjl}h^{kl}=-\Delta _{kl}h^{kl}\hat{K}_{ij}-\Delta _{ij}\hat{K}+\Delta
_{il}\hat{K}_{j}^{l~}+\Delta _{jl}\hat{K}_{i}^{l~}.  \label{LGC}
\end{equation}
Recall that $h_{ij}$ is the induced metric on a hypersurface $r=const.$ with normal $n^i$ such that $g_{ij}=h_{ij}+n_in_j$. It is convenient at this stage to further define $\mu_{ij}=h_{ij}+u_iu_j-\sum_{a=1}^P y^a_iy^a_j$, where $u^i$ is a future-directed timelike vector on $\partial {\cal M}$, associated with a foliation $\Sigma_t=\{t=const.\}$ of the spacetime ${\cal M}$, respectively $(y^a)^i$ is an outward pointing spacelike vector associated with a foliation $\Sigma_{y^a}=\{y^a=const.\}$ of ${\cal M}$. Then $\hat{K}_{ij}=\frac{\mu_{ij}}{r}+\mathcal{O}(r^{-(D-P-5)})$, while $\hat{K}=\frac{D-P-2}{r}+\mathcal{O}(r^{-(D-P-5)})$ and 
replacing these relations in (\ref{LGC}) one obtains:
\beqs
\label{master}
\frac{D-P-4}{r}\Delta_{ij}-\frac{\tilde{\Delta}_iu_j+\tilde{\Delta}_ju_i}{r}+
\sum_{a=1}^P\frac{(\bar{\Delta}_i^ay^a_j+\bar{\Delta}_j^ay^a_i)}{r}
\nonumber 
\\
+
\frac{\Delta}{r}(h_{ij}+u_iu_j-\sum_{a=1}^Py^a_iy^a_j)&=&-R_{ikjl}h^{kl},
\eeqs
where we denote $\tilde{\Delta}_i=\Delta_{ij}u^j$, $\tilde{\tilde{\Delta}}=\Delta_{ij}u^iu^j$, respectively $\bar{\Delta}_i^a=\Delta_{ij}y^{aj}$ and $\bar{\bar{\Delta}}^{ab}=\Delta_{ij}y^{ai}y^{bj}$. 
Quite generally, we shall consider matter sources for which the stress-energy 
tensor at infinity falls-off fast enough such that the Riemann and Weyl tensors agree to leading order at spacelike infinity
(this is the case of the solutions in this work). 
Therefore, one can replace 
in the right-hand side of the above equation the Riemann tensor by the 
Weyl tensor such that the right-hand side  becomes equal to the electric part of the Weyl tensor $E_{ij}$. 

Contracting now (\ref{master}) with $u^j$ one obtains:
\beqs
\frac{D-P-3}{r}\tilde{\Delta}_i-\frac{\tilde{\tilde{\Delta}}u_i}{r}+\sum_{a=1}^P\frac{\tilde{\bar{\Delta}}^ay^a_i}{r}=E_{ij}u^j
\label{star1}
\eeqs
and further contractions with $u^i$, respectively $y^{b i}$ leads to:
\beqs
\tilde{\tilde{\Delta}}&=&\frac{r}{D-P-2}E_{ij}u^iu^j,~~~~~~ \bar{\tilde{\Delta}}^a=\frac{r}{D-P-2}E_{ij}u^iy^{a j}
\eeqs
respectively. Similarly, contracting (\ref{master}) successively with $y^{ai}$ and $y^{b j}$ leads to:
\beqs
\bar{\bar{\Delta}}^{ab}&=&\frac{r}{D-P-2}E_{ij}y^{a i}y^{b j}.
\eeqs
Finally, taking the trace of (\ref{master}) with $h^{ij}$ one obtains:
\beqs
\Delta&=&\frac{\tilde{\tilde{\Delta}}-\sum_{a=1}^P\bar{\bar{\Delta}}^{a a}}{D-P-3}.
\label{fullDelta}
\eeqs

One is now ready to compute the boundary action (for vacuum spacetimes this is the total action):
\beqs
I_{\partial {\cal M}}&=&-\frac{1}{8\pi G}\int_{\partial M}d^{D-1}x\sqrt{-h}(K-\hat{K})
\nonumber
\\
&=&-\frac{1}{8\pi G}\int_{\partial M}d^{D-1}x\sqrt{-h}\frac{r}{(D-P-2)(D-P-3)} 
E_{ij}(u^iu^j-\sum_{a=1}^Py^{a i}y^{aj}).  \label{action}
\eeqs
Now, in order to compute the conserved charges we evaluate the boundary stress-tensor using the formula (see eq. $(4.2)$ of \cite{Mann:2005yr}):
\begin{equation}
T_{ij}=-\frac{1}{8\pi G}(\Delta _{ij}-h_{ij}\Delta ),
\end{equation}%
from which the conserved charges associated with a boundary Killing vector $\xi $ are found by direct integration as in (\ref{charge}). The conserved mass is generally defined using an asymptotic time translation Killing vector $\xi=\partial/\partial t$, while the angular momenta are defined using angular Killing vectors $\xi=\partial/\partial \varphi$. If $\Sigma_t$ is a spatial slice and $\partial \Sigma_t^{\infty}$ is its intersection with the boundary $\partial {\cal M}$ at spatial infinity, then one integrates (\ref{charge}) on $\partial \Sigma_t^{\infty}$ over the brane directions $y^a$ and the $S^{D-P-2}$ sphere at infinity to compute the mass and angular momenta. Therefore, to compute the conserved mass and angular momenta one basically has to evaluate the quantity:
\beqs
8\pi G T_{ij}u^j&=&-(\tilde{\Delta}_i-\Delta u_i)\\
&=&-\frac{r}{(D-P-2)(D-P-3)}\big[(D-P-2)E_{ij}u^j
+\sum_{a=1}^PE_{lm}y^{a l}(y^{a m}u_i-u^my^{a}_i)\big],\nonumber
\eeqs
where we made use of the formulae (\ref{star1}) and (\ref{fullDelta}). As an example, the conserved mass will be given by:
\beqs
M&=&\frac{1}{8\pi G}\int d^Py d^{D-P-2}\theta\sqrt{\mu}T_{ij}\xi^iu^j\\
&=&-\frac{{\cal V}_P}{8\pi G}\int d^{D-P-2}\theta
\sqrt{\mu}\frac{r}{(D-P-2)(D-P-3)}\big[(D-P-2)E_{ij}\xi^iu^j-\sum_{a=1}^PE_{ij}y^{a i}y^{a j}\big],\nonumber
\eeqs
where $\xi=\partial/\partial t$ and ${\cal V}_P=\int d^Py$ is the volume along the brane directions. Similarly, the angular momentum will be computed using:
\beqs
J&=&\frac{1}{8\pi G}\int d^Py d^{D-P-2}\theta\sqrt{\mu}T_{ij}\xi^iu^j\\
&=&-\frac{{\cal V}_P}{8\pi G}\int d^{D-P-2}\theta
\sqrt{\mu}\frac{r}{(D-P-3)}E_{ij}\xi^iu^j,\nonumber
\eeqs
where now $\xi=\partial/\partial\phi$.

The gravitational tension tensor can be defined similarly using the asymptotic spatial translation Killing vectors $\partial/\partial y^a$ and the integral is now taken over all the spatial directions, excluding however the brane direction \cite{Kastor:2006ti}. To compute the tension tensor density one has to evaluate the quantity:
 \beqs
 8\pi G T_{ij}y^{a j}&=&-(\bar{\Delta}_i^a-\Delta y_i^a)\nonumber\\
&=&-\frac{r}{(D-P-2)(D-P-3)}\big[(D-P-2)E_{ij}y^{a j}+E_{lm}y^{a l}(u^{m}u_i-\sum_{b=1}^Py^{b m}y^{b}_i)\nonumber\\
&&-E_{lm}(u^lu^m-\sum_{b=1}^Py^{b l}y^{b m})y^a_i\big].
\eeqs
In consequence, the tension tensor density (defined as in \cite{Kastor:2006ti}) is given by:
\beqs
{\cal T}^{ab}&=&-\frac{1}{8\pi G}\int d^{D-P-2}\theta\sqrt{\mu}T_{ij}y^{a i}y^{b j}
\\
&=&\frac{1}{8\pi G}\int d^{D-P-2}\theta\sqrt{\mu}\frac{r}{(D-P-2)(D-P-3)}
\big[(D-P-3)E_{ij}y^{a i}y^{b j}
\nonumber
\\
&~&{~~~~~~~~~~~~~~~~~~~~~~~~~~~~~~~~~~~~~~~~~~~~~~~~~~~~~~~~~~~~~~}
\delta^{ab}E_{ij}(u^iu^j-\sum_{c=1}^Py^{c i}y^{c j})\big].
\nonumber
\eeqs
Note that ${\cal T}^{ab}$ is a symmetric tensor.

Suppose now that, regardless of the topology of the event horizon (if present), some of the metric functions appearing in (\ref{cycutold}) have the following asymptotic form\footnote{Note that the asymptotics (\ref{asgen}) are more general than (\ref{as2}), 
since extra-diagonal components in the $y^a$-directions 
are allowed. This would cover also the case of KK solutions with an arbitrary torus compactifications,
so far considered in perturbation theory  only  \cite{Kol:2006vu}.}:
\begin{eqnarray}
\label{asgen}
g_{tt}&\simeq& -1+\frac{c_t }{r^{D-P-3}},~~~g_{y^ay^a} \simeq 1+\frac{c_a }{r^{D-P-3}},~~~g_{y^ay^b} \simeq \frac{d_{ab} }{r^{D-P-3}},
\end{eqnarray} 
where $c_t,c_a$ and $d_{ab}$ are real constants, with $a=1.. P$.

 Then it can be checked that to leading order we have the following relations:
\beqs
E^t_t&=&\frac{(D-P-2)(D-P-3)}{2}\frac{c_t}{r^{D-P-1}},~~~E^{y^a}_{y^a}=-\frac{(D-P-2)(D-P-3)}{2}\frac{c_a}{r^{D-P-1}},\nonumber\\
E^a_b&=&-\frac{(D-P-2)(D-P-3)}{2}\frac{d_{ab}}{r^{D-P-1}},
\eeqs
 for any $a,b =1.. P$. One can then easily compute the conserved mass and the diagonal components of the tension tensor
which are given by the relations (\ref{global-charges}) below.
Note that in the general case, the tension tensor has also the non-diagonal components:
\beqs
{\cal T}^{ab}&=&-\frac{\Omega_{D-P-3}}{16 \pi G}(D-P-4)d_{ab}.
\eeqs
Recall now that we are interested in discussing the thermodynamic properties of black objects whose metrics are asymptotic to the form given in (\ref{an1}), which is appropriate when considering nonuniform extended objects with $(p+1)$-compact spatial directions. In order to accommodate this demand one simply has to replace in the above formulae $P$ with $p+1$ and single out one of the coordinates, say the $(p+1)$-coordinate and denote it by $z$, while the $p$ remaining compact directions will be denoted by $Y^i$. Further note that for the ansatz considered in (\ref{an1}) one has $c_a=c_y$, $d_{ab}=0$ for $a, b =1..p$ and $c_{p+1}=c_z$, while ${\cal V}_P={\cal V}_p L$, where $L$ is the length at infinity along the $z$-direction.

%%%%%%%%%%%%%%%%%%%%%%%%%%%%%%%%%%%%%%%%%%%%%%%%%%%%%%%%%%%%%%%%%%%%%%%%%%%%%%%%%%%%
\subsection{Another counterterm choice}
%%%%%%%%%%%%%%%%%%%%%%%%%%%%%%%%%%%%%%%%%%%%%%%%%%%%%%%%%%%%%%%%%%%%%%%%%%%%%%%%%%%%
For comparison and also as a check of the correctness of the results obtained in the previous section, we shall complement these results with those obtained by using the following boundary counterterm:
\begin{eqnarray}
\label{ct}
I_{ct}&=&\frac{c_{(D,p)}}{8 \pi G}\int_{\partial\mathcal{M}} 
d^{D-1} x\sqrt{-h}\sqrt{\mathcal{R}},
\end{eqnarray} 
where $c_{(D,p)}=\sqrt{(D-p-3)/(D-p-4)}$, and $\mathcal{R}$ is the Ricci scalar of the boundary geometry.

Varying the total action with respect to the boundary metric $h_{ij}$, we compute the boundary stress-tensor:
\begin{eqnarray}
\label{Tik}
T_{ij}=\frac{2}{\sqrt{-h}}\frac{\delta I}{\delta h^{ij}}=-
\frac{1}{8\pi G}\Big( K_{ij}-h_{ij}K
-\Psi(\mathcal{R}_{ij}-\mathcal{R}h_{ij})-h_{ij}\Box \Psi+\Psi_{;ij}
\Big),
\end{eqnarray}
where $K_{ij}$ is the extrinsic curvature of the boundary and we denote $\Psi=c_{(D,p)}\sqrt{1/\mathcal{R}}$. 

One should remark here that the counterterm choice is not unique, other choices 
being possible as well depending on the boundary 
topology (see \cite{Kraus:1999di,Lau:1999dp,Mann:1999pc,Mann:2005cx,Astefanesei:2006zd} 
for a related discussion). 
Our choice of using (\ref{ct}) was motivated by the fact that 
the general expression for the boundary stress-tensor is very simple\footnote{For example, we have verified that for $D=5,6,7$, the same expression for the mass and tension 
are found by using a counterterm on the form (\ref{ct}), with $\sqrt{\mathcal{R}}$
replaced by $\mathcal{R}^{3/2}/\sqrt{\mathcal{R}^2-\mathcal{R}_{ij}^{ij}}$. 
However, in that case, the expression of the boundary stress-tensor  is much more 
complicated than (\ref{Tik}), see $e.g.$ ref. \cite{Mann:2005cx}.}. 

Using the counterterm (\ref{ct}) for geometries whose asymptotic form is given by (\ref{an1}), 
to leading order term the relevant components of the boundary stress-tensor are given by: 
\begin{eqnarray} 
T_t^t&=& -\frac{1}{16 \pi G}\frac{(D-p-3)c_t-c_z-pc_y}{r^{D-p-3}},
\nonumber
\\
T_{Y_i}^{Y_i}&=& - \frac{1}{16 \pi G}\frac{c_t- c_z- (D-4)c_y}{r^{D-p-3}},
\nonumber
\\
T_z^z&=&- \frac{1}{16 \pi G} \frac{c_t-(D-p-3)c_z- pc_y}{r^{D-p-3}}.
\end{eqnarray} 
Note that we already made use of the relations $c_a=c_y$, for $a=1..p$ and $c_{p+1}=c_z$. The mass and the gravitational tensions are the conserved charges (\ref{charge}) associated to $\partial/\partial t$, 
$\partial/\partial y^i$ and $\partial/\partial z$, respectively (note that $\partial/\partial z$ is an asymptotic Killing symmetry of the metric). For a straight torus geometry along the brane directions it becomes apparent that the tension-tensor ${\cal T}^{ab}$ is diagonal. Keeping this in mind, when defining the gravitational tension ${\cal T}^{aa}$ along one direction $y^a$ it is also more convenient at this point to extend the integration along all the spatial directions excluding only the direction $y^a$. For a given direction $y^a$ this amounts to multiplying ${\cal T}^{aa}$ with the factor ${\cal V}_P/L_a$.

%%%%%%%%%%%%%%%%%%%%%%%%%%%%%%%%%%%%%%%%%%%%%%%%%%%%%%%%%%%%%%%%%%%%%%%%%%%%%%%%%%%%
\subsection{The global charges}
%%%%%%%%%%%%%%%%%%%%%%%%%%%%%%%%%%%%%%%%%%%%%%%%%%%%%%%%%%%%%%%%%%%%%%%%%%%%%%%%%%%%

For  both the counterterm choices discussed above, one finds the following expressions for the gravitational charges:
\begin{eqnarray} 
\nonumber
&&M=\frac{V_pL\Omega_{D-p-3}}{16 \pi G }\big((D-p-3)c_t-c_z-pc_y\big)~,
\\
\label{global-charges}
&& {\mathcal T}_z=   \frac{V_p \Omega_{D-p-3}}{16 \pi G }\big(c_t-(D-p-3)c_z- pc_y\big)~,
\\
\nonumber
&& {\mathcal T}_{Y_i}= \frac{L}{L_i}\frac{V_p \Omega_{D-p-3}}{16 \pi G }\big(c_t- c_z- (D-4)c_y\big)~,
\end{eqnarray}
where $\Omega_{D-p-3}$ is the area of the unit $(D-p-3)$-dimensional sphere\footnote{Note that these results are consistent with those previously derived (by other means) in 
 the literature \cite{Lu:1993vt}-\cite{Townsend:2001rg}.}.
  
Usually in the literature one also defines a relative tension
\begin{eqnarray} 
n_I=\frac{ {\mathcal T}_IL_I}{M}~,
\end{eqnarray}
(with $I=z,i$) which measures how large the tension along the $I$-direction (or binding energy) is relative
to the mass.

The solutions we construct will also have an electric charge which is computed according to
\begin{eqnarray} 
\label{Qe}
Q_e=\frac{{\cal V}_pL\Omega_{D-p-3}}{16 \pi G }\int e^{a\phi}\star F=\frac{V_pL\Omega_{D-p-3}}{16 \pi G }(D-p-4)q_e,
\end{eqnarray}
(with $q_e$ defined in (\ref{as-V})),
and a dilaton charge $Q_d$ which is read from 
the asymptotic expansion (\ref{as-d}) of the scalar field 
\begin{eqnarray} 
\label{Qd}
Q_d=\frac{{\cal V}_pL\Omega_{D-p-3}}{16 \pi G }(D-p-4)q_d.
\end{eqnarray}
As we shall discuss in the next Section, $Q_d$ is not an independent charge for the solutions generated by using the Harrison transformation and it can be expressed in terms of mass, $z$-tension and electric charge. The same holds for the gravitational tensions along the $Y^i$-directions.

%%%%%%%%%%%%%%%%%%%%%%%%%%%%%%%%%%%%%%%%%%%%%%%%%%%%%%%%%%%%%%%%%%%%%%%%%%%%%%%%%%%%
\section{KK static, vacuum black objects in $d=D-p$ dimensions}
%%%%%%%%%%%%%%%%%%%%%%%%%%%%%%%%%%%%%%%%%%%%%%%%%%%%%%%%%%%%%%%%%%%%%%%%%%%%%%%%%%%%

The charged black objects discussed in the next sections are constructed using as seed 
solutions the known vacuum static BHs and black strings in $d=D-p$ dimensions
($i.e.$ with only one compact direction). 
In absence of matter fields (or of a cosmological constant), these configurations are trivially uplifted to $D-$dimensions by adding 
$p-$flat directions $Y^i$. Therefore it is of interest here to briefly review their basic properties
(see \cite{Obers:2008pj} for a more  detailed overview of the KK solutions).
   
Already in this case, the vacuum Einstein equations admit a variety of interesting solutions, 
approaching asymptotically the ${\cal M}^{d-1}\times S^1$ background. 
Of interest here are the black strings and the black holes, which are distinguished according to their horizon topology.
Both types of solutions are usually described 
within the  following ansatz with three unknown functions\footnote{An ansatz 
involving only two undetermined functions has been proposed in \cite{Harmark:2002tr}, 
for a special choice of coordinates.  
However, it remains an interesting open problem whether one can use that ansatz  for numerical calculations.}
$A,B,C$:
 \begin{eqnarray} 
\label{metric}
ds^2_d=-e^{2A(r,z)}f(r)dt^2+e^{2B(r,z)}\left( \frac{dr^2}{f(r)} 
+dz^2 \right)+e^{2C(r,z)}R^2(r,z)d\omega_{d-3}^2, 
\end{eqnarray}
where $A,B,C$ vanish at large coordinate $r$. $f(r)$ and $R(r,z)$ 
 are suitable background functions.
Usually one takes  $f(r)=1-({r_0}/{r})^{d-4},~~R(r,z)=r$ for black strings, 
 in which case $r_0>0$ corresponds to the horizon radius.
The usual choice for black holes is $f(r)=1$, while $R(r,z)$
has a complicated form (with $R(r,z)\to r$ in the asymptotic region), see Appendix A.
 
Unfortunately the only solutions known in closed form (except for the bubbles, 
which we do not consider in this work) have no $z-$dependence and describe uniform black strings (UBSs) 
with $A=B=C=0$  
($i.e.$ the direct product of the 
Schwarzschild-Tangherlini solution in $(d-1)$-dimensions with a spacelike circle). 
The existence of both caged BHs and  NUBSs has been asserted so far mainly  based on numerical results
(note, however, that analytic approximations of the small BH solutions have been presented in \cite{Gorbonos:2004uc}).
Such solutions have been found at the nonperturbative level by solving a set of 
three partial differential equations for the metric functions $A,B,C$, subject to a suitable set of boundary conditions.
The technical aspects involved
in the numerical construction of these solutions are presented in Appendix A.

The horizon topology of the black string solutions is $S^{d-3}\times S^1$. 
As shown by Gregory and Laflamme \cite{Gregory:1993vy}, a uniform solution is unstable below a 
critical value of the mass. As a result, there also exist nonuniform strings, which 
emerge  smoothly from the UBS branch at a critical point, 
where its stability changes \cite{Gregory:1993vy}. Keeping fixed the horizon coordinate 
$r_0$ and the asymptotic length $L$ of the compact direction, these solutions depend on a 
single parameter, specified via the boundary conditions and physically corresponding to the 
Hawking temperature of the solutions \cite{Gubser:2001ac}. 
Varying this parameter, the nonuniform strings become increasingly deformed  (see Appendix A for details). 
A measure of the deformation of the geometry of these solutions is given by 
the nonuniformity parameter $\lambda$ \cite{Gubser:2001ac}
\begin{equation} 
\lambda = \frac{1}{2} \left( \frac{{\cal R}_{\rm max}}{{\cal R}_{\rm min}}
 -1 \right),
\label{lambda} 
\end{equation}
where ${\cal R}_{\rm max}$ and ${\cal R}_{\rm min}$ represent the maximum radius of a 
$(d-3)$-sphere on the horizon and the minimum radius, being the radius of the `waist'. 
Concerning the thermodynamical stability of the NUBS solutions, it turns out that, at 
least for $d=5, 6$, the entropy decreases with temperature, $i.e.$ they are 
thermally unstable.

Apart from the black strings, KK theory also admits as solutions a branch of 
caged black holes with an event horizon of topology $S^{d-2}$.
For the metric ansatz (\ref{metric}), the event horizon 
is still located at a constant value of  the radial coordinate, $r=r_0$, with $-\infty<r<0$. 
(Note that the functions $A,B,C$ contain in this case suitable 'background' parts \cite{Kudoh:2003ki} 
that enforce  the required behaviour on the boundaries and the topology difference between the
event horizon and infinity, see Appendix A.)

For small values of the mass, these solutions look like the Schwarzschild-Tangerlini solution in $d$ dimensions.
Once one increases their horizon radius for a fixed value of
the asymptotic length $L$
of the extra-direction, they deform as they start to `feel' the compact $S^1$ direction
and their mass increases. 
Analogous to the NUBS case, the geometrical properties of the black holes can be described in terms of 
two parameters denoted here by $R_{eq}$, the equatorial radius of the horizon, and $L_{axis}$, 
the proper length along the exposed symmetry axis. 
Again, the entropy of the caged BHs is a decreasing function of temperature, $i.e.$ the considered
BHs are thermally unstable\footnote{Note however, that the recent results in \cite{Headrick:2009pv}
indicate the existence of a more complicated picture, with
the possible existence of caged BHs possessing a positive specific heat.}.
 
Although a number of details remain to be clarified, the numerical results  
suggest that the black hole and the NUBS branches merge at a topology changing transition \cite{Kol:2004ww}, \cite{Kudoh:2004hs}. 
This scenario is supported by all numerical data available. 
For NUBS, the normalized thermodynamic quantities change between one (the uniform solution) 
and some critical values.
The normalized  mass, relative tension and entropy of the caged black holes vary between zero and 
some critical values that seem to correspond to those where the NUBS branch stops to exist. 
At the same time, the Hawking temperature of the BHs varies between infinity 
(for vacuum state with $M={\mathcal T}=0$) and a minimal value.
% which is also approached by the black string solutions.  
The set of black hole and the set of NUBS solutions then appear to merge at critical
values of these quantities.
 In a $(n,M)$ or $(n,T)$ diagram, this feature can be seen looking at the vacuum data ($U=0$) in Figure 5.
(For convenience, the thermodynamic quantities there are normalized by their value for the critical UBS solution.)
 This scenario also implies that the `waist' of the NUBSs would shrink to nothing, 
 locally forming a cone geometry. At that point, both $L_{axis}$ and $R_{min}$ go to zero, while
$R_{eq}$ tends to $R_{max}$.

For both black holes and black strings, the following Smarr relation holds
\begin{eqnarray}
\label{smarrform1}
\frac{(d-3)M-L{\mathcal T}}{d-2}  = T_H S,
\end{eqnarray} 
where $T_H$ is the Hawking temperature and $S$ is the entropy of the solutions (which is one quarter of the horizon area). The first law of thermodynamics, which for the KK black objects takes the form: 
\begin{eqnarray}
\label{fl1}
dM=T_HdS+{\mathcal T}dL
\end{eqnarray} 
 is also satisfied.
 
%%%%%%%%%%%%%%%%%%%%%%%%%%%%%%%%%%%%%%%%%%%%%%%%%%%%%%%%%%%%%%%%%%%%%%%%%%%%%%%%%%%%
\section{Charged black objects in KK theory}
%%%%%%%%%%%%%%%%%%%%%%%%%%%%%%%%%%%%%%%%%%%%%%%%%%%%%%%%%%%%%%%%%%%%%%%%%%%%%%%%%%%%
The vacuum configurations discussed above can be uplifted to $D$-dimensions and used to generate solutions 
of the Einstein-dilaton-antisymmetric form theory by applying a Harrison transformation,
 as described in Section $2$. 
The asymptotic structure of the metric and the topology of the event horizon remain unchanged in the process.

%%%%%%%%%%%%%%%%%%%%%%%%%%%%%%%%%%%%%%%%%%%%%%%%%%%%%%%%%%%%%%%%%%%%%%%%%%%%%%%%%%%%
\subsection{The solutions and the geometry}
%%%%%%%%%%%%%%%%%%%%%%%%%%%%%%%%%%%%%%%%%%%%%%%%%%%%%%%%%%%%%%%%%%%%%%%%%%%%%%%%%%%%
In our approach, the two metric parts which enter (\ref{ansa0}) are
\begin{eqnarray}
\label{an2}
&& g_{ij}^{(0)}(x)dy^i dy^j=-e^{2A_0(r,z)}f(r)dt^2+\sum_{i=1}^{p}(dY^i)^2,
\\
\nonumber
&&g_{\alpha \beta}^{(0)}(x)dx^\mu dx^\nu=
e^{2B_0(r,z)}(\frac{dr^2}{f(r)}+dz^2)+e^{2C_0(r,z)}R^2(r,z)d \omega_{D-p-3}^2,
\end{eqnarray}
where $A_0(r,z),B_0(r,z)$ and $C_0(r,z)$ are the functions of a vacuum seed solution.

The simplest charged solution one can construct by employing the Harrison transformation 
described in Section $2$, 
uses a vacuum uniform black string solution as seed configuration. The resulting solution
 corresponds to a $(p+1)-$brane solution with the line element:
\begin{eqnarray} 
\nonumber
&ds^2=-\left(1+(\frac{r_0}{r})^{D-p-4}\frac{U^2}{1-U^2}\right)^{-2\alpha(D-p-3)}
\big[-(1-(\frac{r_0}{r})^{D-p-4})dt^2+\sum_{i=1}^{p}(dY^i)^2\big]
\\
\label{uniform}
&+(1+(\frac{r_0}{r})^{D-p-4}\frac{U^2}{1-U^2})^{2\alpha (p+1)}
 \left(\frac{dr^2}{1-(\frac{r_0}{r})^{D-p-4}}+dz^2+r^2d\omega_{D-p-3}^2\right),
\end{eqnarray}
and matter fields
\begin{eqnarray} 
\nonumber
& {\cal A}_{t y_1\dots y_p}=\sqrt{2(D-2)\alpha}  
\frac{(1-(\frac{r_0}{r})^{D-p-4}) }
{1+(\frac{r_0}{r})^{D-p-4}\frac{U^2}{1-U^2}}U~,
\\
\nonumber
&\phi =-2a(D-2)\alpha\log(1+(\frac{r_0}{r})^{D-p-4}\frac{U^2}{1-U^2}),
\end{eqnarray}
with $\alpha$ given by (\ref{alpha}) and $0\leq U <1$.
For any value of $a,U$, the surface $r=r_0$ is still an event horizon, while
 $r=0$ is a curvature singularity.
As we have previously mentioned, the geometric  structure outside 
the event horizon and  also  the asymptotic structure are preserved by the solution generating procedure. 
The inclusion of the matter fields affects the geometry in both $z-$ and $Y^i-$directions, however in different ways. 
 One can see that the proper length of the $z$ circle increases for any finite $r$, while the opposite holds for  a $Y^i-$direction. Moreover, for a given value of $r$, the area $\Omega_{D-p-3}$ of the ${D-p-3}$-sphere increases with $U$.

The extremal limit of (\ref{uniform}) is found by taking $U \to 1$ together with a rescaling 
of $r_0$  such that $r_0^{D-p-4}=(1-U^2)c^{D-p-4}$ and it reads:
\begin{eqnarray} 
\label{ext-uniform}
&&ds^2=-\left(1+(\frac{c}{r})^{D-p-4} \right)^{-2\alpha(D-p-3)}
\left(- dt^2+\sum_{i=1}^{p}(dY^i)^2\right)
\\
\nonumber
&&{~~~~~~~~}+(1+(\frac{c}{r})^{D-p-4} )^{2\alpha (p+1)}
 \left( dr^2 +dz^2+r^2d\omega_{D-p-3}^2\right),
 \\
&& 
{\cal A}_{t y_1\dots y_p}=\sqrt{2(D-2)\alpha}  
 \left(1+(\frac{c}{r})^{D-p-4}\right)^{-1} ,
 ~
 \phi=  -2a(D-2)\alpha\log(1+(\frac{c}{r})^{D-p-4} ), ~~{~~~~}
\end{eqnarray} 
$c$ being a positive constant. 
%However, one can verify that, for $p\neq 0$, this solution has
%a naked singularity at $r=0$. 

The basic features of the nonextremal uniform configurations are shared by the
charged solutions generated starting from NUBS and caged black holes.
In both cases the  geometry  is described by the metric form:
 \begin{eqnarray} 
\label{metric-new}
&&ds^2 =-e^{2A(r,z)}f(r)dt^2
+e^{2B(r,z)}\left( \frac{dr^2}{f(r)} +dz^2 \right)
\\
\nonumber
&&~~{~~~~~~~~~~~~~~~~~~~~~~~~~~~~}+e^{2C(r,z)}R^2(r,z)d\omega_{D-3}^2
+e^{2C_1(r,z)}\sum_{i=1}^{p}(dY^i)^2, 
\end{eqnarray}
where
\begin{eqnarray}
\nonumber
&&A(r,z)=A_0(r,z)-\alpha (D-p-3)\log \Omega(r,z),~~
B(r,z)=B_0(r,z)+ \alpha (p+1)  \log \Omega(r,z),
\\
\label{new-funct}
&&C(r,z)=C_0(r,z)+ \alpha (p+1) \log \Omega(r,z),~~
C_1(r,z)=  - \alpha (D-p-3)  \log \Omega(r,z),
\end{eqnarray} 
and  
\begin{eqnarray} 
\label{m1}
\Omega(r,z)=\frac{1}{1-U^2}(1-U^2 e^{2A_0(r,z)}f(r)).
\end{eqnarray}
The only nonvanishing component  of the $(p+1)$-form and the dilaton are given by
\begin{eqnarray} 
\label{matter-new}
{\cal A}_{ty_1..y_p}=\sqrt{2(D-2)\alpha}  e^{2A_0(r,z)}f(r)~\frac{\Omega(r,z)}{U} ,~~
\phi= -2a (D-2)\alpha \log \Omega(r,z).
\end{eqnarray}
Again, the position of the horizon is not affected by the Harrison transformation.
Once one knows the expressions of the functions $A_0(r,z)$, $B_0(r,z)$, $C_0(r,z)$
for the vacuum solution,  one can derive all the geometric properties of 
the corresponding charged configurations. However, in absence of an explicit  analytical solution, 
it is unclear how to derive the extremal version of the  black objects with an explicit $z$-dependence starting from 
the non-extremal solution (\ref{metric-new}), (\ref{matter-new}). 
Therefore, for the rest of this work we shall consider nonextremal solutions only.

%%%%%%%%%%%%%%%%%%%%%%%%%%%%%%%%%%%%%%%%%%%%%%%%%%%%%%%%%%%%%%%%%%%%%%%%%%%%%%%%%%%%
\subsection{The charges of the solutions}
%%%%%%%%%%%%%%%%%%%%%%%%%%%%%%%%%%%%%%%%%%%%%%%%%%%%%%%%%%%%%%%%%%%%%%%%%%%%%%%%%%%%
A straightforward computation based on the generic asymptotic form of 
the metric functions (\ref{as2})
 leads to the simple relations:
 \begin{eqnarray}
\label{r1} 
c_t&=&c_t^{(0)}(1+2\alpha(D-p-3)\frac{U^2}{1-U^2}),~~~
c_z=c_z^{(0)}+2 c_t^{(0)} \alpha(p+1)\frac{U^2}{1-U^2},
\nonumber
\\
c_y&=&- 2\alpha(D-p-3) c_t^{(0)}\frac{U^2}{1-U^2},
\end{eqnarray}
 $c_t^{(0)}$, $c_z^{(0)}$ being the constants in the 
 asymptotic expansion of the seed solution with $U=0$.
 The expressions for the electric charge $Q_e$ 
and the dilaton charge $Q_d$, as derived from (\ref{as2}),(\ref{Qe}),(\ref{Qd}),  are:
%The expression of the dilaton and $p+1$-form potential can be read
%from (\ref{dilaton}).
%These functions decay asymptotically as $1/r^{D-p-4}$,
%the electric charge $Q_e$ and the dilaton charge $Q_d$ being
%and the  potential difference $\Phi$ between the event horizon and infinity
\begin{eqnarray}
\label{Qe-fin1}
&&Q_e=\frac{LV_p\Omega_{D-p-3}}{16 \pi G}(D-p-4)
\sqrt{2(D-2)\alpha}~\frac{U}{1-U^2}~c_t^{(0)},
\\
\label{Qd-fin1}
&&Q_d=\frac{LV_p\Omega_{D-p-3}}{16 \pi G}(D-p-4)
 \frac{U^2}{1-U^2}~c_t^{(0)}.
\end{eqnarray} 
There is also a one-to-one correspondence between the solution generating parameter 
$U$ and the electrostatic potential difference $\Phi$ between the event horizon and
infinity
\begin{eqnarray}
\Phi=\sqrt{2(D-2)\alpha} U.
\end{eqnarray} 
As argued below, in a thermodynamical description of
the system, $\Phi$ 
corresponds to the chemical potential.

One can see that the mass and tensions
of a charged solution can be expressed as linear combinations of the
corresponding quantities of the vacuum  configuration: 
\begin{eqnarray}
\label{tr-ch}
\left[
\begin{array}{c }
M\\
{\mathcal T}_{y^i}\\
{\mathcal T}_z\\
\end{array}
\right]\ 
=
\left[
\begin{array}{ccc}
a_{11} & a_{12}& a_{13}\\
a_{21} & a_{22}& a_{23}\\
a_{31} & a_{32}& a_{33}\\
\end{array}
\right]\ 
\cdot
\left[
\begin{array}{c }
M_0\\
{\mathcal T}^0_{y^i}\\
{\mathcal T}^0_z\\
\end{array}
\right]\ 
, 
\end{eqnarray}
  where 
\begin{eqnarray}
\nonumber
&&a_{11}=1+\frac{2(D-2)(D-p-3) \alpha}{(D-p-2)}\frac{ U^2}{1- U^2},
~~a_{13}=-\frac{2(D-2) \alpha L}{(D-p-3)} \frac{ U^2}{1- U^2},
\\
\nonumber
&&a_{21}= \frac{2(D-2)(D-p-3)\alpha}{(D-p-2) L_i} \frac{ U^2}{1- U^2},
~~a_{23}=-\frac{2(D-2)(D-p-3)\alpha L_i}{(D-p-2) L} \frac{ U^2}{1- U^2},
\end{eqnarray}
 while $a_{12}=a_{31}=a_{32}=0,~~a_{22}=a_{33}=1$.
(Note that the tension along the $z-$direction does not change when adding an electric charge to the system).

The electric charge can also be expressed as a function of mass and $z-$tension of the vacuum solution:
\begin{eqnarray}
\label{Qe-fin}
 Q_e=\frac{\sqrt{2 (D-2)\alpha}}{ (D-p-2)} 
 \left( (D-p-3)M_0-\ L{\mathcal T}_z^0  \right) \frac{U}{1-U^2},
\end{eqnarray}
with a similar relation  holding for the dilaton charge.

The initial vacuum configuration has two global charges, $M_0$
and ${\mathcal T}_z^0$.
The Harrison transformation introduces via the parameter $U$
a new charge. This implies that $Q_d$ and ${\mathcal T}_i$
are not new global independent charges and therefore they do not enter the 
thermodynamical description of the system since they cannot be varied at will. Indeed, 
one finds  the following quadratic relation 
\begin{eqnarray}
\frac{Q_e^2}{\kappa_1\left(
M(D-p-3)-L {\mathcal T}_z
\right)
+Q_d}
=
\left(
1-\frac{2(D-2)(D-p-3)\alpha}{D-p-2}
\right)
Q_d,
 \end{eqnarray}
with
\begin{eqnarray}
\kappa_1=\frac{\sqrt{2(D-2)\alpha}}{(D-p-4)
\left(
D-p-2-2(D-2)(D-p-3)\alpha
\right)
},
\end{eqnarray}
which determines the dilaton charge in terms of the mass, tension along the $z-$direction
and the electric charge.
The tensions along the $Y^i$-directions can also be written as
\begin{eqnarray}
 L_i {\mathcal T}_i=\frac{1}{D-p-3}
\left(
\frac{2(D-2)(D-p-3)U^2 \alpha+1-U^2}{2(D-2)U^2 \alpha}\Phi Q_e +L {\mathcal T}_z
\right).
\end{eqnarray}

%%%%%%%%%%%%%%%%%%%%%%%%%%%%%%%%%%%%%%%%%%%%%%%%%%%%%%%%%%%%%%%%%%%%%%%%%%%%%%%%%%%%
\subsection{The thermodynamics}
%%%%%%%%%%%%%%%%%%%%%%%%%%%%%%%%%%%%%%%%%%%%%%%%%%%%%%%%%%%%%%%%%%%%%%%%%%%%%%%%%%%%
%%%%%%%%%%%%%%%%%%%%%%%%%%%%%%%%%%%%%%%%%%%%%%%%%%%%%%%%%%%%%%%%%%%%%%%%%%%%%%%%%%%%
%\subsubsection{General results}
%%%%%%%%%%%%%%%%%%%%%%%%%%%%%%%%%%%%%%%%%%%%%%%%%%%%%%%%%%%%%%%%%%%%%%%%%%%%%%%%%%%%
The Hawking temperature $T_H$ of a black object is computed either by requiring the 
absence of conical singularities on the Euclidean section (which is found by analytical continuation
$t\to i \tau$) or from the surface gravity of the solutions in the Lorentzian section. 
As  usual with black objects in Einstein gravity, 
the KK solutions considered in this work have an entropy which is one quarter of the event horizon area.

One can easily prove that the relation between Hawking temperature $T_H$ and the entropy $S$ 
of the charged solutions and the corresponding quantities $T_{H(0)}$ and $S_0$ of the vacuum seed solution  is
\begin{eqnarray} 
\label{TH-ch}
%T_H=T_0(\cosh \beta)^{-2\alpha(D-2)},~~
%S=S_0(\cosh \beta)^{2\alpha(D-2)},
 T_H=T_{H(0)} (1-U^2)^{\alpha(D-2)},~~~
 S=S_0(1-U^2)^{-\alpha(D-2)},
\end{eqnarray}
such that the product between the entropy and temperature remains invariant under the
 Harrison transformation. Thus adding an electric charge to a system with a fixed event horizon radius
  would decrease 
 its temperature and increase the entropy, which agrees with the experience based on Reissner--Nordstr\"om solution.

Considering the thermodynamics of these solutions,
they should satisfy the first law of thermodynamics 
\begin{eqnarray}
\label{firstlaw}
dM=T_HdS+ \Phi dQ_e+{\mathcal T}_zd L .
\end{eqnarray}
One may regard the parameters $S,~Q_e$ and $L$ as a complete set of extensive parameters
for the mass  $M(S,Q_e,L)$ and define the intensive parameters
conjugate to them.
These quantities are the temperature, the chemical potential and the $z$-tension.

For any type of charged black object, the following Smarr relation holds
\begin{eqnarray}
\label{charged-Smarr}
\frac{(D-p-3)M-{\mathcal T}_z L}{D-p-2}=T_H S+\frac{D-p-3}{D-p-2}\Phi Q_e,
\end{eqnarray}
which results from (\ref{smarrform1}) together with (\ref{tr-ch}), 
(\ref{Qe-fin}) and  (\ref{TH-ch}).
%(Note that, as expected the dilaton charge does not enter the thermodynamics.)

In the canonical ensemble, we study solutions holding the temperature
$T_H$, the electric charge $Q_e$
and the length $L$ of the extra-dimension fixed.
The associated thermodynamic potential is the Helmholz free energy 
\begin{eqnarray}
\label{F}
F[T_H,Q_e,L]=M-T_HS,~~~ d F = -S d T_H  + \Phi dQ_e +{\mathcal T}_z dL \ .  
\end{eqnarray}
The situation of KK black objects in a grand canonical ensemble is also of interest,
 in which case we keep 
the temperature, the chemical potential and the tension fixed.
 In this case the thermodynamics is obtained from the Gibbs potential 
\begin{eqnarray}
\label{G}
G[T_H,\Phi,{\mathcal T}_z]=E-T_H S- \Phi Q_e -{\mathcal T}_z L .
\end{eqnarray}
By using the relations (\ref{tr-ch}), (\ref{Qe-fin}), one finds that both $F$ and $G$ can also be expressed 
in terms of the global charges of the vacuum seed solutions and the generating parameter $U$.

It is well-known that different thermodynamic ensembles are not exactly
equivalent and may not lead to the same conclusions regarding the thermodynamic stability
as they correspond to different
physical situations (see e.g. \cite{Caldarelli:1999xj}, \cite{Chamblin:1999hg}). 
Mathematically, thermodynamic stability is equated with the subadditivity 
of the entropy function.
In the canonical ensemble, the subadditivity of the entropy
is exactly equivalent to the positivity
of the specific heat at constant $(Q_e,L )$ 
\begin{equation}
\label{specheat} 
C_{Q,L} 
\equiv \left( \frac{\partial M}{\partial T_H}
\right)_{Q,L} = T_H \left( \frac{\partial S}{\partial T_H} \right)_{Q,L} > 0 .
\end{equation}
Remarkably, one can prove that 
the specific heat of a charged configuration
can be expressed in terms of the specific heat and entropy of 
the vacuum seed solution\footnote{In evaluating $C_{Q,L}$ one uses (\ref{TH-ch})
together with the expression (\ref{Qe-fin1}) for the electric charge.
The Smarr relation (\ref{smarrform1}) implies also $c_t^{(0)}\sim T_{H(0)} S_0$.
Thus, one finds $T_{H(0)}S_0d(U/(1-U^2))+U/(1-U^2)d(T_{H(0)}S_0)=0$ and
substitutes back in (\ref{specheat}).}:
\begin{eqnarray}
\label{CQ}
 C_{Q,L}=
 \frac{1}{(1-U^2)^{2\alpha(D-2)}}
\frac{C_{Q,L}^0(1+U^2-2\alpha(D-2)U^2)-2\alpha(D-2)U^2S_0}
{U^2+1+2\alpha(D-2)U^2(1+\frac{C_{Q,L}^0}{S_0})} .
\end{eqnarray}
The condition for thermodynamic stability in the grand canonical 
ensemble is \cite{Chamblin:1999hg}, \cite{Cvetic:1999ne}:  
\begin{eqnarray}
\label{cgca}
 C_{Q,L}>0,~~\epsilon_T \equiv \left( \frac{\partial Q}{\partial \Phi} \right)_{T_H} > 0,
\end{eqnarray}
where $\epsilon_T$ is the isothermal electric permittivity, which measures the stability of solutions to 
electrical fluctuations. 
This quantity can also be expressed in terms of the 
quantities of the vacuum seed solution\footnote{Here one uses
the relation 
$0=dT_{H(0)}-2\alpha(D-2)U/(1-U^2)T_{H(0)}dU$, which follows
from  (\ref{TH-ch}).}
\begin{eqnarray}
\label{eps}
\epsilon_T =\frac{T_{H(0)}S_0}{(1-U^2)^2}\left(U^2+1+2\alpha(D-2)U^2(1+\frac{C_{Q,L}^0}{S_0})\right).
\end{eqnarray}

We have seen in the previous Section that the specific heat of the vacuum solutions $C_{Q,L}^0$ 
is a strictly negative quantity
for both black holes and black strings.
However, one can see from (\ref{CQ}), (\ref{eps}) that  $C_{Q,L}$ and $\epsilon_T$
may become positive for some range of $U$, $a,p$. 
 
 Here we would like to point out that for any black  object constructed by using the $d=5,6$
vacuum solutions discussed in Section 4, the condition (\ref{cgca}) 
for thermodynamic stability in the grand canonical ensemble
cannot be satisfied. 
To prove that, we express  $C_{Q,L}$ and $\epsilon_T$ as 
\begin{eqnarray}
& C_{Q,L}=\frac{1}{(1-U^2)^{2\alpha(D-2)}}
  S_0\left(\frac{x+1}{1+\bar \alpha (x+1)}-1\right),~~~
  \epsilon_T=\frac{T_0S_0}{(1-U^2)^2}(1+U^2)(1+\bar \alpha (x+1)),
 \end{eqnarray}
 where we denote
 $x=C_{Q,L}^0/S_0$, $\bar \alpha=\alpha(D-2)2U^2/(1+U^2)$.
 Moreover, we  have verified\footnote{For the solutions with a $z-$dependence, we have used the 
numerical data in \cite{Kudoh:2004hs}, \cite{Kleihaus:2006ee}.} that $C_{Q,L}^0+S_0<0$ for all vacuum seed solutions with 
$d=5,6$
($e.g.$ this quantity is
$C_{Q,L}^0/S_0+1=-(D-p-4)$ for UBSs).
 Therefore $x+1<0$ and, as a result, $\epsilon_T$ and  $C_{Q,L}$ cannot be both positive  at the same time.
 (This is a general result, valid for any values of $a,p,U$.)
 Thus we conclude that all charged KK solutions generated from $d=5,6$ vacuum solutions 
 are unstable in a grand canonical ensemble.

 Of course, this does not exclude the existence of stable solutions in a canonical ensemble.
 In the next subsection we shall find such configurations 
 for both black holes and black strings ($i.e.$ with $p=0$). 
A similar analysis for a generic $p$ does not seem to be possible, except
  for the 
 case of uniform  brane solutions (thus without an explicit dependence on $z$),
 which we shall consider next.
 The quantities which enter the thermodynamics can be derived\footnote{The 
 thermodynamics of uniform
 black strings and $p-$branes has been discussed in \cite{Cai:1994jh} 
 (see also \cite{Reall:2001ag} and references therein). }
 easily in this case from (\ref{uniform})
  \begin{eqnarray}
&&M=\frac{\Omega_{D-p-4}V_pL}{16 \pi G}
\left(
(D-p-3)(1-U^2)+2(D-2)(D-p-4)U^2\alpha
\right)
\frac{r_0^{D-p-4}}{1-U^2},
\\
\nonumber
&&{\mathcal T}_z=\frac{\Omega_{D-p-4}V_p}{16 \pi G}r_0^{D-p-4},~~
Q_e=\frac{\Omega_{D-p-4}V_pL}{16 \pi G}\sqrt{2(D-2)}(D-p-4)\frac{U}{1-U^2}r_0^{D-p-4},
 \end{eqnarray} 
 where $r_0$ is the event horizon radius.
One can see that the charge to mass ratio of these solutions has
a nontrivial dependence on $a,p,U$:
\begin{eqnarray}
\label{ratio}
\frac{Q_e}{M}=\frac{\sqrt{2(D-2)\alpha}(D-p-4)U}{(D-p-3)(1-U^2)+2(D-2)(D-p-4)U^2\alpha}.
\end{eqnarray} 
 The specific heat and isothermal electric permittivity
 of the uniform solutions are given by
 \begin{eqnarray}
&&C_{Q,L}=\frac{ S_0}{(1-U^2)^{2\alpha(D-2)}}
\frac{ 2\alpha(D-2)(D-p-4)U^2-(1+U^2)(D-p-3) }
{1+U^2-2\alpha (D-2)(D-p-4)U^2},
 \\
&&\epsilon_T=\frac{T_0S_0}{(1-U^2)^2}\left(U^2+1-2\alpha(D-2)(D-p-4)U^2\right).
 \end{eqnarray} 
 For given $a,p$ this implies the existence of a phase transition for a critical value of $U$
 \begin{eqnarray}
 U_c=\frac{1}{\sqrt{2\alpha(D-2)(D-p-4)-1}},
 \end{eqnarray} 
 where $C_{Q,L}$  changes the sign. 
 All solutions with $U>U_c$ (thus with a large enough electric charge) are
 stable in a canonical ensemble. 
 Also, one can easily show that $C_{Q,L}$ is always negative for large  enough values of $a$
 (this property is shared by the solutions with a dependence on the extra-dimension $z$).

 This issue is especially interesting in 
 connection with the Gubser-Mitra (GM) conjecture \cite{Gubser:2000ec}, 
 that correlates
the dynamical and thermodynamical stability 
for systems with translational symmetry and infinite extent.
According to this conjecture, the appearance of a negative specific heat
of a black brane is related to the onset of a classical instability.
Supposing that this conjecture holds also for the charged solutions discussed in this paper,
one would expect all charged, nonextremal black branes to be unstable, 
irrespective of the value of the electric charge\footnote{The Gregory-Laflamme instability of $D=5$ 
electrically charged black strings has been discussed in the recent work \cite{Frolov:2009jr}.}.  
This is indeed the case when considering charged solutions in a 
grand canonical ensemble.
This agrees also with the results based on the map (\ref{tr-ch}).
As shown in \cite{Gregory:1993vy}, 
the vacuum uniform solutions with a given $L$ are unstable below a  critical mass $M_c^{(0)}$.
A consequence of the Harrison transformation is that the unstable mode of the vacuum
solutions would be present also for the nonextremal charged solutions.
Thus, for a given $U$, one expects a branch of  charged, nonuniform solutions to emerge 
for a critical value of the mass of the uniform solution 
$M_c=M_c^{(0)}(1+\frac{2(D-2)(D-p-4)}{(D-p-3)(1-U^2)}U^2\alpha).$

%%%%%%%%%%%%%%%%%%%%%%%%%%%%%%%%%%%%%%%%%%%%%%%%%%%%%%%%%%%%%%%%%%%%%%%%%%%%%%%%%%%%
\subsection{The $p=0$ case: charged black strings and caged black holes}
%%%%%%%%%%%%%%%%%%%%%%%%%%%%%%%%%%%%%%%%%%%%%%%%%%%%%%%%%%%%%%%%%%%%%%%%%%%%%%%%%%%%
To simplify the general picture,
we shall analyse for the rest of this section the simplest case of 
charged nonuniform black strings  and black holes \footnote{These are the only cases
where data is available for both the BHs and NUBSs branches.} 
 in $D=5,6$
KK theory, $i.e.$ with a single extra-direction, $p=0$.
As we shall see, the solutions have a rich structure already in this case.

Starting with the geometrical properties, we note that for the approach in this work,
the charged BHs and the NUBSs are described by the ansatz 
(\ref{metric}) (we recall that $f(r)=1-({r_0}/{r})^{d-4},~~R(r,z)=r$
for black strings, while $f(r)=1$ and $R(r,z)$ has a complicated expression 
for caged BHs, see Appendix A).
As seen from (\ref{new-funct}), the inclusion of an electric charge changes the shape of the functions $A,B,C$. 

In Figure 1 we exhibit
the metric functions  $-g_{tt}/f=e^{2A}$, $g_{zz}=e^{2B}$, and
$g_{\theta \theta}/r^2=e^{2C}$ for a typical
$D=6$ vacuum NUBS solution ($U=0$) and a charged one with $U=0.85$ in Einstein-Maxwell
theory $(a=0)$. The nonuniformity parameter of these solutions as defined by (\ref{lambda}) 
is $\lambda=0.65$. 
 
 To obtain a more quantitative picture of the NUBS metric functions,
we exhibit $A$, $B$, and
$C$ in Figure 2 for three values
of $U$ and two fixed values of $z$.  
The solutions there have a dilaton coupling constant $a=1$ and $\lambda=0.93$.
The corresponding
matter functions for the same configurations are shown in Figure 3.
One can see that although the shape of the functions is qualitatively the same in both cases, 
the precise values
taken by $A,B,C$ in the region close to the horizon are very different.
However, in both cases the metric functions exhibit an extremum at the horizon for $z=L/2$.
 
%%%%%%%%%%%%%%%%%%%%%%%%%%%%%%%%%%%%%%%%%%%%%%%%%%%%%%
\setlength{\unitlength}{1cm}
\begin{picture}(15,18)
\put(-1,0){\epsfig{file=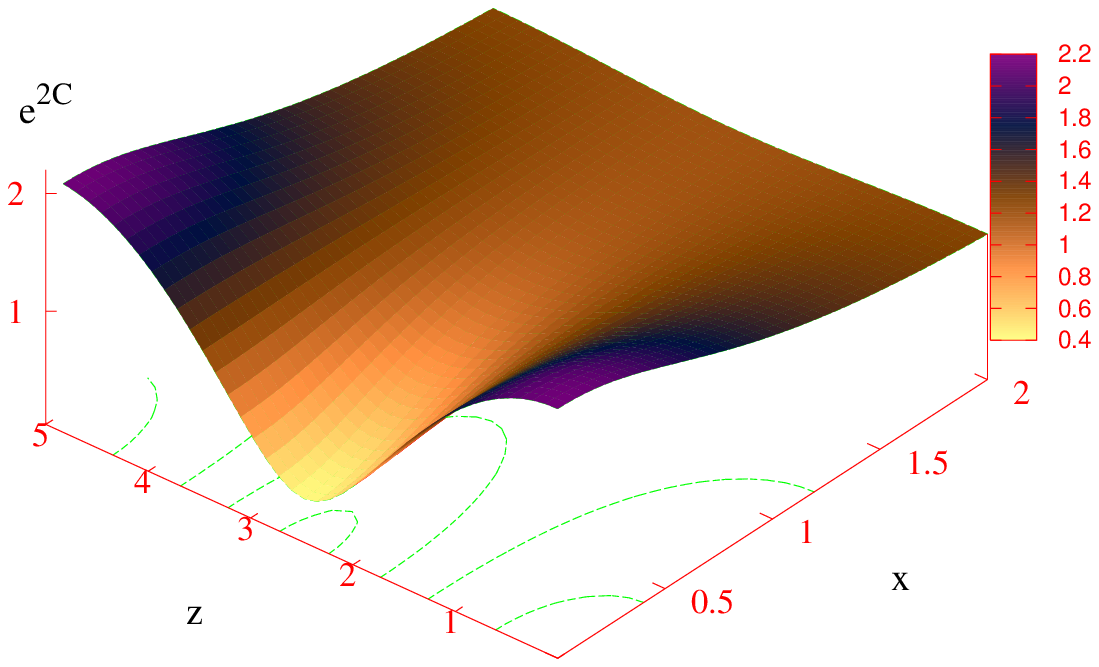,width=8cm}}
\put(7,0){\epsfig{file=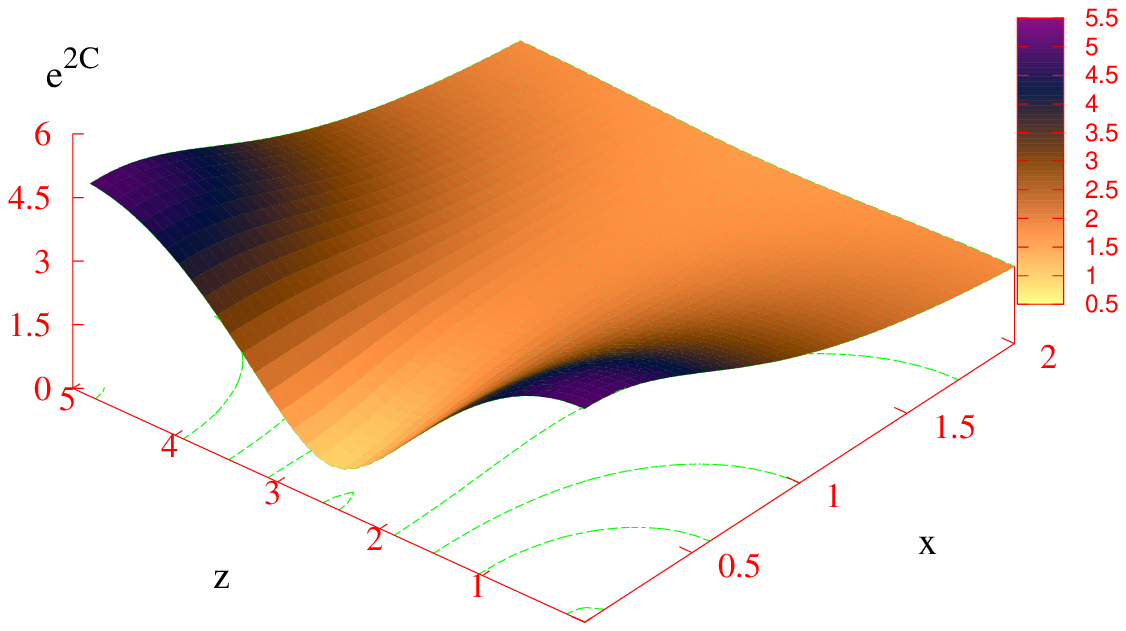,width=8cm}}
\put(-1,6){\epsfig{file=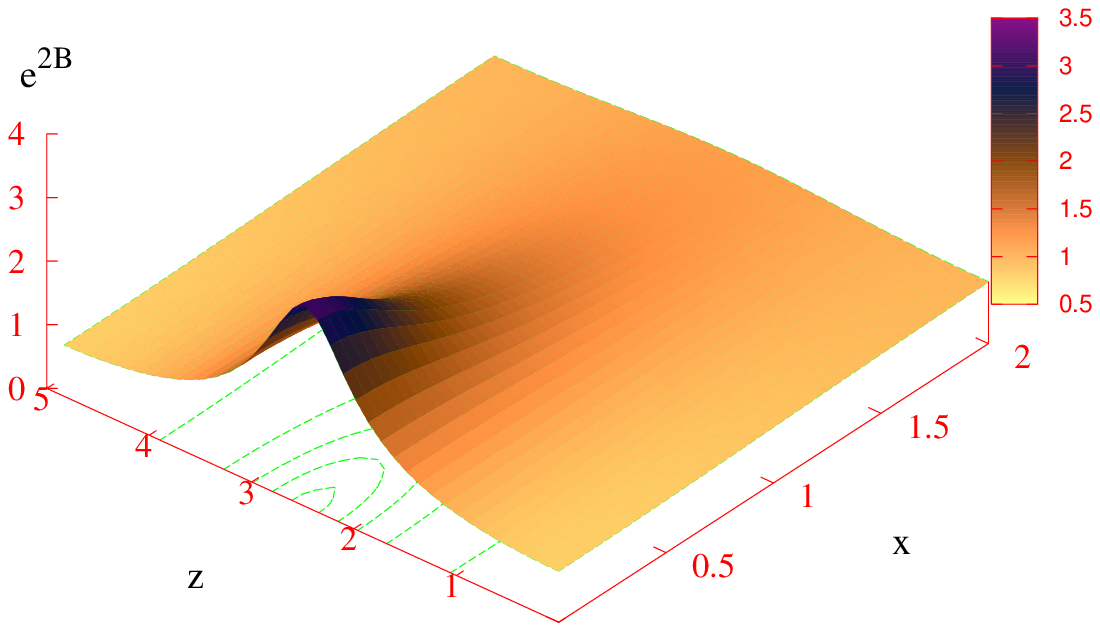,width=8cm}}
\put(7,6){\epsfig{file=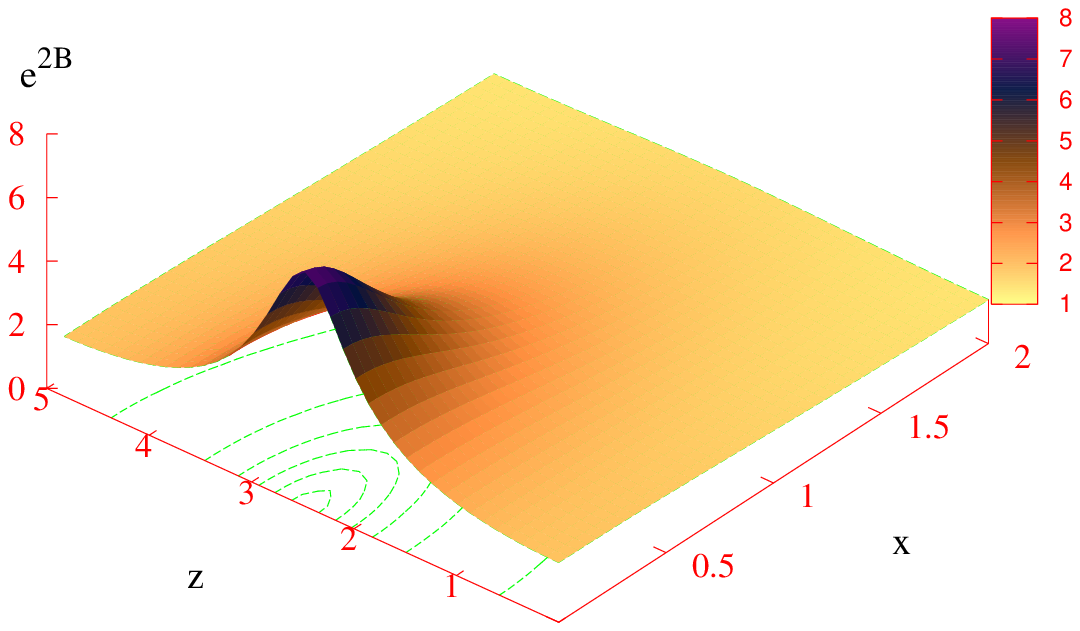,width=8cm}}
\put(-1,12){\epsfig{file=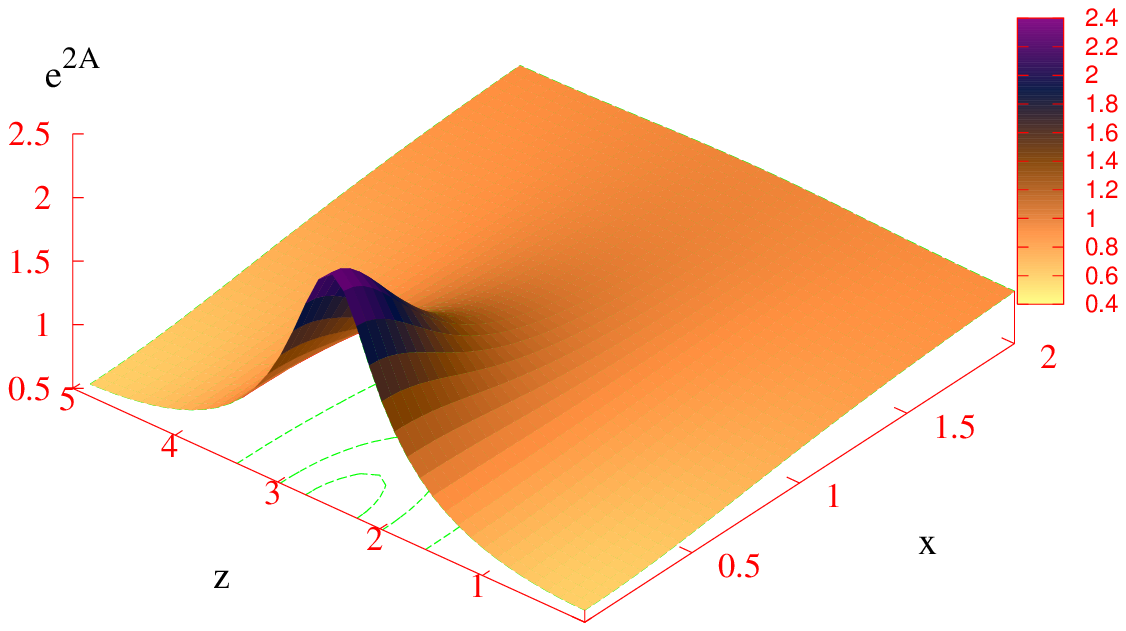,width=8cm}}
\put(7,12){\epsfig{file=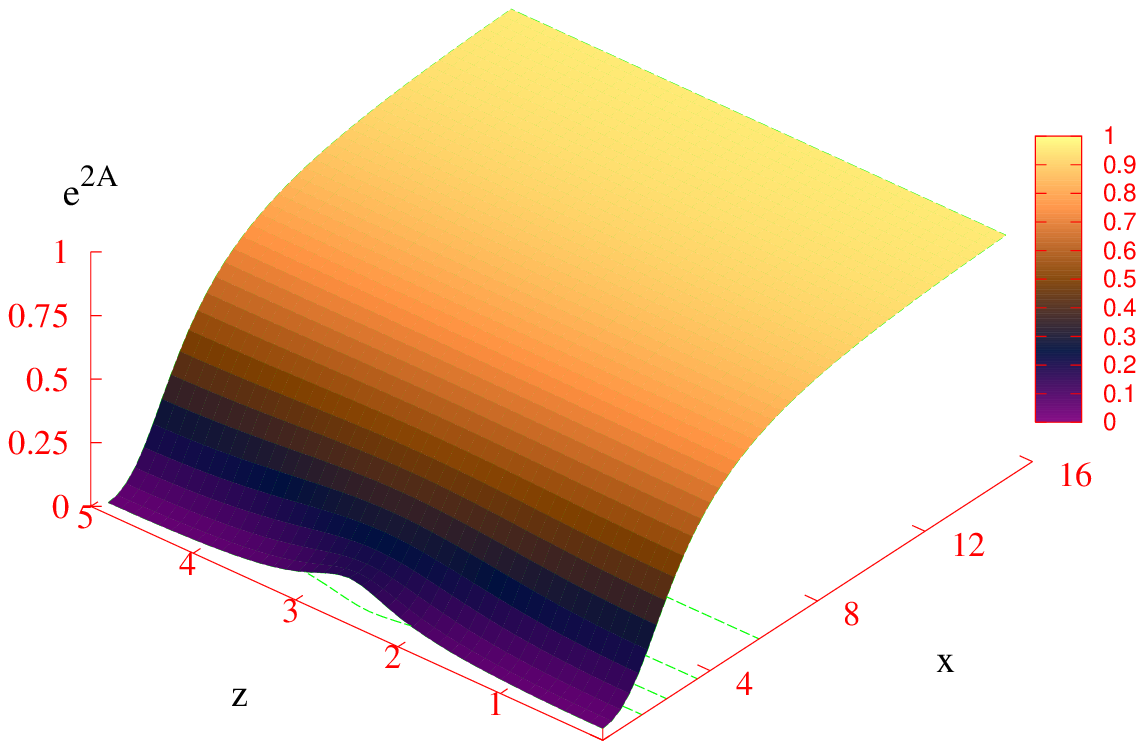,width=8cm}}
\end{picture}
\\
\\
{\small {\bf Figure 1.}
The metric functions 
$e^{2A}$, $e^{2B}$, and $e^{2C}$ of a typical $D=6$ nonuniform string solution  
are shown as functions of the   coordinates $x= \sqrt{r^2-r_0^2},z$
for a vacuum solution (left) and a charged solution with  $U=0.85$ (right).
 (Note the different scale of the functions in these cases.)
 }
%%%%%%%%%%%%%%%%%%%%%%%%%%%%%%%%%%%%%%%%%%%%%%%%%%%%%%
\\
\\
The proper length of the $z-$circle increases with $U$, for any finite value of $U$.
(The input parameters of the solutions in Figures 1-3 are $r_0=1,L=4.9516$).

An interesting point here concerns
 the deformation of the horizon of the NUBS solutions.
One can see from (\ref{metric-new}) that although the charged solutions  have 
different values of $R_{max}$, $R_{min}$
from the vacuum case, the nonuniformity parameter $\lambda$ as defined in
(\ref{lambda}) does not depend on the value of the chemical potential (or of the electric charge).

%%%%%%%%%%%%%%%%%%%%%%%%%%%%%%%%%%%%%%%%%%%%%%%%%%%%%%
\setlength{\unitlength}{1cm}
\setlength{\unitlength}{1cm}
\begin{picture}(15,18)
\put(-1,0){\epsfig{file=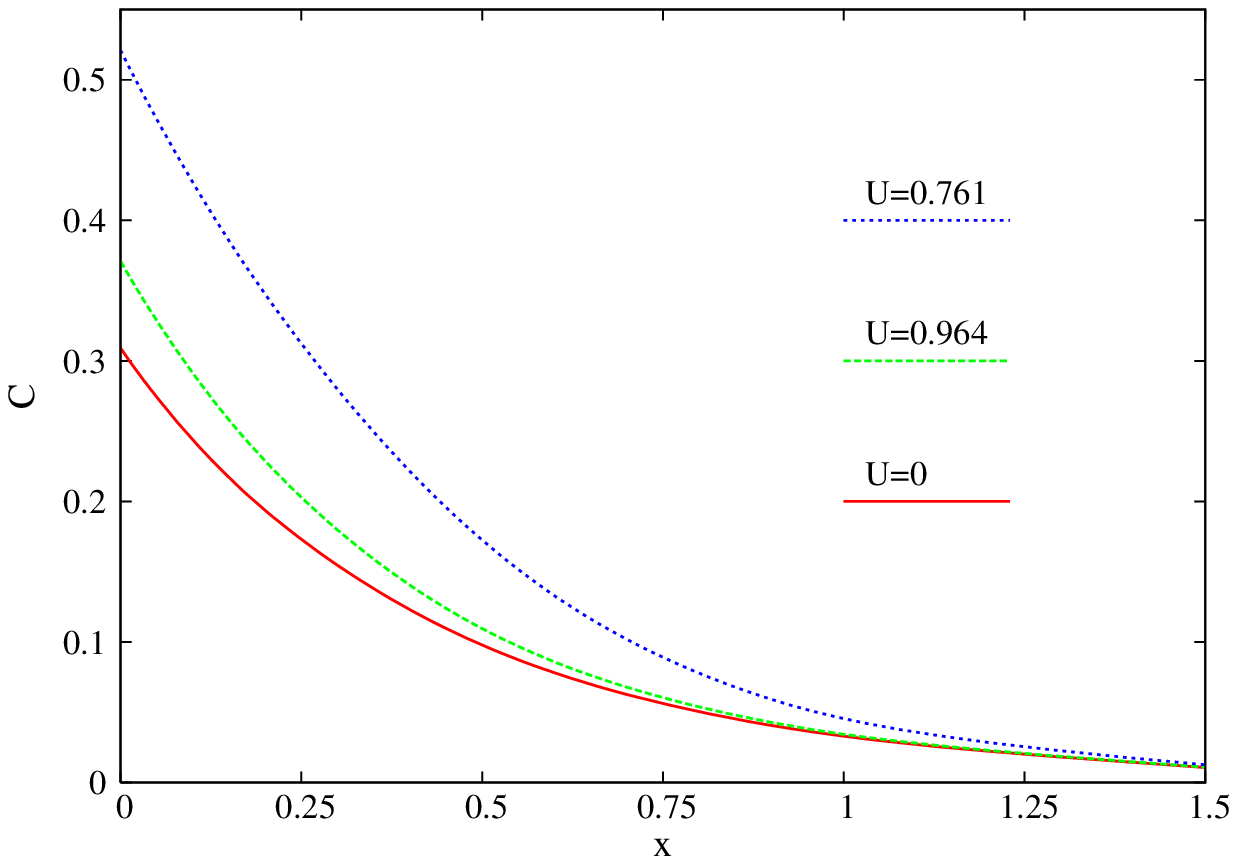,width=8cm}}
\put(7,0){\epsfig{file=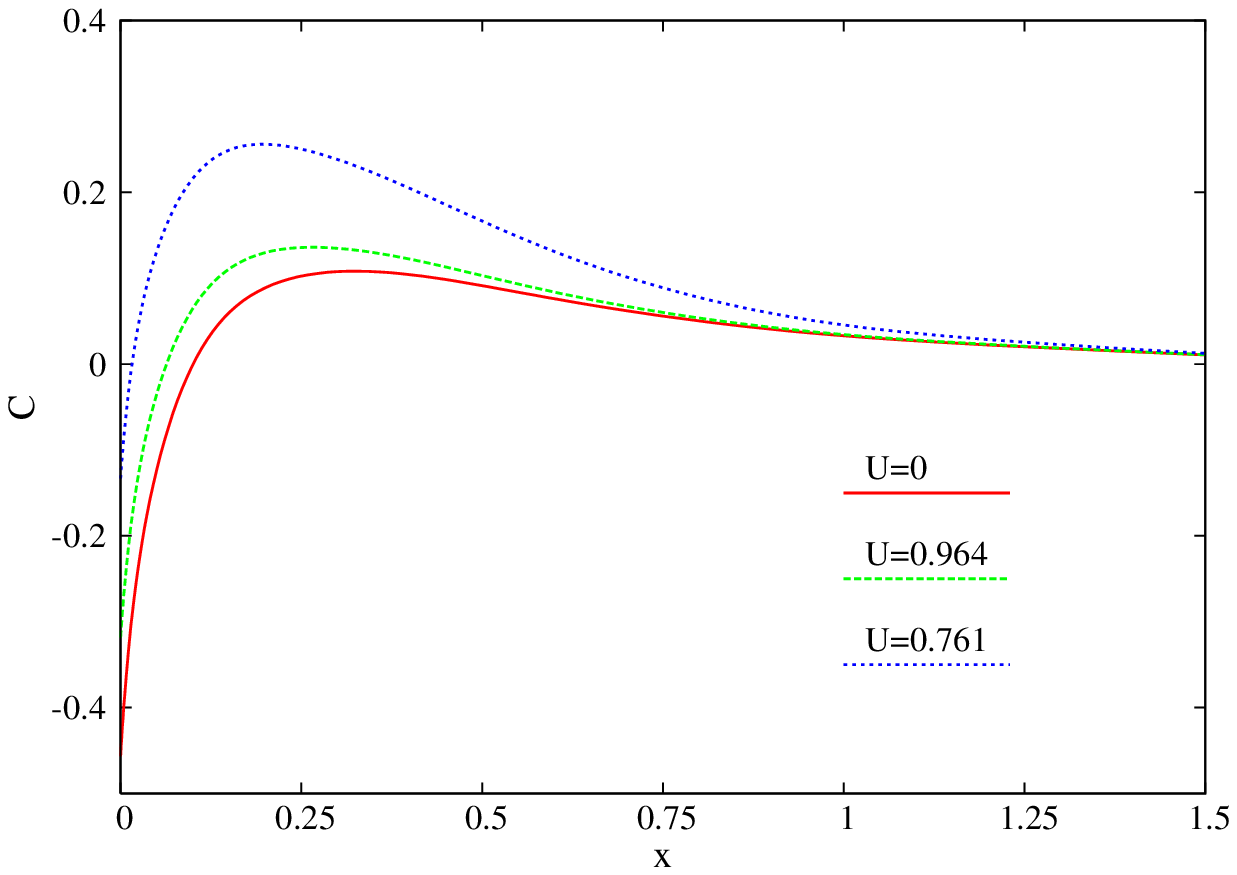,width=8cm}}
\put(-1,6){\epsfig{file=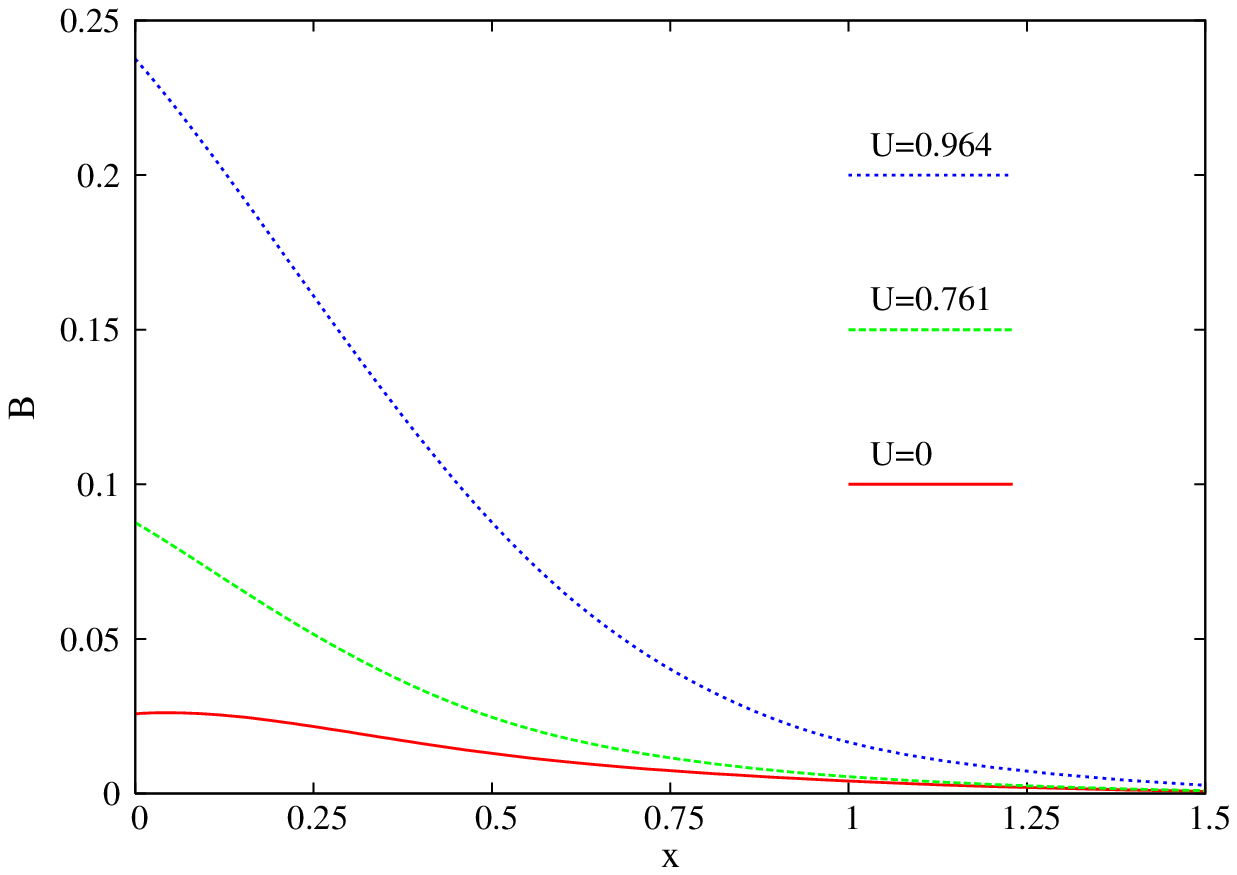,width=8cm}}
\put(7,6){\epsfig{file=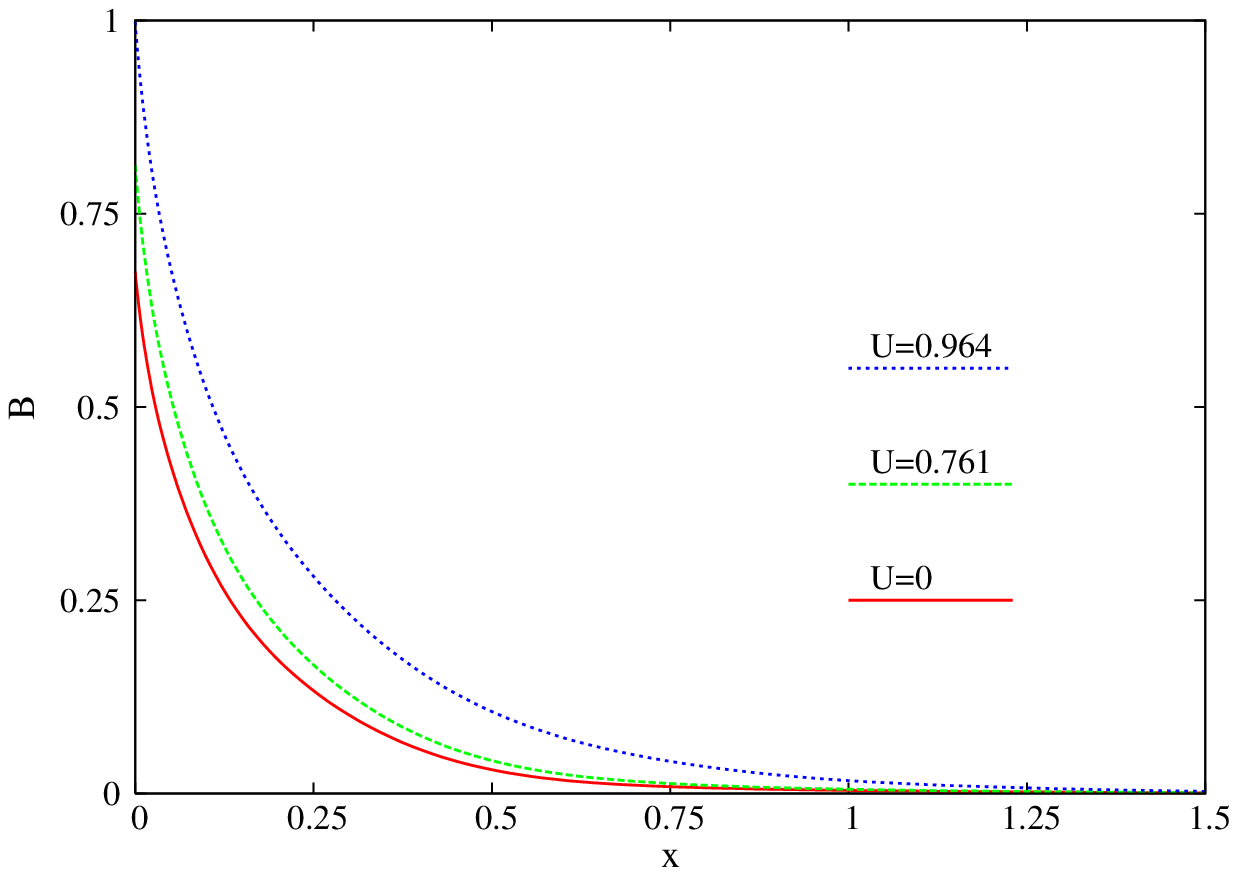,width=8cm}}
\put(-1,12){\epsfig{file=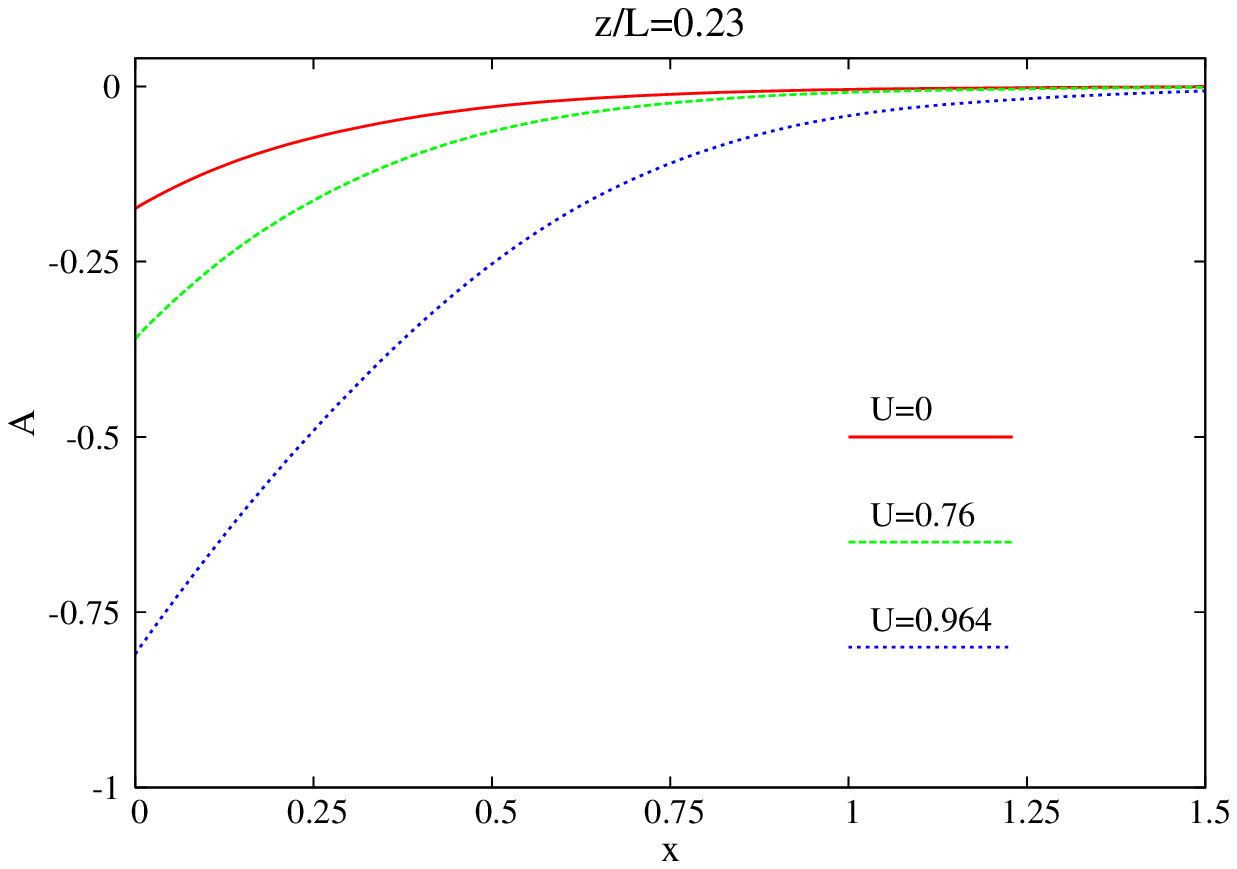,width=8cm}}
\put(7,12){\epsfig{file=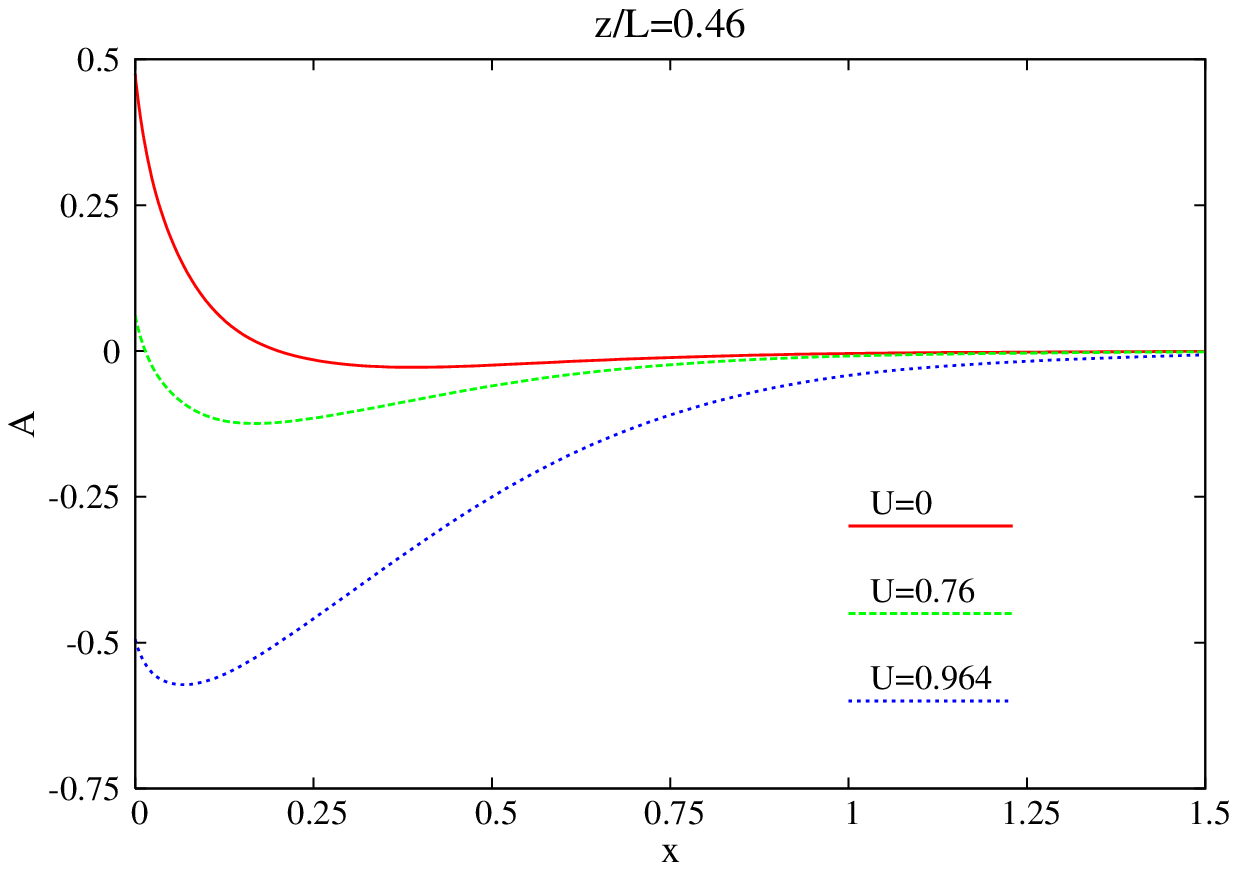,width=8cm}}\end{picture} 
\\
\\
{\small {\bf Figure 2.}
The metric functions 
$A$, $B$, and $C$
of $D=6$ charged nonuniform black string solutions 
are shown as functions of the  radial coordinate $x= \sqrt{r^2-r_0^2}$,
for  three values of the parameter $U$.
} 
 \vspace{0.4cm}
%%%%%%%%%%%%%%%%%%%%%%%%%%%%%%%%%%%%%%%%%%%%%%%%%%%%%% 
 \\
Therefore, the
qualitative features of the horizon geometry remain the same as in the vacuum case.
In particular, the parameter $\lambda$ approaches very large values 
for solutions constructed using seed NUBS close to the transition point
(this occurs for any value of $U$).

%%%%%%%%%%%%%%%%%%%%%%%%%%%%%%%%%%%%%%%%%%%%%%%%%%%%%%%%%%%%%%%%%%%%
\setlength{\unitlength}{1cm}
\begin{picture}(8,6)
\put(-1,0.0){\epsfig{file=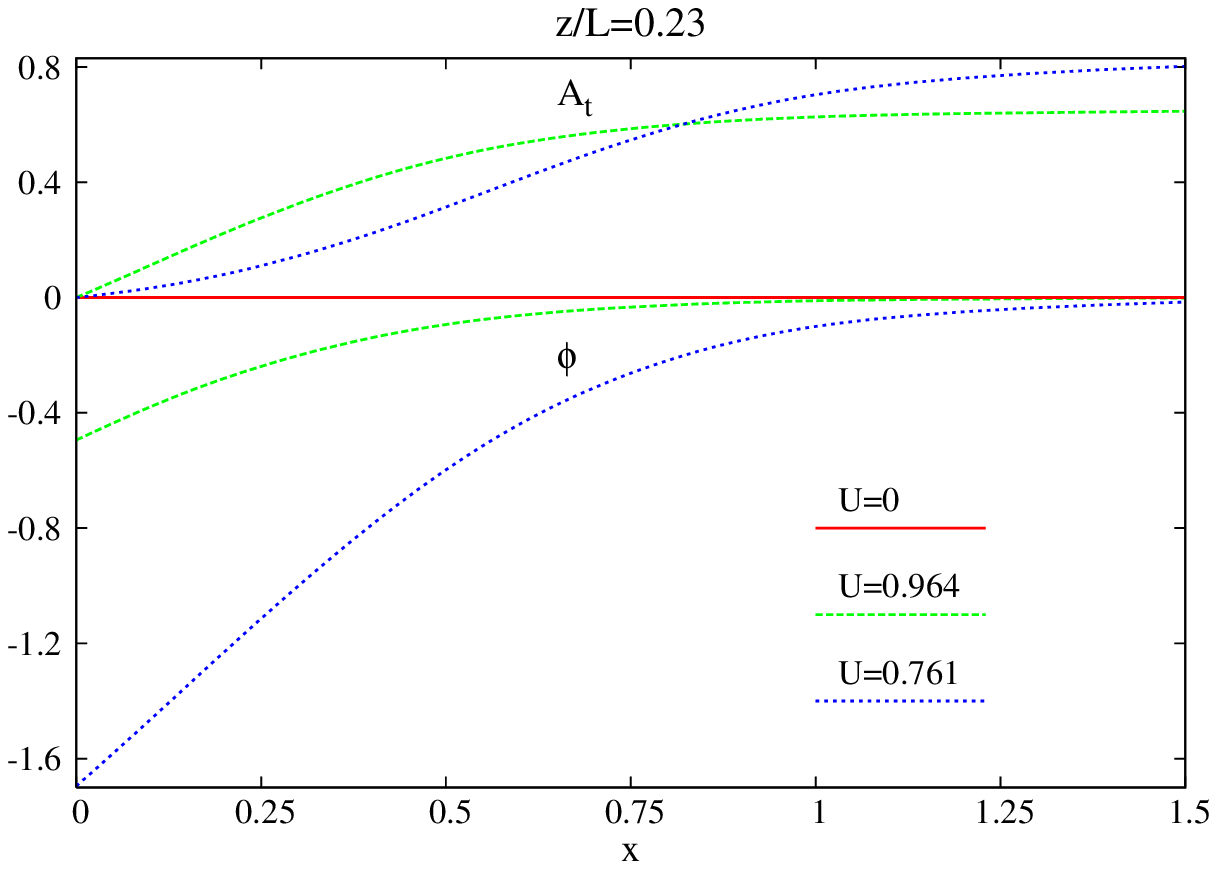,width=8cm}}
\put(7,0.0){\epsfig{file=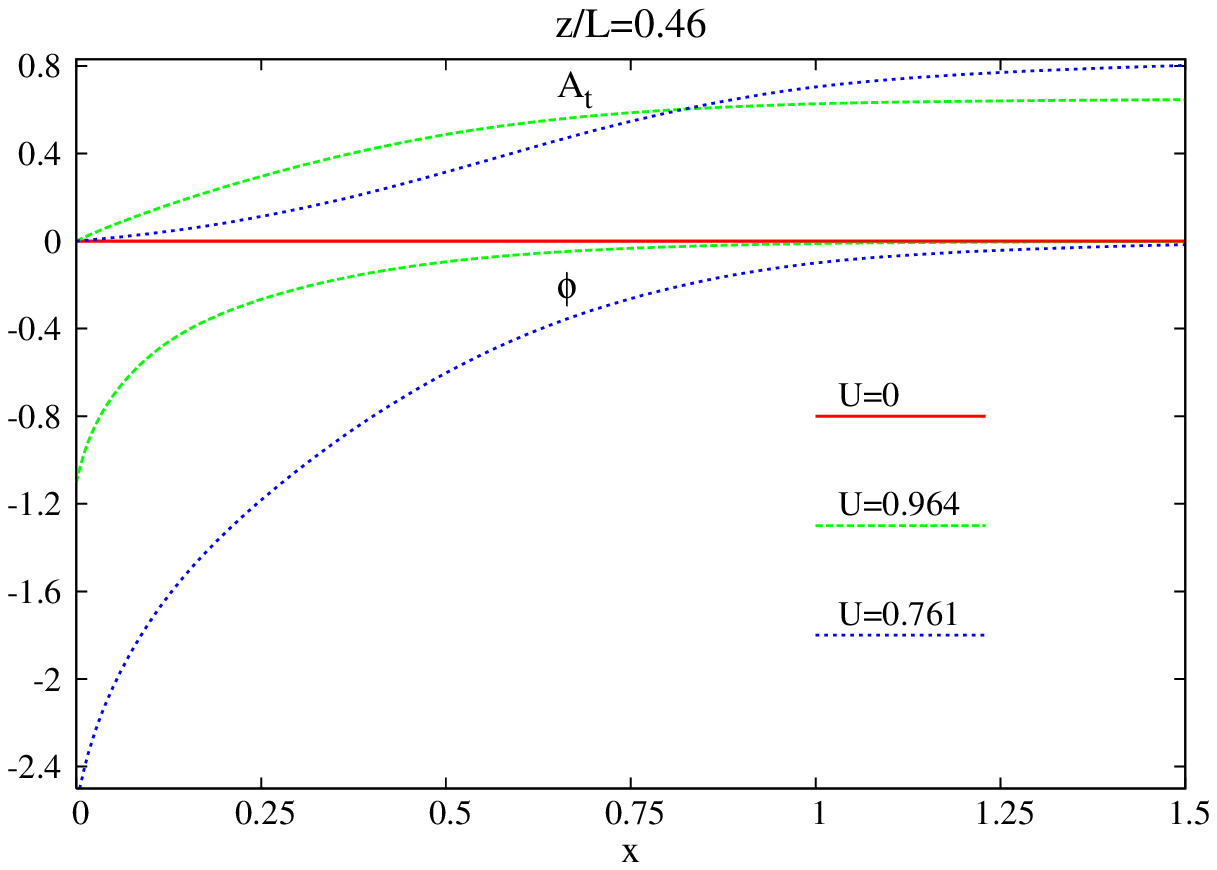,width=8cm}}
\end{picture}
\\
\\
{\small {\bf Figure 3.}
 The dilaton $\phi$ and the electric potential $A_t$
 are shown for the same solutions exhibited in Figure 2.
}
\vspace{0.5cm}
%%%%%%%%%%%%%%%%%%%%%%%%%%%%%%%%%%%%%%%%%%%%%%%%%%%%%%%%%%%%%%%%%%%
\\

The geometrical properties of the 
caged BH solutions can be discussed in a similar way.
In Figure 4  we exhibit
the metric functions  $e^{2A}$, $e^{-2B}$  and
$e^{2C}$ for a typical
$D=6$ vacuum BH solution ($U=0$) and a charged one with $U=0.85$ in EM
theory $(a=0)$. 
The position of the event horizon there was $r_0=-1$ (see Appendix A for more details on 
 the seed solutions).
As a generic feature, for any value of the parameter $U$,
the ratio between $R_{eq}$ and $L_{axis}$ remains the same as
in the vacuum case.
At a technical level, this is a consequence of the fact that the Harrison 
transformation (\ref{ansa1}) involves just a conformal transformation of the 
spacelike part $g_{\alpha\beta}dx^\alpha dx^\beta$ of the seed geometry.

This observation is relevant in connection with
the issue of a topology changing transition between 
the black hole and the NUBS branches. 
Since, from (\ref{tr-ch}),
 adding an electric charge implies just a recombination of the global charges of the system,
we argue that the picture one finds here would be qualitatively
 similar to that valid for vacuum solutions.

To clarify this point, we show
in Figure 5 the mass $M$ and temperature $T$
versus the relative string tension $n$, 
for the nonuniform string branch as well as for the
black hole branch\footnote{The same picture is found when considering the $(n,M)$ diagram in
$D=5$ and the  $(n,T)$ plot in $D=6$, respectively.
For all values of $U$, we note qualitative agreement of the shape
and the relative position
of the charged nonuniform string branch and the charged  black hole branch
in five dimensions with the shape and relative position
of the corresponding branches in six dimensions. The backbending noticed 
in \cite{Kleihaus:2006ee} for the vacuum case is present also for charged NUBS,
 both in five and in six dimensions.}, for several values of the parameter $U$.
The black hole data are taken from \cite{Kudoh:2004hs}.
(For convenience,  we normalize in this subsection the thermodynamic quantities 
by their values for the corresponding charged 
UBS with the same value of $U$.)

%%%%%%%%%%%%%%%%%%%%%%%%%%%%%%%%%%%%%%%%%%%%%%%%%%%%%%
\setlength{\unitlength}{1cm}
\begin{picture}(15,18)
\put(-1,0){\epsfig{file=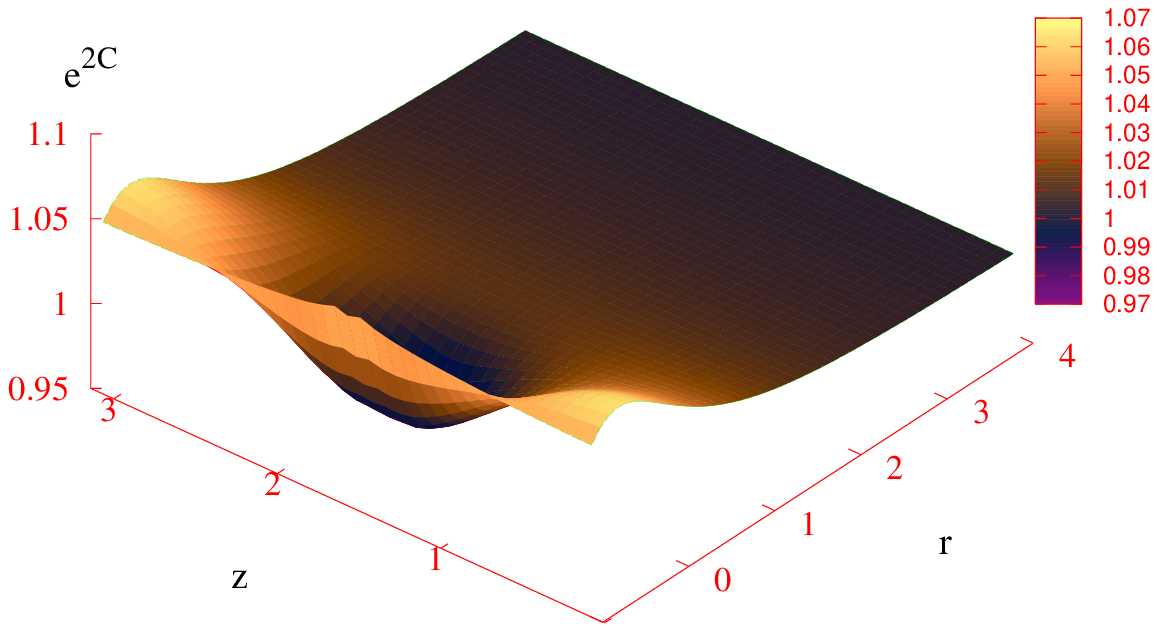,width=8cm}}
\put(7,0){\epsfig{file=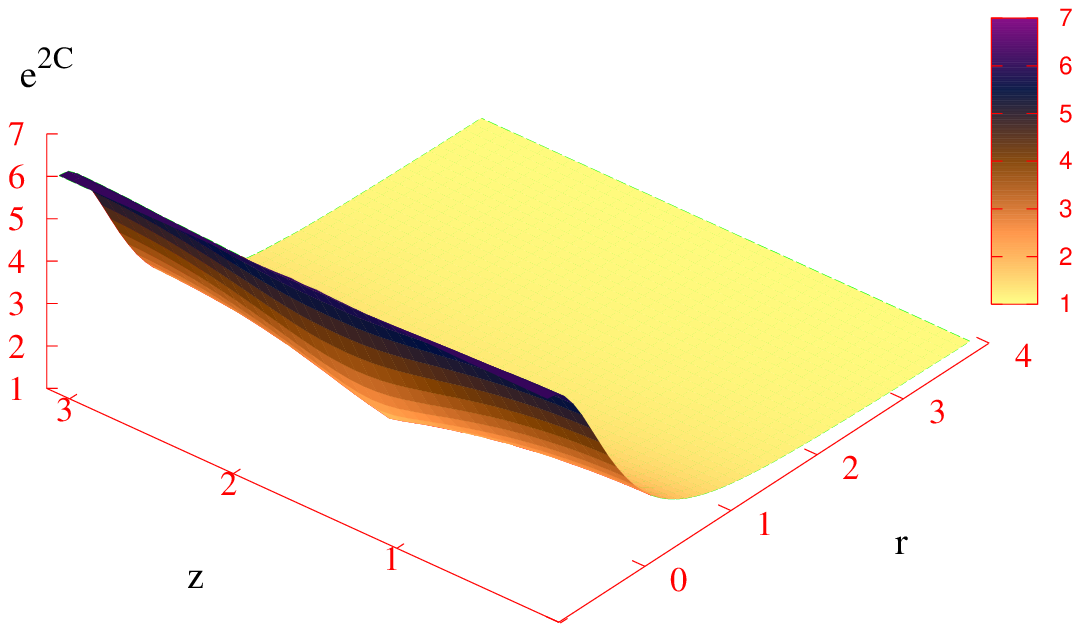,width=8cm}}
\put(-1,6){\epsfig{file=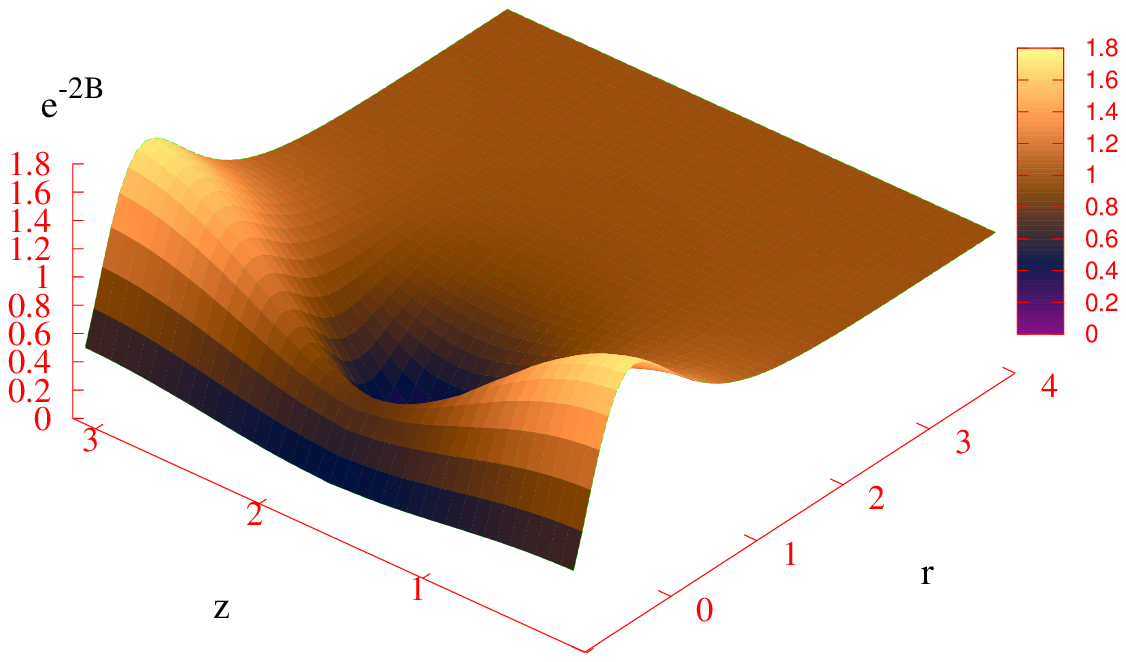,width=8cm}}
\put(7,6){\epsfig{file=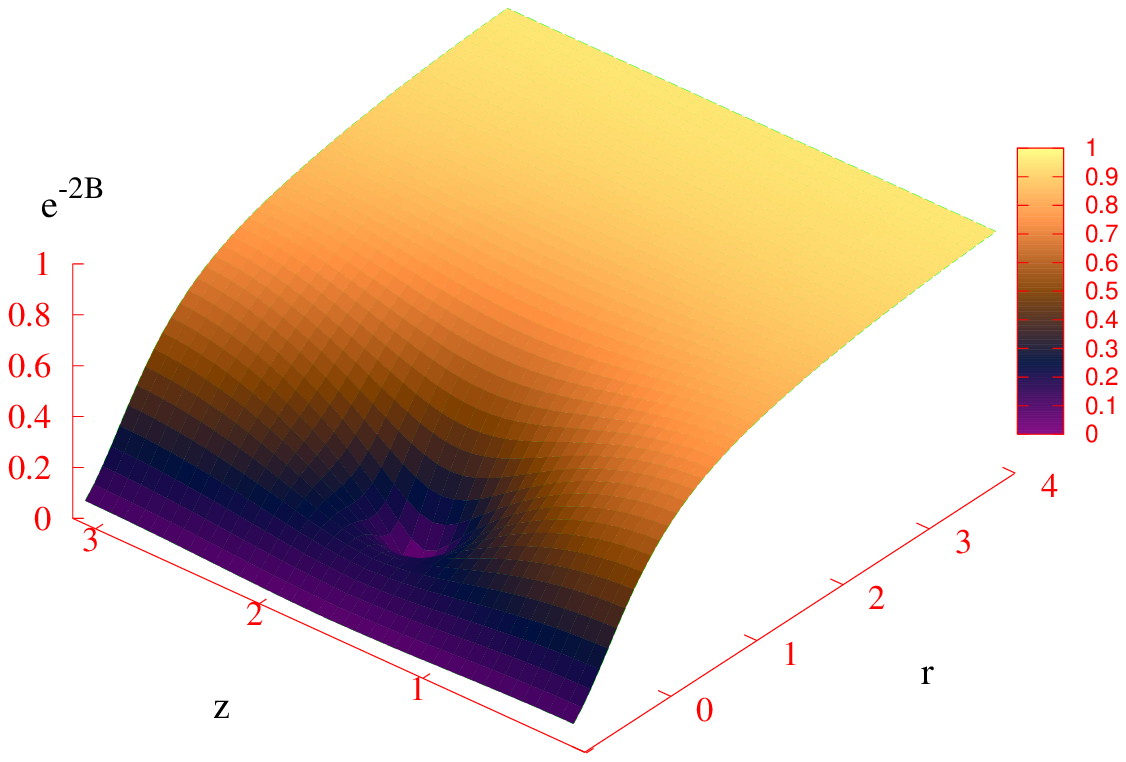,width=8cm}}
\put(-1,12){\epsfig{file=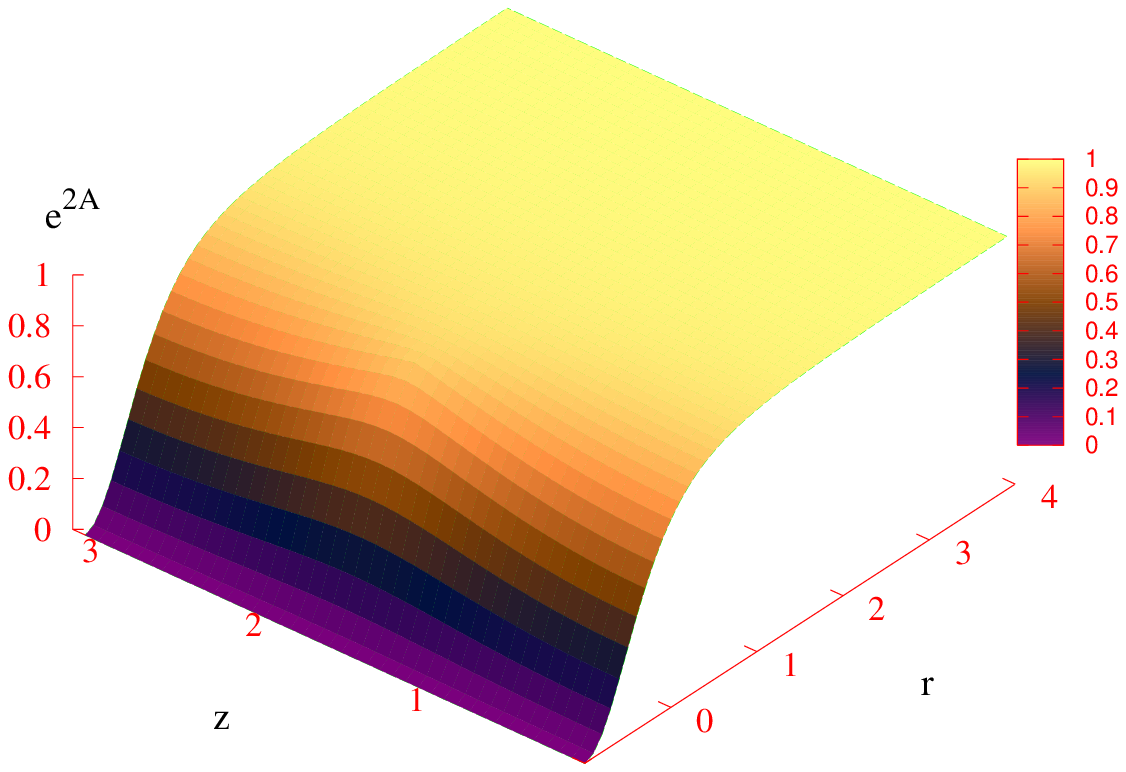,width=8cm}}
\put(7,12){\epsfig{file=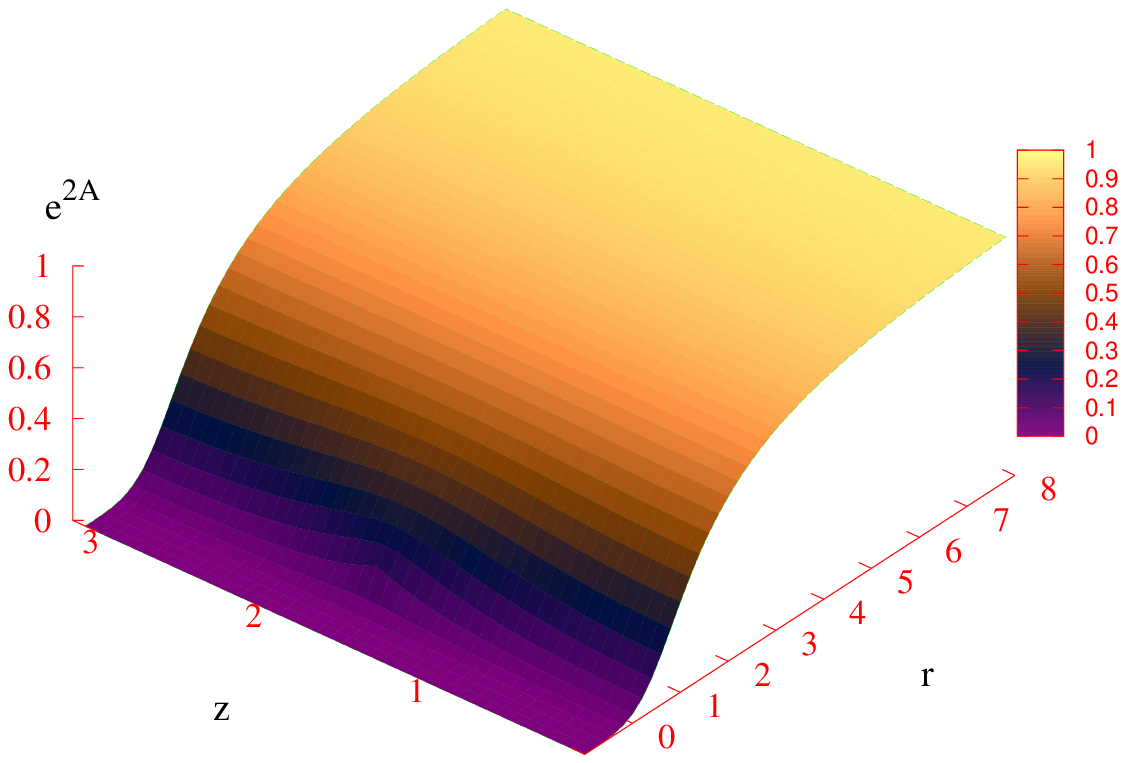,width=8cm}}
\end{picture}
\\
\\
{\small {\bf Figure 4.}
The metric functions 
$e^{2A}$, $e^{-2B}$, and $e^{2C}$ of a typical $D=6$ KK black hole solution  
are shown as functions of the   coordinates $r,z$
for a vacuum solution (left) and a charged solution with  $U=0.964$ (right).
}
\vspace{0.5cm}
%%%%%%%%%%%%%%%%%%%%%%%%%%%%%%%%%%%%%%%%%%%%%%%%%%%%%%

One can see that we have all reasons to expect that the topology changing transition scenario
proposed in the vacuum case, holds here also.
However, as seen in Figure 5, a nonzero electric charge  translates into a different position in the
($\bar M, \bar n$) plane (and ($\bar T_H, \bar n$),

%%%%%%%%%%%%%%%%%%%%%%%%%%%%%%%%%%%%%%%%%%%%%%%%%%%%%%
\vspace{0.5cm}
\setlength{\unitlength}{1cm}
\begin{picture}(8,6)
\put(-1,0.0){\epsfig{file=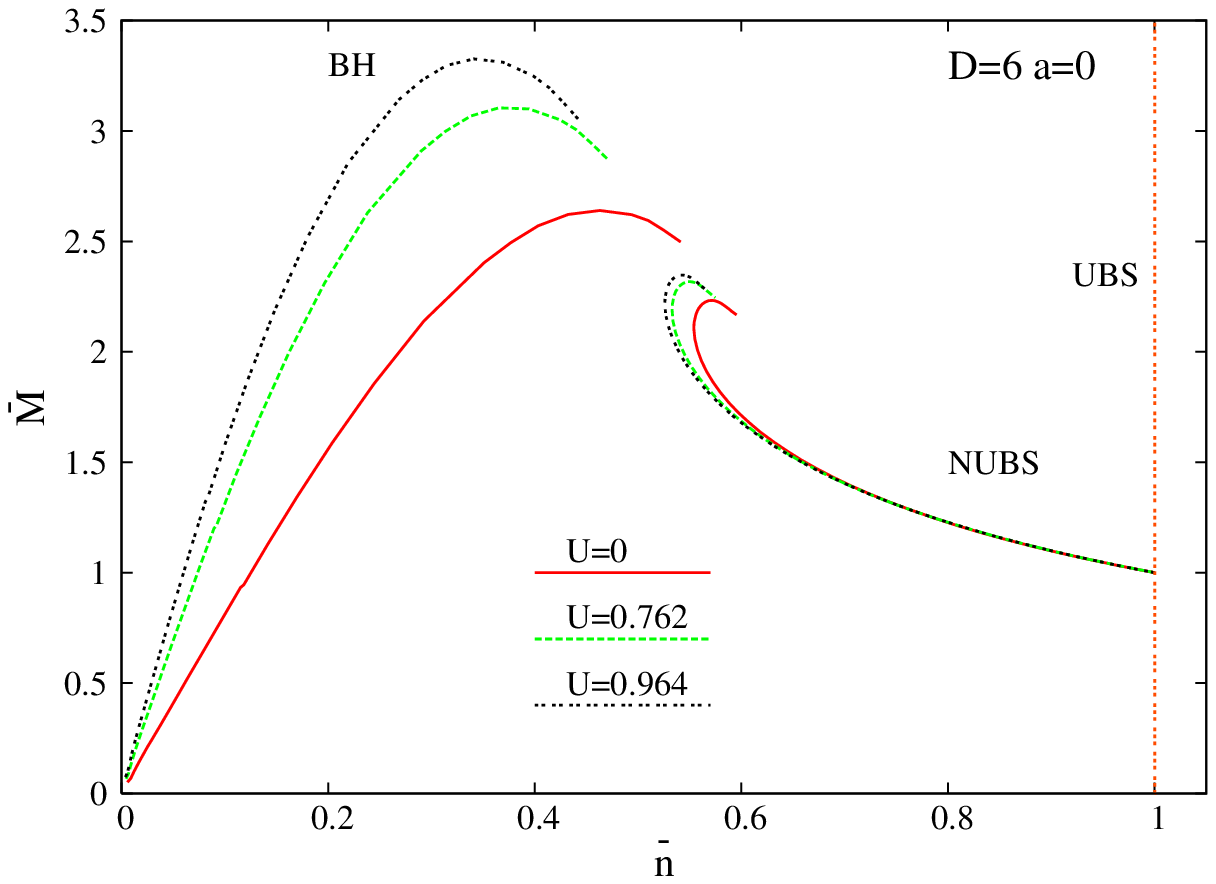,width=8cm}}
\put(7,0.0){\epsfig{file=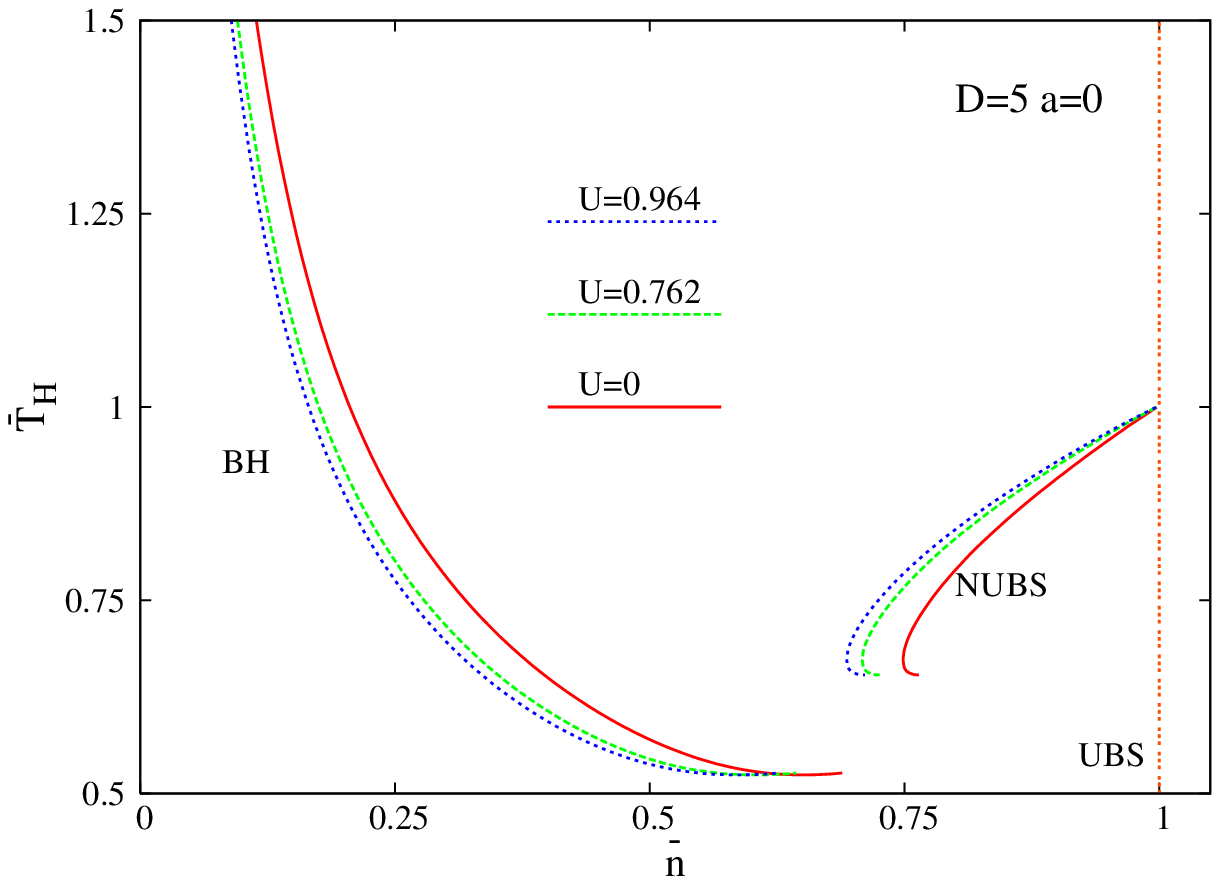,width=8cm}}
\end{picture}
\\
\\
{\small {\bf Figure 5.}
 {\it Left:} The mass $M$ of the $D=6$ charged nonuniform string
and black hole branches
is shown versus the relative string tension $n$. 
 {\it Right:} The Hawking temperature $T_H$ 
of the $D=5$  charged nonuniform string
and black hole branches
is shown versus the relative string tension $n$ (these plots are given for three values of $U$).
$\bar M$, $\bar n$ and $\bar T_H$ correspond to the quantities normalized by the corresponding 
values of the
 uniform string solutions with the same values of $U$.  
}
\vspace{0.5cm}
%%%%%%%%%%%%%%%%%%%%%%%%%%%%%%%%%%%%%%%%%%%%%%%%%%%%%%
\\
respectively) of the {\it putative} transition point. We have also noticed that as $U$
increases towards its maximal value,
the $\bar M(\bar n)$ curves converge to a critical one,
which is close to that corresponding to $U=0.964$ in Figure 5.
The same holds for the $(\bar n,\bar T_H)$ diagram.

The general thermodynamical properties of the charged solutions
are rather complicated as compared to the vacuum case, 
with a nontrivial dependence on the dilaton coupling constant $a$.
However, the qualitative features are again the same for both $D=5$ and $D=6$ solutions.

%%%%%%%%%%%%%%%%%%%%%%%%%%%%%%%%%%%%%%%%%%%%%%%%%%%%%%%%%%%%%%%%%%%%%%

\setlength{\unitlength}{1cm}
\begin{picture}(8,6)
\put(-1,0.0){\epsfig{file=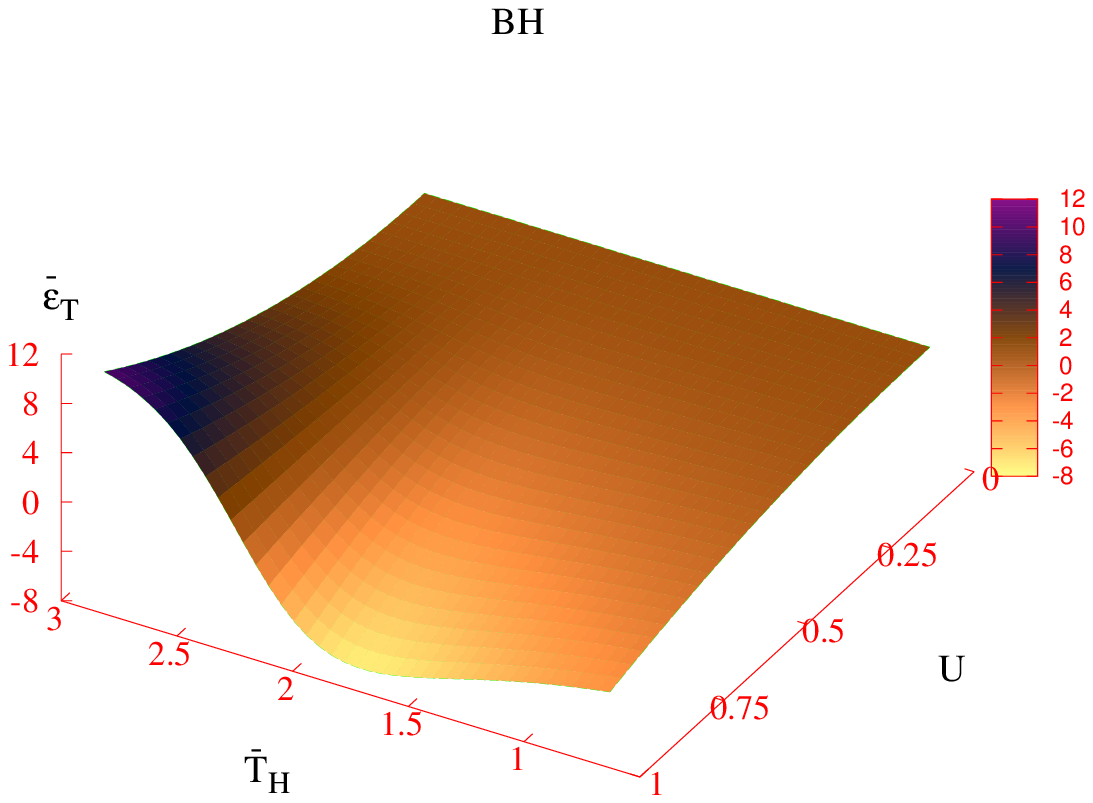,width=9cm}}
\put(7,0.0){\epsfig{file=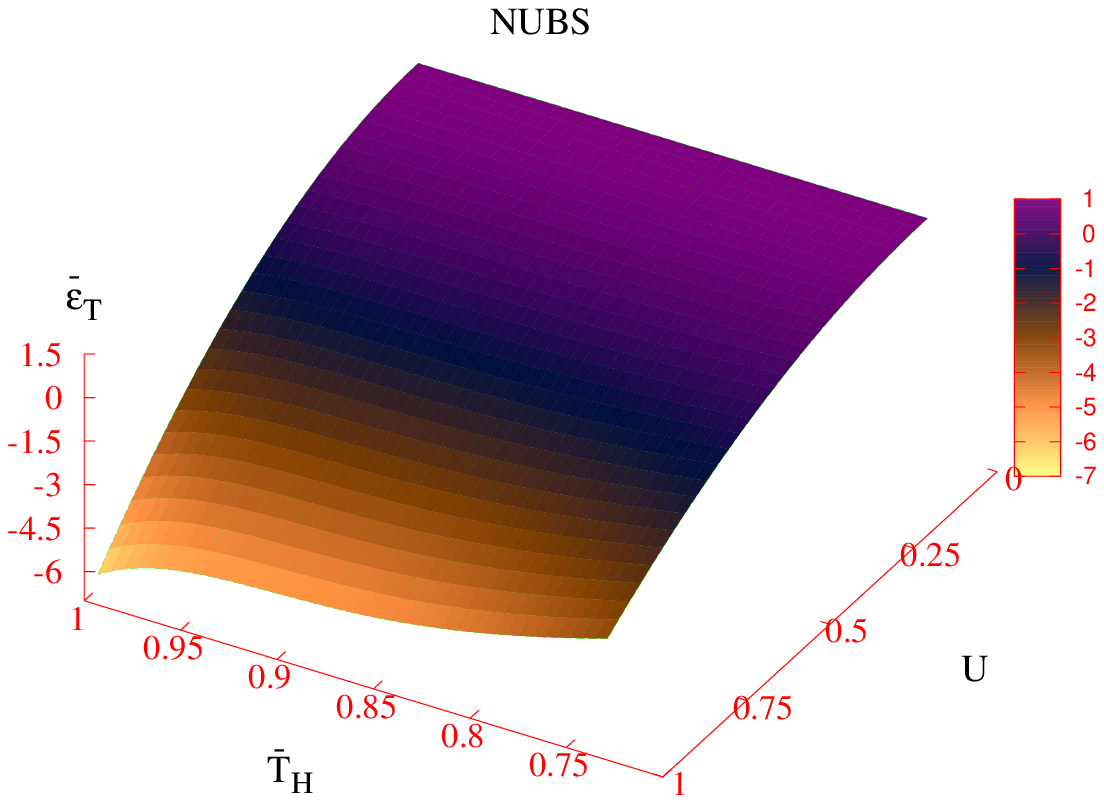,width=9cm}}
\end{picture}
\\
\\
{\small {\bf Figure 6.}
 The scaled isothermal electric permitivity 
	is shown as a function of the normalized temperature and the parameter $U$ 
	for the $D=6$ charged
	caged black holes (left) and nonuniform black string strings (right). 
}
\vspace{0.5cm}
%%%%%%%%%%%%%%%%%%%%%%%%%%%%%%%%%%%%%%%%%%%%%%%%%%%%%%%%%%%%%%%

\setlength{\unitlength}{1cm}
\begin{picture}(8,6)
\put(-1,0.0){\epsfig{file=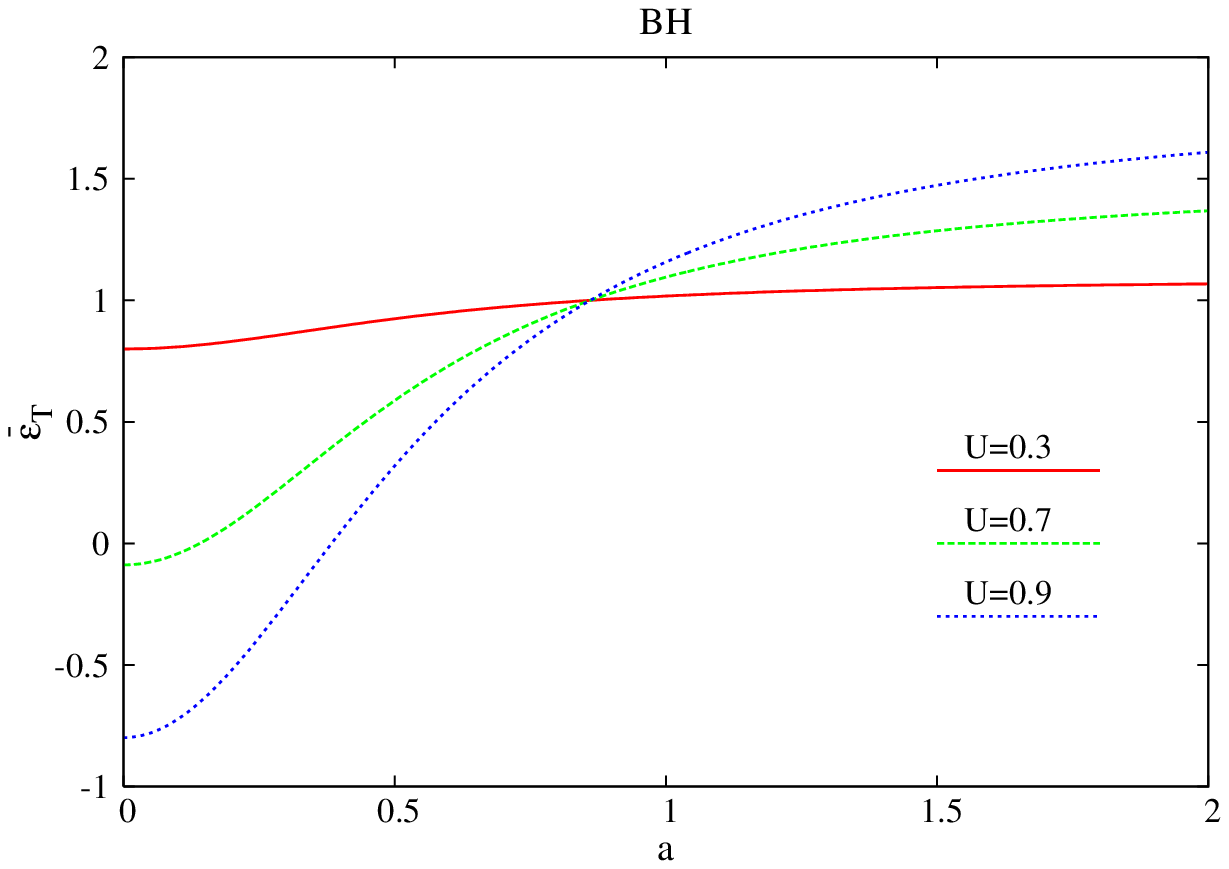,width=8cm}}
\put(7,0.0){\epsfig{file=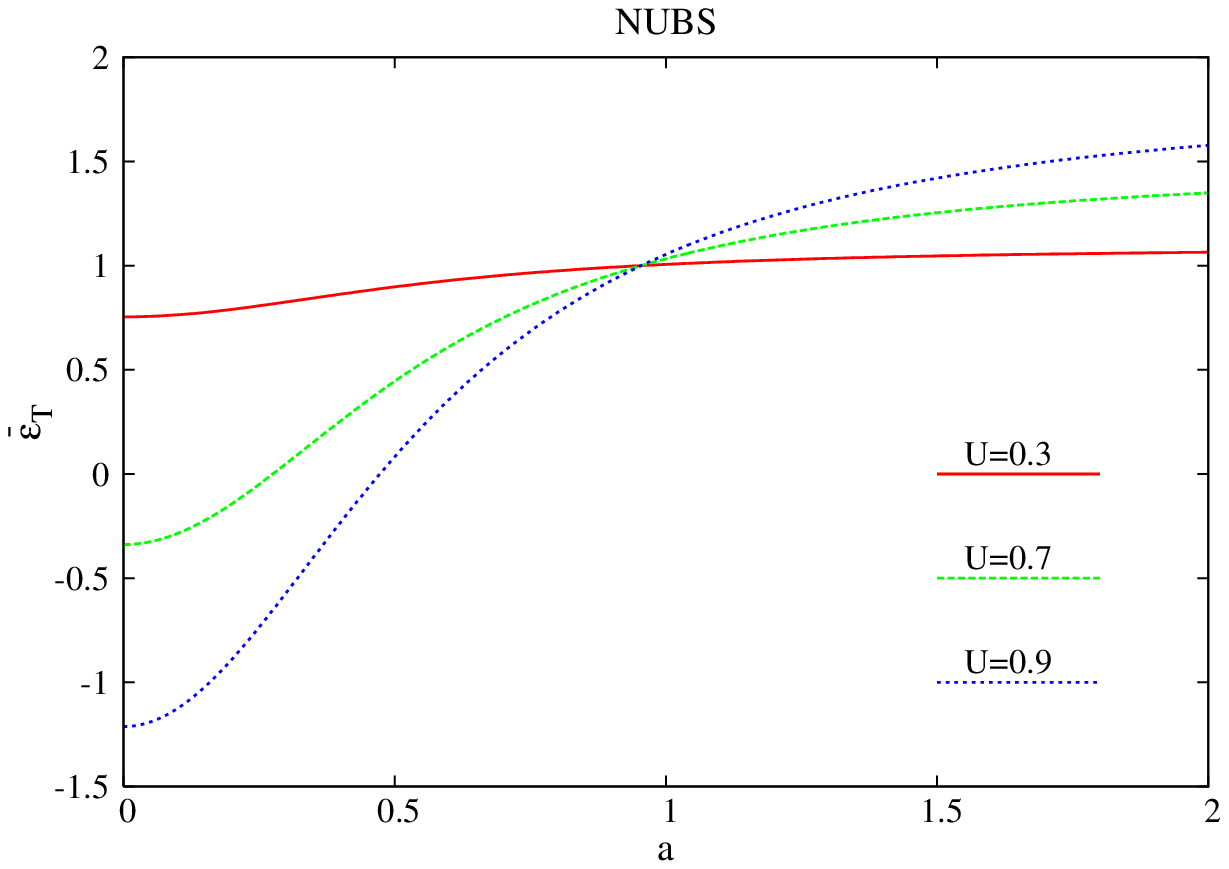,width=8cm}}
\end{picture}
\\
\\
{\small {\bf Figure 7.}
  The scaled isothermal electric permitivity 
	is shown as a function of the dilaton coupling  constant $a$
 for the $D=5$ charged
	caged black holes (left) and nonuniform black strings (right). 
 The plots here correspond to a fixed 
	 normalized temperature $\bar T_H=0.7$ and several values of the chemical potential $U$.
}
\vspace{0.5cm}
%%%%%%%%%%%%%%%%%%%%%%%%%%%%%%%%%%%%%%%%%%%%%%%%%%%%%%%%%%%%%%%%%%%%%%%

Also, as argued above, both the BHs and the NUBSs are thermodynamically 
unstable in a grand canonical ensemble\footnote{ As discussed in \cite{Kleihaus:2006ee}, an unexpected feature of the NUBS
 solutions is the occurence of a new branch at a critical (minimal) value of the relative string tension, 
close to transition point. Along this  branch (which has a small extension in the $(M,n)-$plane, see Figure 5),
the mass decreases with $n$, while the specific heat is still negative.
 In what follows, to simplify the general picture, we shall 
 ignore this secondary branch for all plots of  the specific heat and isothermal electric permitivity.
The same holds for caged BH solutions, 
the last points for the data in \cite{Kudoh:2004hs} indicating the existence of
a (small) secondary branch  close to the transition point, see Figure 5 
(this feature was confirmed by the recent results in \cite{Headrick:2009pv}). 
This secondary BH branch was also ignored when discussing $C_{Q,L}$ and $\epsilon_T$.
}. 
Here we start by showing in Figure 6 the scaled
isothermal electric permitivity\footnote{$\bar \epsilon_T$ has the same sign as the 
isothermal electric permitivity and stays finite as $U\to 1$.}
\begin{eqnarray}
\label{eps-b}
\bar \epsilon_T = U^2+1+2\alpha(D-2)U^2(1+\frac{C_{Q,L}^0}{S_0}) 
\end{eqnarray}
for caged BHs and NUBS solutions as a function of the normalized temperature and 
the parameter $U$.  
These plots correspond  to a value $a=0$ of the dilaton coupling constant, $i.e.$ the EM
theory.
One can see that, for NUBS,  $\epsilon_T$ is a decreasing function of $U$ for any allowed value  of the temperature.
However,
it becomes negative
for large enough values of the chemical potential.
The NUBS solutions close to the transition point have a positive  $\epsilon_T$ for
small $U$ only.
As seen in Figure 6 (left), the picture is more complicated for caged BHs.
The small charged black holes ($i.e.$ with large values of $T_H$) have always a positive $\epsilon_T$.
However, larger black holes have a negative 
isothermal electric permitivity, for large enough $U$.   

This picture depends however on the value of the dilaton coupling constant $a$.
Our study indicates that for both types of black objects $\epsilon_T$ becomes positive
for large enough values of $a$, irrespective of the values of $U,T_H$ (see Figure 7).

%%%%%%%%%%%%%%%%%%%%%%%%%%%%%%%%%%%%%%%%%%%%%%%%%%%%%%%%%%%%%%%%
\setlength{\unitlength}{1cm}
\begin{picture}(8,6)
\put(-1,0.0){\epsfig{file=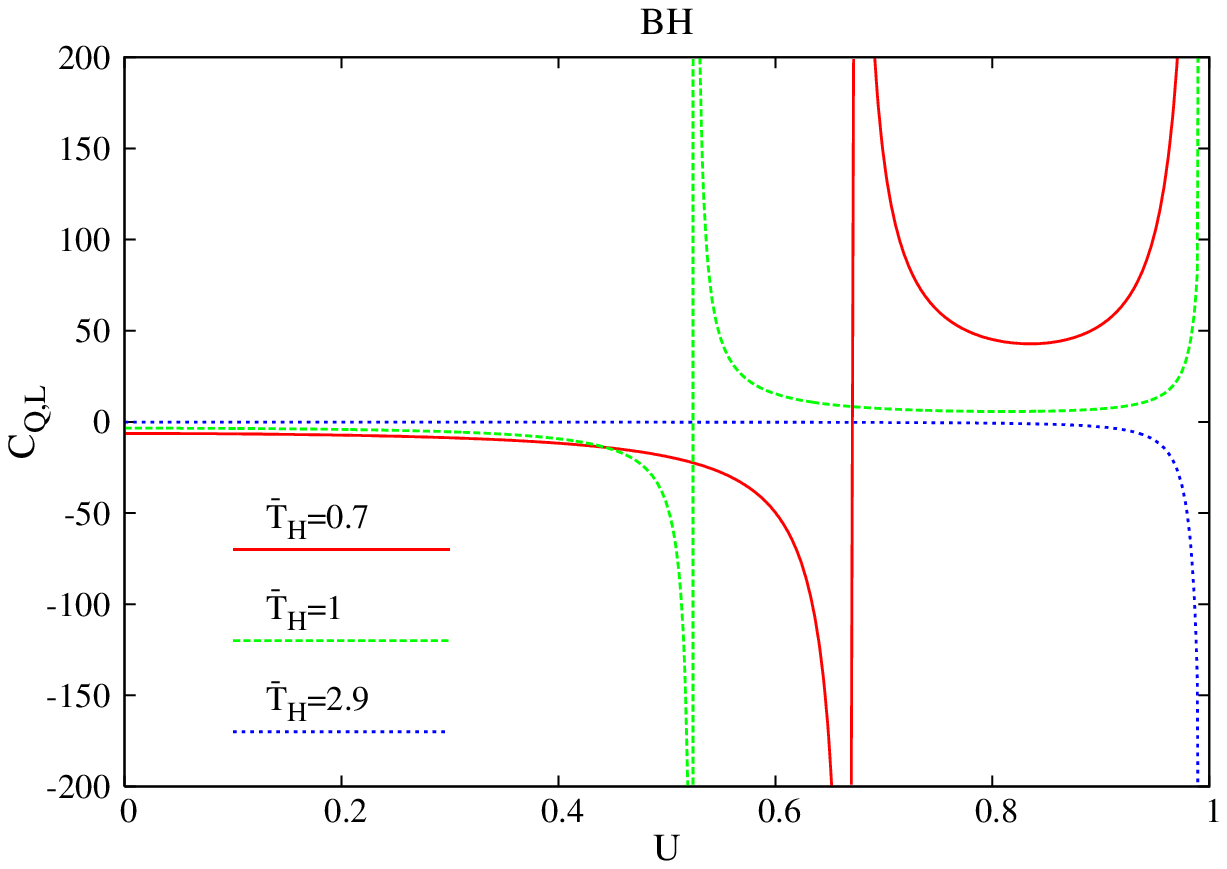,width=8cm}}
\put(7,0.0){\epsfig{file=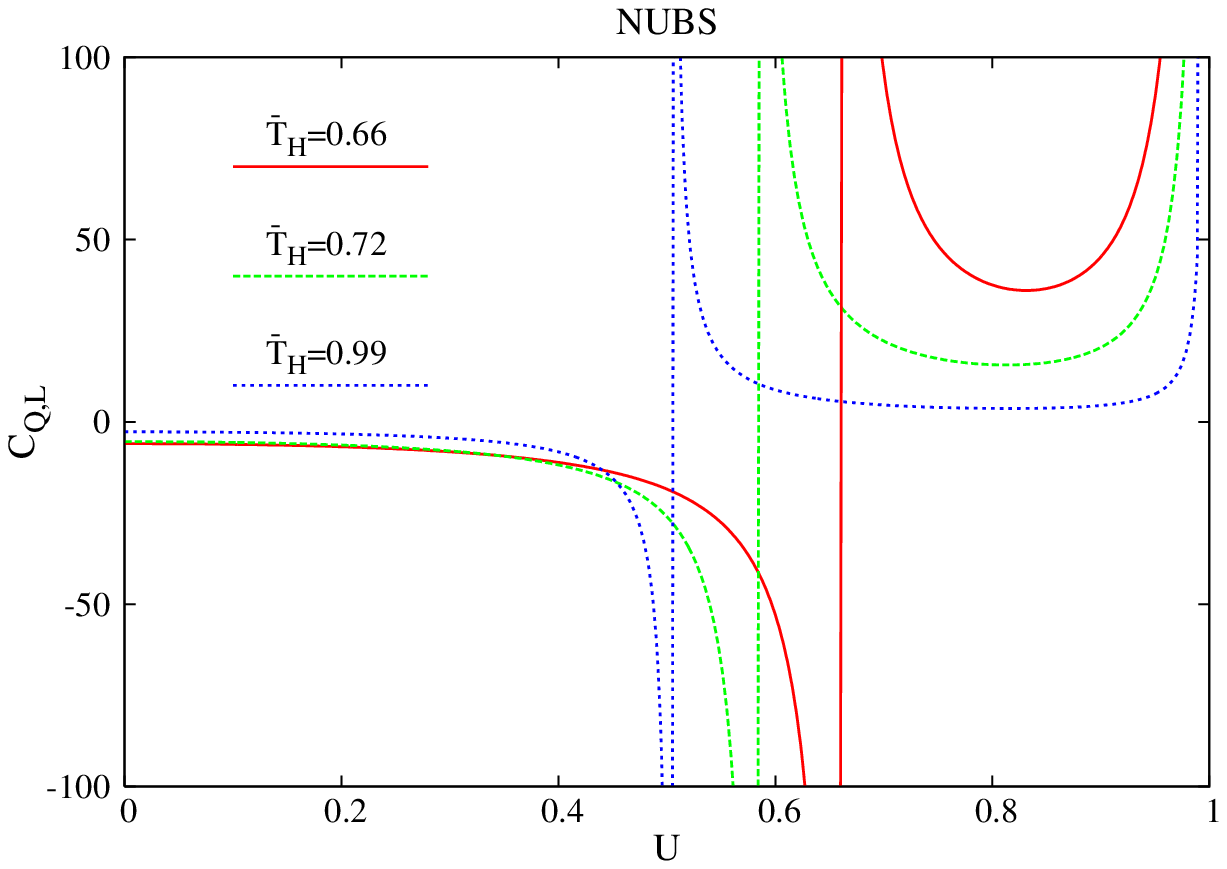,width=8cm}}
\end{picture}
\\
\\
{\small {\bf Figure 8.}
  The specific heat
	is shown as a function of   $U$
 for the $D=5$ charged
	caged black holes (left) and nonuniform black string strings (right). 
 The plots here correspond  to  a vanishing dilaton coupling constant $a=0$
 and several values of the normalized temperature.
}
\vspace{0.5cm}
%%%%%%%%%%%%%%%%%%%%%%%%%%%%%%%%%%%%%%%%%%%%%%%%%%%%%%%%%%%

%%%%%%%%%%%%%%%%%%%%%%%%%%%%%%%%%%%%%%%%%%%%%%%%%%%%%%%%%%%
\setlength{\unitlength}{1cm}
\begin{picture}(8,6)
\put(-1,0.0){\epsfig{file=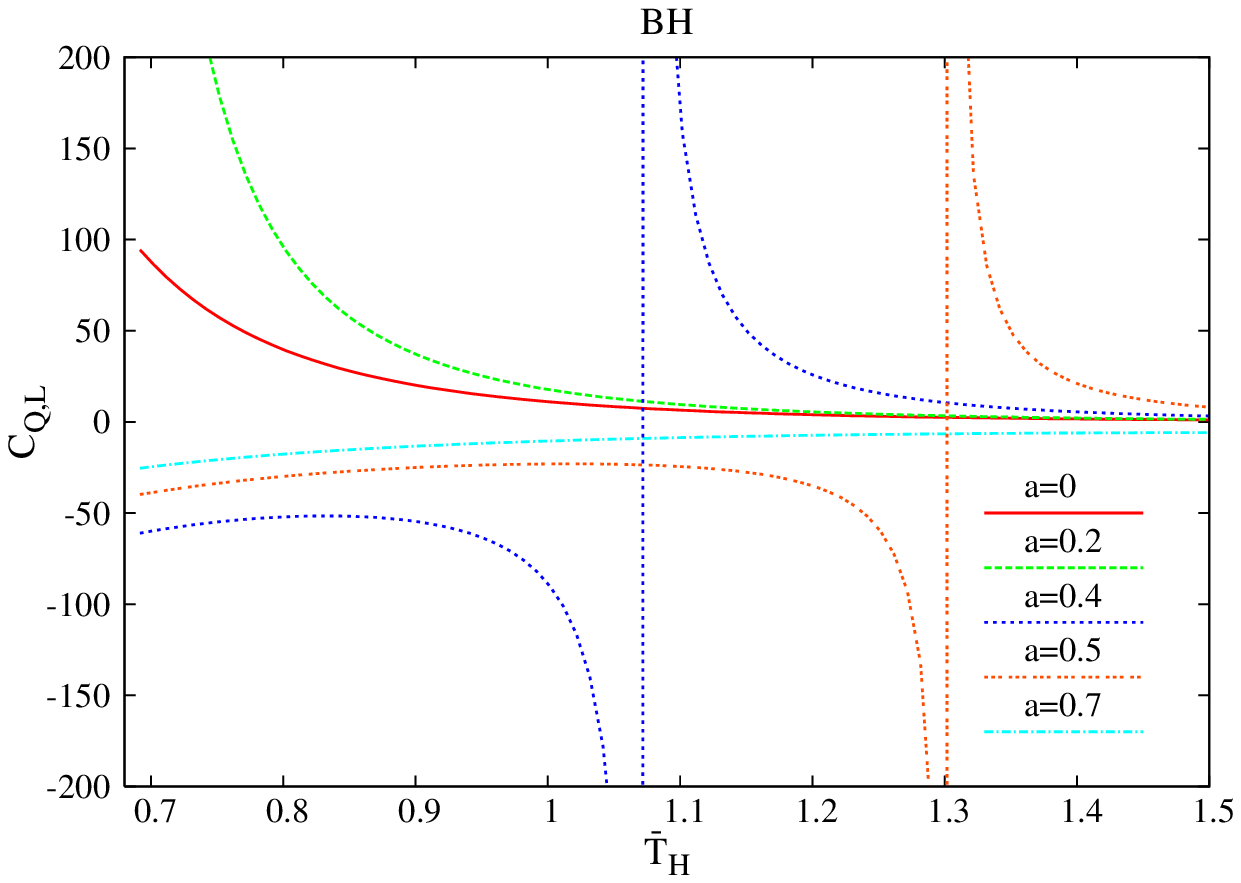,width=8cm}}
\put(7,0.0){\epsfig{file=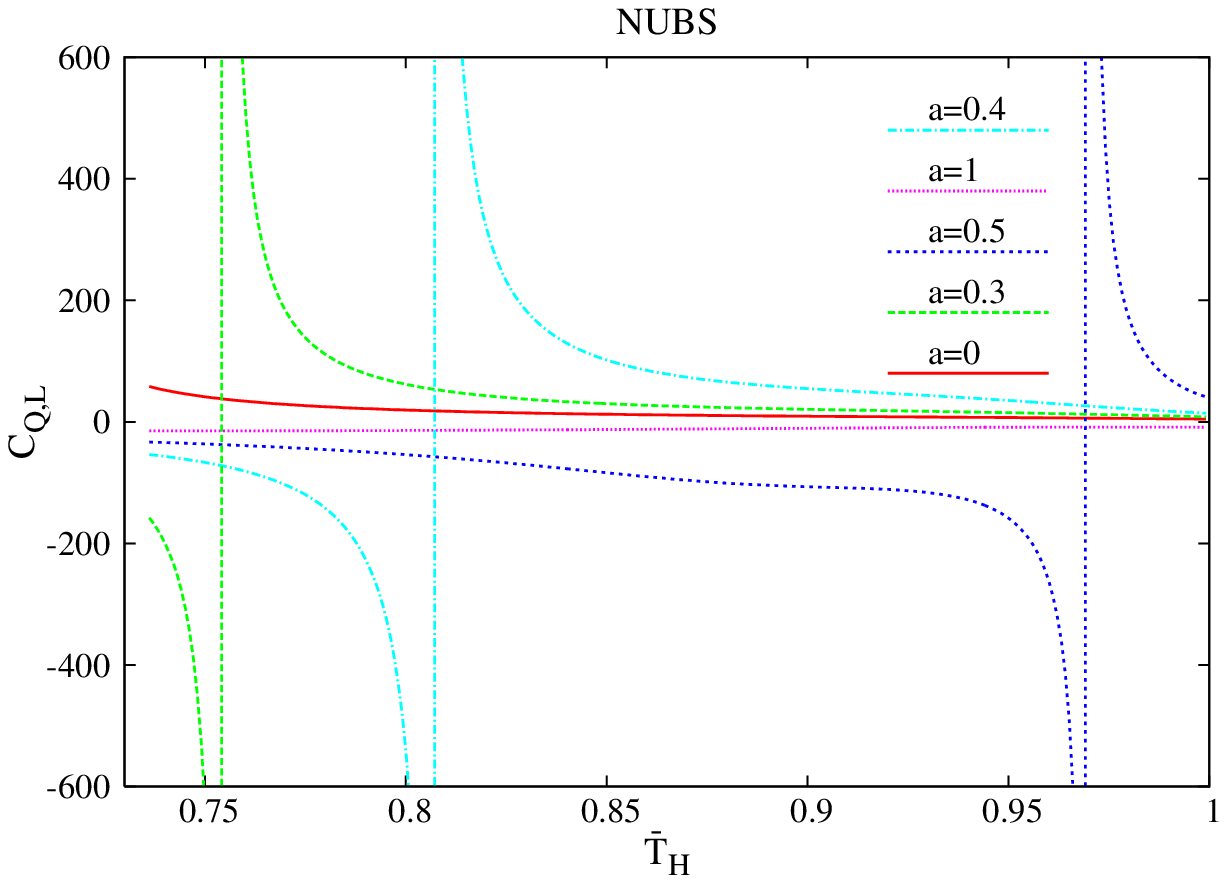,width=8cm}}
\end{picture}
\\
\\
{\small {\bf Figure 9.}
  The specific heat
	is shown as a function of the scaled temperature
 for the $D=6$ charged
	caged black holes (left) and nonuniform black strings (right). 
 These plots are for a fixed value of the chemical potential
 $U=0.5$ and several values of the dilaton coupling constant.
}
\vspace{0.5cm}
%%%%%%%%%%%%%%%%%%%%%%%%%%%%%%%%%%%%%%%%%%%%%%%%%%%%%%%%%%%

This can be understood by observing that for large $a$, the term
$U^2+1$ becomes dominant in the numerator of the isothermal electric permitivity.

In Figure 8 we present the specific heat as a function of the 
parameter $U$ for several values of the
rescaled temperature
for black strings and caged black holes in $D=5$ EM theory 
(a similar picture is found in $D=6$).
The specific heat of the black strings is always negative for small
values of $U$.
However, for any $T_H$, it changes the behaviour for a critical value of the
chemical potential $ U=U_c$ (which is fixed by the condition $\epsilon_T(U_c)=0$).
$C_{Q,L}$ exhibits a discontinuity at $U_c$, indicating the existence of 
a phase transition in the thermodynamical system.
The value of $U_c$ decreases as the  scaled   temperature  increases. 

%%%%%%%%%%%%%%%%%%%%%%%%%%%%%%%%%%%%%%%%%%%%%%%%%%%%%%%%%%%%%%%%%%%%%%%%%%%%%%%%%%%%%
\setlength{\unitlength}{1cm}
\begin{picture}(8,6)
\put(-1,0.0){\epsfig{file=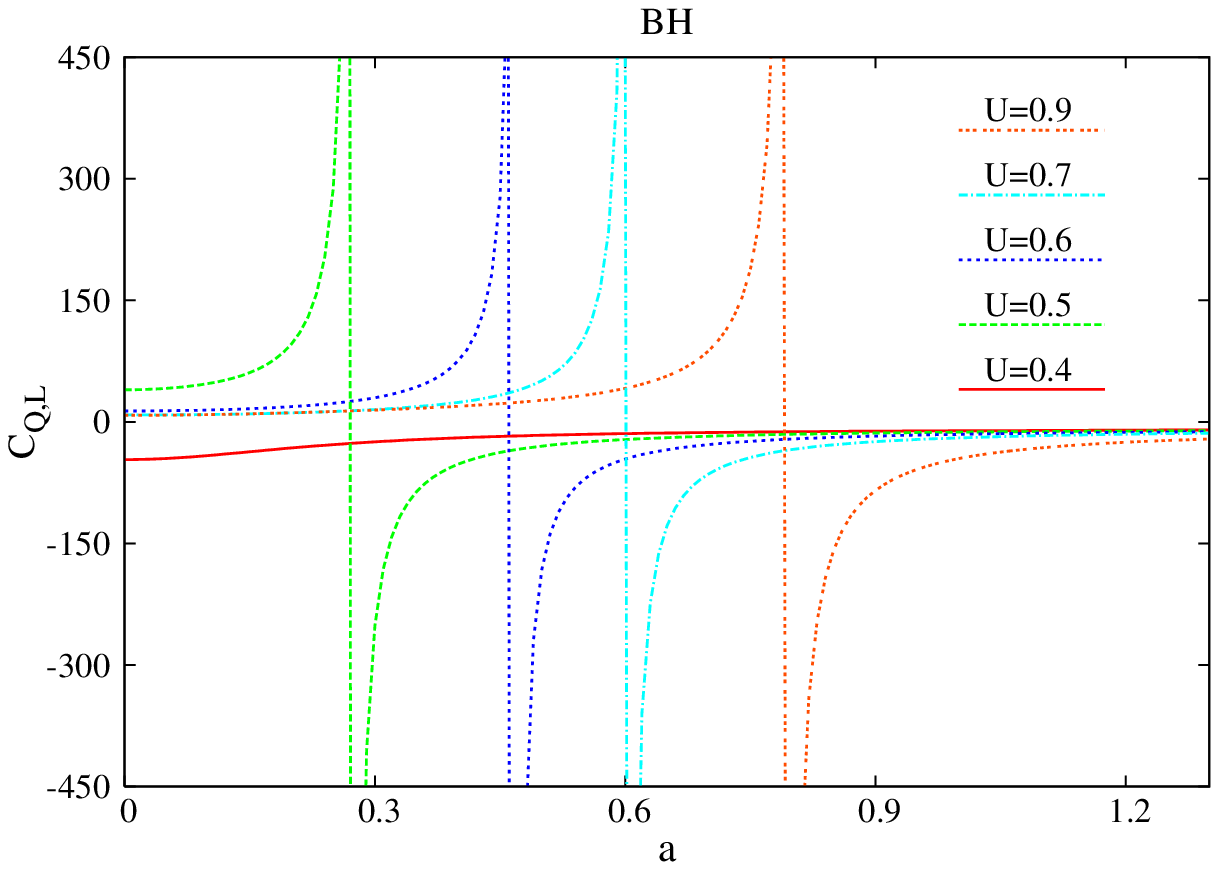,width=8cm}}
\put(7,0.0){\epsfig{file=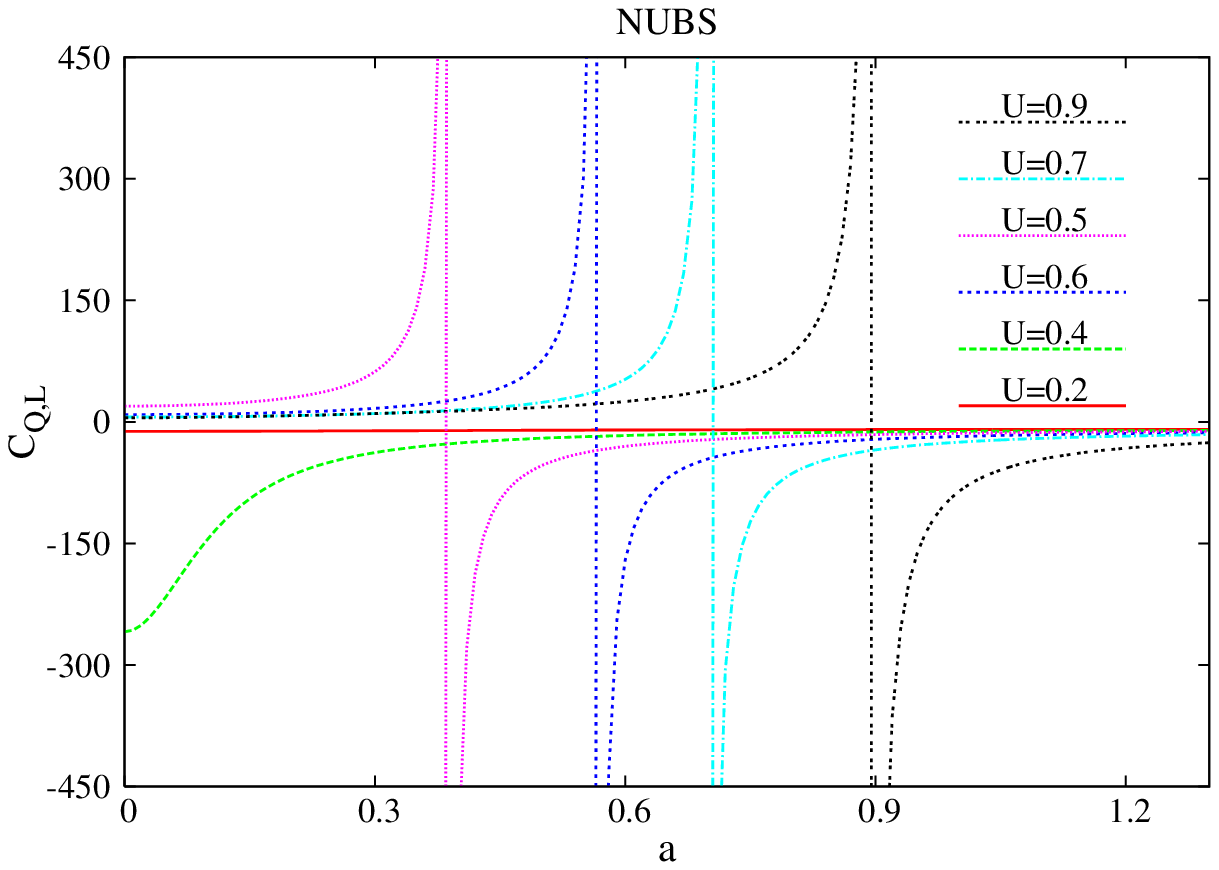,width=8cm}}
\end{picture}
\\
\\
{\small {\bf Figure 10.}
The specific heat is shown as a function of the dilaton coupling constant
 for the $D=6$ charged
	caged black holes (left) and nonuniform black strings (right). 
 The plots here correspond  to a fixed value of the scaled temperature $\bar T_H=0.8$
 and several values of the chemical potential.
}
\vspace{0.5cm}
%%%%%%%%%%%%%%%%%%%%%%%%%%%%%%%%%%%%%%%%%%%%%%%%%%%%%%%%%%%%%%%%%%%%%%%%%%%%%%%%%%%
\\
The picture is slightly different for caged black holes.
The scenario we found for NUBSs applies only for solutions close  enough to the critical region
where the topology changing transition is expected to occur.
The solutions generated from small caged BHs with a large
enough  scaled  temperature have a negative specific heat (see Figure 8 left).
 
Again this picture depends on the value of the dilaton coupling constant.
In Figure 9, the  specific heat is shown  as a function of the scaled temperature
for $D=6$ solutions with a fixed value of $U$
and a range of $a$.

The dependence of the solutions on the dilaton coupling constant $a$
is further illustrated in Figure 10, where we show the specific heat as a function of
$a$ for several values of $U$.
For small values of $a$, one notices the possible existence
of configurations with a positive specific heat, for both caged BHs and NUBSs.
One can see that for large enough values of the dilaton coupling constant,
the specific heat becomes negative irrespective of the value of $U$, for both black holes and black strings.
This can by understood by noticing that, since $\alpha\to 0$ for large enough $a$,
the sign of the specific heat (\ref{CQ}) of the charged solutions is fixed by $C_{Q,L}^0$, which is a negative quantity.

%%%%%%%%%%%%%%%%%%%%%%%%%%%%%%%%%%%%%%%%%%%%%%%%%%%%%%%%%%%%%%%%%%%%%%%%%%%%
\section{Slowly rotating charged NUBS}
%%%%%%%%%%%%%%%%%%%%%%%%%%%%%%%%%%%%%%%%%%%%%%%%%%%%%%%%%%%%%%%%%%%%%%%%%%%%
It would be interesting to generalize the charged KK configurations discussed 
in the previous sections by adding 
a nonvanishing angular momentum.
Unfortunately, there is no simple procedure to spin a given static system. 
Moreover, all known techniques fail for the case of KK solutions with a dependence on 
the extra-dimension. Some of the reasons for this are that the great majority of the 
solution generating techniques rely heavily either on high degrees of symmetry in the metric 
(such as the geometries in the generalized Weyl class) or on the 
possibility of dimensional reduction down to $3$-dimensions where 
target space symmetries can be used to add rotation to a given static solution. 
As we have previously stated, solutions with a dependence on the 
extra-dimension clearly do not belong to these classes of geometries 
and therefore any rotating non-uniform solution has to be found by 
brute force solving of the field equations.

To the best of our knowledge, the only rotating solutions 
with a dependence on the extra-dimension correspond to the NUBSs  
discussed in 
\cite{Kleihaus:2007dg}.
The solutions there have been constructed in $D=6$ dimensions,
for the special case of two equal magnitude angular momenta.
The angular dependence factorizes in this case, and the problem reduces
to solving a set of partial
differential equations with dependence only
on the radial variable $r$ and the extra-dimension variable $z$.
$D=6$ charged nonuniform black string configurations in heterotic 
string theory were also constructed in \cite{Kleihaus:2007dg} 
by employing a solution-generating technique.
It would be interesting to extend these results for other values of $D$
and non-equal angular momenta.
The case of a black string with a single angular momenta is of particular interest,
given the complicated picture one finds already for $D>4$
asymptotically Minkowski black hole solutions \cite{Emparan:2008eg}.
However, this leads to a difficult numerical problem, since the equations would depend 
on three variables.
Moreover, as far as we know,  rotating EM-dilaton 
black strings do not exist in closed 
form even for the uniform case\footnote{Except for special values of 
the dilaton coupling constant where the solutions can be generated
from the Myers-Perry black holes following the techniques in \cite{Kunz:2006jd},
\cite{Llatas:1996gh}.}, 
and one has to rely on numerical methods to construct them.

As an extension of the results in the previous sections,
 we present in what follows a numerical study of the NUBS solutions
with a single angular momentum,
in the limit of slow rotation (the case of UBSs is discussed in Appendix B, where 
a closed form solution is presented).
For such configurations, the equations  depend only
on the radial variable $r$ and the extra-dimension  variable $z$
and are solved to first order in the angular momentum parameter.  
 To simplify the general picture, we shall restrict here to the case of EM theory,
 although the results can easily be generalized to include a dilaton field.

In this approach, one starts with the following general ansatz depending on $r,z$ and 
the angular variable $\theta$,
%(with $0\leq \theta \leq \pi/2$)
describing charged rotating NUBS solutions with a single angular momentum:
\begin{eqnarray} 
\nonumber
&&ds^2=-e^{2\bar A(r,z,\theta)}f(r) dt^2+e^{2\bar B(r,z,\theta)}\left( \frac{dr^2}{f(r)} +e^{2S (r,z,\theta)}dz^2\right)
+e^{2\bar C(r,z,\theta)}r^2 \bigg( d\theta^2
\\
\nonumber
&&{~~~~~~~~~~~}+e^{2F(r,z,\theta)} \sin^2\theta (d\varphi+J(r,z,\theta)dt)^2+
e^{2H(r,z,\theta)}\cos^2\theta d\omega_{D-5}^2 \bigg),
\\
\label{gen-ans-rot}
&&{\cal A}=A_\mu(r,z,\theta) dx^{\mu},
\end{eqnarray}
(we recall that $f(r)=1-({r_0}/{r})^{D-4}$ in this case).
Then we suppose the following expansion of the functions in (\ref{gen-ans-rot}) in terms of a small parameter $a$
(which should not be confused with the dilaton coupling constant):
\begin{eqnarray} 
\nonumber
X_i( r,z,\theta) =X_i^{(0)}(r,z)+a  X_i^{(1)}(r,z,\theta),
\end{eqnarray}
with $X_i^{(0)}(r,z)$ being the functions which enter the static solution ($e.g.$ $\bar A(r,z,\theta)=A(r,z)+a A_1(r,z,\theta)$, 
$\bar B(r,z,\theta)=B(r,z)+a~B_1(r,z,\theta)$, $C(r,z,\theta)=C(r,z)+ a~C_1(r,z,\theta)$ etc.).

A straightforward but cumbersome computation shows that for
infinitesimal $a$,
the only term in the metric that changes  is $g_{t\varphi}$ and
 we find the following consistent metric form:
\begin{eqnarray} 
\nonumber
&&ds^2=-e^{2A(r,z)}f(r) dt^2+e^{2B(r,z)}\left( \frac{dr^2}{f(r)} +dz^2\right)
 +e^{2C(r,z)}r^2 (d\theta^2+
\sin^2\theta d\varphi^2+\cos^2\theta d\omega_{D-5}^2)
\\
&&{~~~~~~~~}~
\label{ans-NUBS-rot}
-2a \sin^2\theta  j(r,z)d\varphi dt,
\end{eqnarray}
the angular dependence of the nondiagonal component being factorized.
Similarly,  the only component of the vector potential that changes is $A_\varphi$,
the corresponding expression of the Maxwell field being
\begin{eqnarray} 
{\cal A}=A_t(r,z)dt+a \sin^2 \theta h(r,z)d \varphi.
\end{eqnarray}
Thus the problem is reduced to solving the equations for $j(r,z), h(r,z)$
in the background of the
static charged solutions.
These equations read (to simplify the relations, we denote here $A_t=V$):
\begin{eqnarray} 
\label{eq-rot}
\nonumber
&&j_{,rr}+\frac{j_{,zz}}{f}
-(V_{,r}h_{,r}+\frac{V_{,z}h_{,z}}{f})
-(A_{,r}j_{,r}+\frac{A_{,z}j_{,z}}{f})
+2j (A_{,r}C_{,r}+\frac{A_{,z}C_{,z}}{f})
\\ 
&&{~~~~~~~~~~~~}
+(D-5)(C_{,r}j_{,r}+\frac{C_{,z}j_{,z}}{f})-2 j (C_{,rr}+\frac{C_{,zz}}{f}) 
-2(D-3)j(C_{,r}^2+\frac{C_{,z}^2}{f})
\\~
\nonumber
&&{~~~~~~~~~~~~}
+\frac{2j}{r}(A_{,r}-2(D-3)C_{,r}-\frac{D-4}{r})
 +\frac{(D-5)j_{,r}}{r}=0,
 \\
 \nonumber
 &&h_{,rr}+\frac{h_{,zz}}{f}
 +(A_{,r}h_{,r}+\frac{A_{,z}h_{,z}}{f})
 +(D-5)(C_{,r}h_{,r}+\frac{C_{,z}h_{,z}}{f})
 +\frac{e^{-2A}}{f}\big(2j \big(V_{,r}C_{,r}+\frac{V_{,z}C_{,z}}{f})
 \\
 \label{eq-rot-NUBS}
&&{~~~~}-(V_{,r}j_{,r}+\frac{V_{,z}j_{,z}}{f})
 \big)
 +\frac{2e^{-2A}j}{rf}V_{,r}
 +(\frac{D-5}{r}+\frac{f'}{f})h_{,r}
 -\frac{2(D-4)e^{2B-2C}}{r^2f}h=0~.
\end{eqnarray}
The equation for $j$ has a total derivative structure which helps us relate the 
total angular
momentum to event horizon quantities:
\begin{eqnarray} 
&\frac{1}{L}\int_0^L dz e^{-A+(D-3)C}r^{D-3}
\left( 
j_{,r}-\frac{2j}{r} -2j C_{,r}
- h V_{,r} 
\right)
\big|_{r=r_h}
={\rm lim}_{r\to\infty}r^{D-3}\left(j_{,r}-\frac{2j}{r}\right).{~~}
\end{eqnarray}
%where $j_1$ is the constant in the asymptotic expansion of $j(r)\sim j_1/r^{D-4}$.

For the slowly rotating solutions, the horizon radius is still determined by the equation $f(r_0)=0$.
Also, the surface gravity and
area of the event horizon do not change to order ${\cal O}(a)$. However, the angular momentum $J$
and the magnetic dipole moment are evinced already at this order. Namely, the  angular momentum is
related to the asymptotic form of the metric by
\begin{eqnarray} 
\label{as-j}
g_{t\varphi}=-\frac{J}{2\Omega_{D-3}L}\frac{\sin^2\theta}{r^{D-4}},
\end{eqnarray}
while the magnetic dipole moment $\mu$ is read from the asymptotic
form of the magnetic gauge potential $h(r,z)$
\begin{eqnarray} 
\label{as-h}
h(r,z)=-\frac{\mu}{(D-4)\Omega_{D-3}L}\frac{1}{r^{D-4}}.
\end{eqnarray}

%%%%%%%%%%%%%%%%%%%%%%%%%%%%%%%%%%%%%%%%%%%%%%%%%%%%%%%%%%%
\setlength{\unitlength}{1cm}
\begin{picture}(8,6)
\put(-1,0.0){\epsfig{file=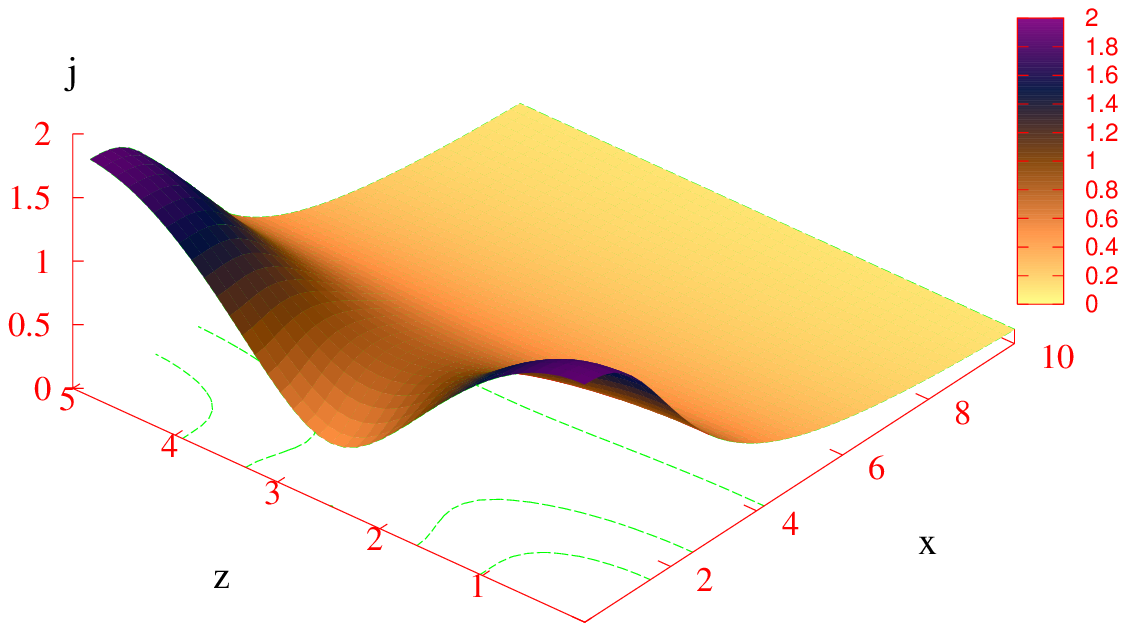,width=8cm}}
\put(7,0.0){\epsfig{file=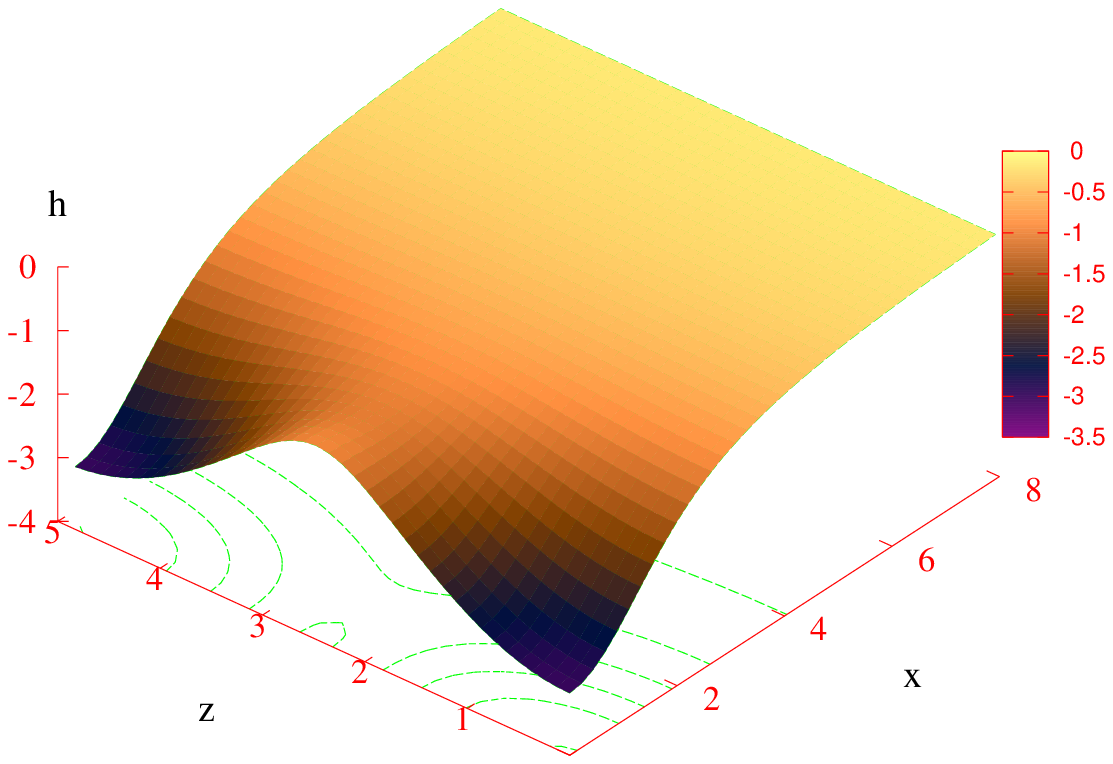,width=8cm}}
\end{picture}
\\
\\
{\small {\bf Figure 11.}
  The functions $j(r,z)$ and $h(r,z)$ are shown for 
  a typical $D=6$ slowly rotating NUBS in Einstein-Maxwell theory (with $x=\sqrt{r^2-r_0^2}$).
  The parameters of this configuration  are $r_0=1,L=4.9516$, $\lambda=0.47$, $U=0.833$. 
}
\vspace{0.5cm}
%%%%%%%%%%%%%%%%%%%%%%%%%%%%%%%%%%%%%%%%%%%%%%%%%%%%%%%%%%%%

In practice, the idea is to start with a known vacuum 
NUBS solution and to charge it by using the 
Harrison transformation described in the previous sections.
This generates static charged NUBS solutions of the 
EM equations ($i.e.$ $a=0$, no dilaton field).
Then the eqs. (\ref{eq-rot-NUBS}) are solved with the background functions $A,B,C$.

We have performed this computation for $D=6$ NUBS solutions\footnote{We expect slowly rotating 
NUBS solutions to exist 
for any value of $D$.
However, the case $D=6$ is more convenient for numerics when dealing with KK solutions with a 
dependence on the extra-dimension, see $e.g.$ the comments on that in \cite{Wiseman:2002zc}, \cite{Kudoh:2003ki}.}
 with several values of the
nonuniformity parameter $\lambda$ (with $\lambda$ defined by the equation (\ref{lambda})) and a range of $U$.  
The equations (\ref{eq-rot}) were solved by using the same methods as those described in Appendix A for the vacuum NUBS case.
In this scheme, one introduces a new radial coordinate $x=\sqrt{r^2-r_0^2}$
such that the event horizon is located at $x=0$.
Then the equations (\ref{eq-rot-NUBS}) are solved subject 
to the following set of boundary conditions:
\begin{eqnarray} 
\label{bc-rot}
j|_{x=0}=j_H e^{2C(0,z)},~~\partial_x h|_{x=0}=0,~~j|_{x=\infty}=~h|_{x=\infty}=0, 
~~\partial_{z}j |_{z=0,L}=\partial_{z}h |_{z=0,L}=0,~~{~~}
\end{eqnarray}
with $j_H$ a constant fixing the  angular velocity of the horizon, $\Omega_H=a j_H/r_0^2$.

The profiles of the functions $J,h$ for a typical solution are 
presented in Figure 11.
One can see that they present a strong dependence on the extra-coordinate $z$,
possessing a peak around $z=L/2$.

A  quantity of interest here is the gyromagnetic ratio $g$
which is defined in the usual way by
\begin{eqnarray} 
\label{gyro-ratio}
g=\frac{2M\mu}{Q_eJ}.
\end{eqnarray}
In $D=4$ classical electrodynamics, the gyromagnetic ratio of a charged rotating body is always
$g=1$. However, the gyromagnetic ratio of an electron in Dirac theory is $g=2$, which is
precisely the value found for a rotating Kerr-Newmann black hole.
When considering the higher dimensional version of these EM black holes\footnote{Note that 
the $D>4$ counterparts  of the Kerr-Newmann black hole  are not known in closed form.},  the
lowest order perturbation theory gives for
the gyromagnetic ratio the result $g = D - 2$, which
seems a natural higher dimensional generalization of the
gyromagnetic ratio in $D = 4$ dimensions \cite{Aliev:2004ec}, \cite{Aliev:2006yk}.
However,
numerical calculations revealed that in higher dimensions,
in a nonperturbative approach, the gyromagnetic ratio is not constant,
but deviates from $D - 2$ for finite values of the charge  \cite{Kunz:2005nm}, \cite{Kunz:2006eh},
which was also confirmed by higher order perturbation theory \cite{NavarroLerida:2007ez}.

%%%%%%%%%%%%%%%%%%%%%%%%%%%%%%%%%%%%%%%%%%%%%%%%%
\setlength{\unitlength}{1cm}
\begin{picture}(8,6)
\put(3,0.0){\epsfig{file=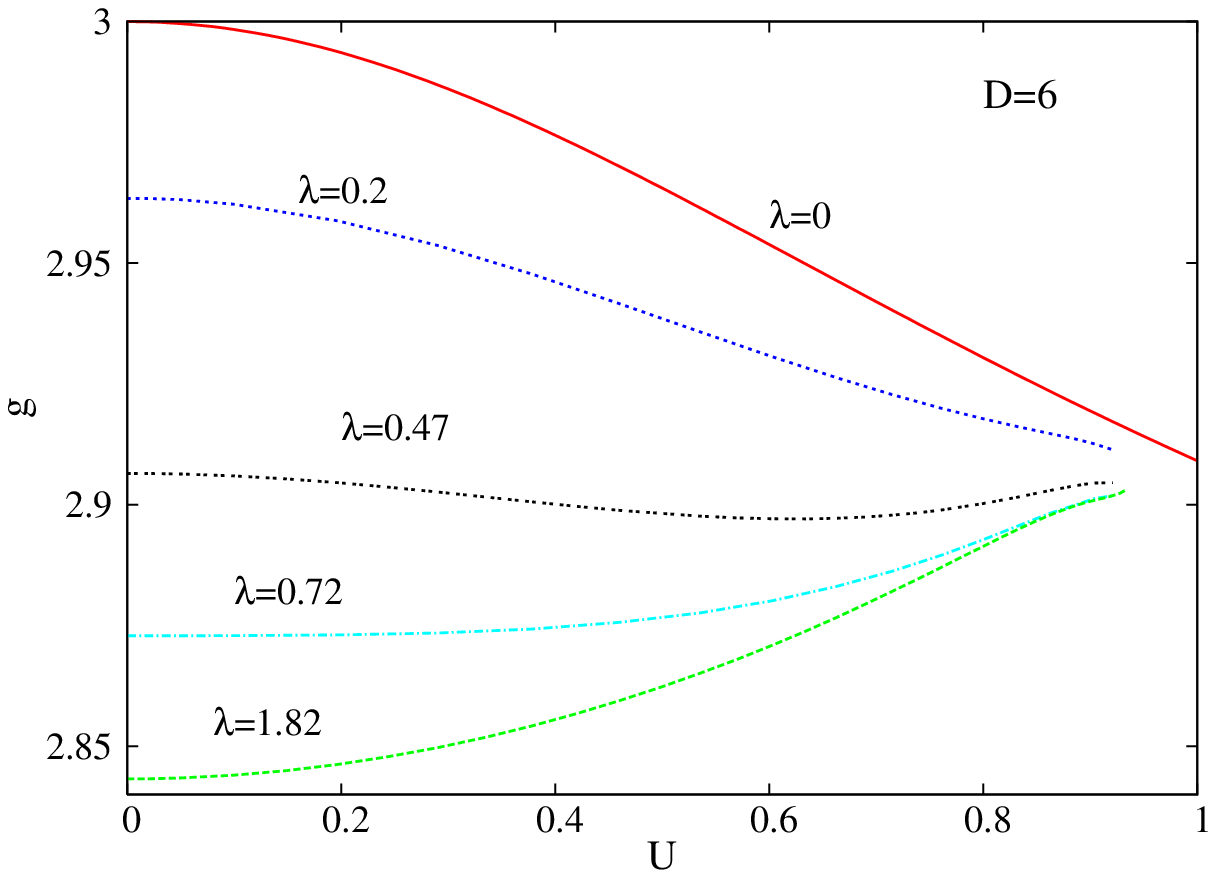,width=8cm}}
\end{picture}
\\
\\
{\small {\bf Figure 12.}
  The gyromagnetic ratio $g$ is shown as a function of
$U$ for several values of the nonuniformity parameter $\lambda$
for $D=6$ slowly rotating black string solutions in Einstein-Maxwell theory.
}
\vspace{0.4cm}
%%%%%%%%%%%%%%%%%%%%%%%%%%%%%%%%%%%%%%%%%%%%%
\\
 
However, so far most of the studies in the literature were 
limited to the case of black holes in asymptotically flat spacetime,
the gyromagnetic ratio of solutions with compact extra-dimensions being not explored.
As found in Appendix B, $g$ varies for uniform black string solutions between $D-3$ (no charge, $U=0$)
and $\frac{(D-2)^2(D-4)}{D(D-5)+5}$ (in the extremal limit, $U\to 1$).
One can see in Figure 12 that the picture is much more complicated for $D=6$ solutions
with an explicit dependence on the extra-coordinate.
For small values of $\lambda$, the gyromagnetic ratio $g$ still decreases with $U$.
However, the situation changes for large enough 
values of the nonuniformity parameter $\lambda$.
Our numerical results suggest that, for any $\lambda$, in the extremal limit, the gyromagnetic
ratio of NUBS
 solutions approaches the same value $g_c=32/11$.

It would be interesting to perform a similar computation for caged black holes and to consider
the issue of topology change for spinning solutions.
 
%%%%%%%%%%%%%%%%%%%%%%%%%%%%%%%%%%%%%%%%%%%%%%%%%%%%%%%%%%%%%%%%%%%%%%%%%%%%
\section{Conclusions}
%%%%%%%%%%%%%%%%%%%%%%%%%%%%%%%%%%%%%%%%%%%%%%%%%%%%%%%%%%%%%%%%%%%%%%%%%%%%
  
The  charged solutions in string theory and M-theory are usually  obtained from the vacuum KK black objects 
in $d$-dimensions by sequences of 
boosts and dualities \cite{Harmark:2004ws}.
This gave a precise connection between the rich phase structure
of KK black objects and that of the corresponding branes in string/M-theory and provided
predictions for the thermodynamics of supersymmetric
Yang-Mills theories on a circle.
Our proposal here was to use instead 
the Harrison transformation derived in \cite{Gal'tsov:1998yu}.
This construction is more general since it holds for any value
of the dilaton coupling constant $a$ and any value of  $p$ (the brane extra-dimensions), in particular
 for the Einstein-Maxwell (EM) theory ($p=0$).
 
  The main purpose of this paper was to use this technique
 in order to investigate the issue of charged black objects 
 in KK theory for an arbitrary value of $a$, by using the known 
   vacuum NUBS  and the KK localized BHs as seed solutions.
 We have observed that both the geometrical and thermodynamical
 properties of the  charged solutions can be derived from the
corresponding $D-$dimensional vacuum configurations.
Asymptotically flat black objects generated in this way from vacuum fields
have a geometrical structure similar to that of the vacuum seed solutions.

 A natural question that  arrises at this point is 
 how to compute the various conserved quantities
 and the total action of the charged black objects.
 One particularly powerful approach to this problem is given by the counterterms method. 
 For asymptotically flat spaces there have been proposed many such 
 counterterms and their form relies heavily on the topology at infinity
 \cite{Kraus:1999di}, \cite{Lau:1999dp}-\cite{Mann:2005cx}. 
 However, in certain situations, these counterterms fail to regularize 
 the gravitational action. One recent proposal put forward by Mann and Marolf \cite{Mann:2005yr} 
 seems to avoid these problems for asymptotically flat spaces. 
A secondary aim of this paper was to propose a direct extension of the Mann-Marolf 
counterterm to `{\it transverse asymptotically flat}' spaces in KK theory. 
In Section $3$ we showed that the total action and conserved charges can be expressed 
 in a novel way in terms of the electric part of the Weyl tensor and that, moreover, 
 the final results are consistent with 
 those previously derived in the $p$-brane literature.

The thermodynamics of charged black objects in KK theory has been considered
in Section 5.
There we have argued that
 these solutions are
thermodynamically unstable in a grand canonical ensemble.  
Also, our results show that the thermodynamics of these black objects 
 is affected by the value of the dilaton coupling constant. 
A general discussion of the charged black strings and black holes for the simplest case 
of a KK theory in $D=5,6$ dimensions
with a single compact extra-dimension has also been studied there. 
These solutions  display a rich phase-structure  already in this case,
some of them being thermally stable in a canonical ensemble, for large enough values of the electric charge.

The  charged solutions we have discussed are likely to possess spinning generalization. 
Unfortunately, the Harrison transformation considered in our work is restricted
to the case of static configurations.  
In Section 6 we have presented  a numerical study of the uniform and NUBS solutions
with a single angular momentum,
in the limit of slow rotation.
Our results provide evidence for the existence of such solutions
and show that the gyromagnetic ratio depends on the nonuniformity
parameter of the black strings.

Although we have restricted our study to the (physically more relevant)
region outside the horizon, it would be interesting to consider also the 
interior of these charged black objects.
In the vacuum case, the ref. \cite{Kleihaus:2007cf} presents numerical evidence,
that the singularity touches the horizon as the horizon topology changing transition is reached.
We expect that the inclusion of the electric charge would lead to a more complicated
 picture, at least in the absence of a dilaton field.

In this paper we have considered the simplest case of a single gauge field.
In principle, one can easily add secondary gauge fields to the action (\ref{actioni})
by considering sequences of boosts and dualities  (this would also generate supplementary scalars,
with a specific coupling to the gauge fields).
The methods in this work can easily be extended to that case.
On general grounds, we expect the features we have found to be generic;
in particular, all solutions are likely to be unstable in a grand canonical ensemble.

Also, as an avenue for future research, it would be interesting to study
KK solutions with more general matter fields than the model (\ref{actioni}).
The case of configurations with nonabelian gauge fields
is of particular interest, given the rich structure
of these systems and the interplay  in this case between internal group symmetries and
the spacetime symmetries \cite{Volkov:1998cc}.

\vspace{0.6cm}

%%%%%%%%%%%%%%%%%%%%%%%%%%%%%%%%%%%%%%%%%%%%%%%%%%%%%%%%%%%%%%%%%%%%%%%%%
{\bf\large Acknowledgements} 
\\
We are grateful to T. Wiseman for kindly providing us with the $D=5$ and $D=6$ black hole data
in \cite{Kudoh:2004hs}.
BK gratefully acknowledges support by the DFG. 
The work of ER was supported by a fellowship from the Alexander von Humboldt Foundation. 
The work of CS was supported by the Natural Sciences and Engineering Council of Canada. 
%%%%%%%%%%%%%%%%%%%%%%%%%%%%%%%%%%%%%%%%%%%%%%%%%%%%%%%%%%%%%%%%%%%%%
%%%%%%%%%%%%%%%%%%%%%%%%%%%%%%%%%%%%%%%%%%%%%%%%%%%%%%%%%%%%%%%%%%%%%

\appendix
%%%%%%%%%%%%%%%%%%%%%%%%%%%%%%%%%%%%%%%%%%%%%%%%%%%%%%%%%%%%%%%%%%%%%%%%%
\section{The construction of vacuum Kaluza-Klein solutions: 
technical aspects}
%%%%%%%%%%%%%%%%%%%%%%%%%%%%%%%%%%%%%%%%%%%%%%%%%%%%%%%%%%%%%%%%%%%%%%%%% 

%%%%%%%%%%%%%%%%%%%%%%%%%%%%%%%%%%%%%%%%%%%%%%%%%%%%%%%%%%%%%
\subsection{NUBS solutions}
The NUBS solutions were constructed by several different groups 
 \cite{Kudoh:2004hs}, 
\cite{Kleihaus:2006ee}, \cite{Wiseman:2002zc}, \cite{Sorkin:2006wp},
 with different numerical methods, and are known for 
 all dimensions up to eleven.
 
The basic steps in the nonperturbative construction of these solutions are as follows.
First, the numerics is performed by using a new radial coordinate $x$, which is 
defined such that the $r,z$ part of the metric becomes conformally 
flat\footnote{For example, in $D=6$, $x=\sqrt{r^2-r_0^2}.$}, $i.e.$
\begin{eqnarray} 
 x=\int \frac{dr}{\sqrt{f(r)}}+x_0,
\end{eqnarray} 
and 
\begin{eqnarray} 
ds^2_d=-e^{2A(x,z)}f(r(x))dt^2+e^{2B(x,z)}\left( dx^2 
+dz^2 \right)+e^{2C(x,z)}r^2(x)d\omega_{d-3}^2.~~{~~~~}
\end{eqnarray}
The constant $x_0$ in the above relation is fixed by imposing that
the horizon is located at $x=0$, where $f(r_0)=0$ (moreover, $x/r\to 1$ in the
asymptotic region).

The Einstein equations $G_t^t=0,~G_x^x+G_z^z=0$ and $G_\theta^\theta=0$ 
(where $\theta$ denotes an angle of the $D-3$ dimensional sphere and $G_a^b$
the Einstein tensor) 
 yield three elliptic equations for the metric functions $A,~B,~C$  
 (these equations are given $e.g.$ in Section 2 of ref. \cite{Kleihaus:2006ee}).
The remaining Einstein equations $G_z^x=0,~G_x^x-G_z^z=0$
yield two constraints. 
As first observed in \cite{Wiseman:2002zc}, 
these constraints satisfy  Cauchy-Riemann relations
\begin{eqnarray}
&\partial_z\left(\sqrt{-g} G^x_z \right) +
\partial_{x}\left(  \sqrt{-g} \frac{1}{2}(G^x_x-G^z_z) \right)
 =  0 ,~~
%\nonumber 
\partial_{x}\left(\sqrt{-g} G^x_z \right)
-\partial_z\left(  \sqrt{-g} \frac{1}{2}(G^x_x-G^z_z) \right)
 =  0  .~~{~~~~}
\end{eqnarray}
Thus the weighted constraints satisfy Laplace equations,
and they are fulfilled,
when one of them is satisfied on the boundary 
and the other at a single point
\cite{Wiseman:2002zc}. 

All known KK solutions with a dependence on the extra-dimension
have the  reflection symmetry 
w.r.t.~$z=L/2$. Thus, in numerics it is enough to consider
only half of the $z$-domain.  

The NUBS solutions are then constructed subject to the following set of 
boundary conditions
\begin{eqnarray}
\nonumber
&&A\big|_{x=\infty}=B\big|_{x=\infty}=C\big|_{x=\infty}=0,
\\
\label{bc2}
&&B\big|_{x=0}-A\big|_{x=0}=d_0,~\partial_{ x} 
A\big|_{x=0}=\partial_{x} C\big|_{ x=0}=0,
\\
\nonumber
&&\partial_z A\big|_{z=0,L/2}=\partial_z B\big|_{z=0,L/2}
=\partial_z C\big|_{z=0,L/2}=0,
\end{eqnarray}
where the constant $d_0>0$ is related to the Hawking 
temperature of the solutions.
Regularity further requires that the condition 
$\partial_{x} B\big|_{x=0}=0$ holds for the solutions.
%Moreover, these boundary conditions 
%guarantee that the constraints are satisfied.
%since $\sqrt{-g} G^x_z=0$ everywhere on the boundary,
%and $ \sqrt{-g}( G^x_x-G^z_z ) =0$ on the horizon.

The Hawking temperature and entropy of the NUBS solutions are given by
\begin{eqnarray}
\label{temp} 
T_H=e^{A_0-B_0}T_H^0,~~~S=S_0\frac{1}{L}\int_0^L e^{B_0+(D-3)C_0}dz,
\end{eqnarray}
where $T_H^0,~S_0$ are the corresponding quantities of the uniform solution
with $A=B=C=0$
\begin{eqnarray}
\label{temp0} 
T_H^0=\frac{D-4}{4 \pi r_0},~~~S_0= \frac{1}{4G}L\Omega_{D-3}r_0^{D-3},
\end{eqnarray}
and $A_0(z),B_0(z),C_0(z)$ are the values of the metric functions on the event horizon $r=r_0$ $(x=0)$.
The mass and tension of these solutions are given by (\ref{global-charges}),
where the constants $c_t$, $c_z$ are read from the asymptotics
of the metric functions $A$ and $B$, respectively.

All known NUBS solutions were found by solving numerically the set of three coupled non-linear
elliptic partial differential equations for $A,B,C$,
subject to the above boundary conditions.
For a better numerical accuracy, ref. \cite{Kleihaus:2006ee} 
proposed to employ a new radial coordinate
%\begin{equation}
$ \bar r = x/(1+x)  , $
%  \label{barx2} \end{equation}
which
maps spatial infinity to the finite value $\bar r=1$.

The branch of nonuniform strings is obtained by
starting at the critical point of the uniform string branch
and varying the boundary parameter $d_0$, which enters the Eq.~(\ref{bc2}).
The resulting solutions have a nontrivial dependence on $z$, 
the parameter  $\lambda$ as defined by the equation (\ref{lambda}) 
increasing with $d_0$ for the main NUBS branch.
Usually, the NUBS solutions are found with good accuracy (as verified $e.g.$ by using the
Smarr law (\ref{smarrform1})), except for the 
configurations close to the
topology changing point ($i.e.$ for very large values of the nonuniformity
parameter $\lambda$).

%%%%%%%%%%%%%%%%%%%%%%%%%%%%%%%%%%%%%%%%%%%%%%%%%%%%%%%
\subsection{Caged black holes}
%%%%%%%%%%%%%%%%%%%%%%%%%%%%%%%%%%%%%%%%%%%%%%%%%%%%%%%
Given the mismatch between the horizon topology and the asymptotic structure of 
spacetime\footnote{It is interesting to notice
the analogy with the $D>4$ asymptotically Minkowski black holes with a nonspherical topology of the horizon
which present the same feature.
For example, the horizon topology of a black ring is $S^{D-3}\times S^1$ ($i.e.$
it looks like a black string in the horizon region), the ${\cal M}^{D}$ background
being approached in the far field \cite{Emparan:2006mm}.
In the vacuum case, the only  exact solution here corresponds to the $D=5$ black ring in \cite{Emparan:2001wn}.
However, some of the black objects with a nonspherical topology of the horizon
can be studied in a nonperturbative approach 
 by employing a suitable version of the 
Weyl formalism \cite{Kudoh:2006xd}, \cite{Kleihaus:2009wh}.
This leads to a much simpler numerical problem as compared to the caged BHs in KK theory.
} 
the numerics is much more difficult in this case, the only nonperturbative results available so far
in the literature being for $D=5, 6$ dimensions \cite{Kudoh:2003ki}, \cite{Kudoh:2004hs}, \cite{Sorkin:2003ka}.
The ansatz (\ref{metric}) 
corresponds to that used
by Kudoh and  Wiseman in \cite{Kudoh:2003ki}, \cite{Kudoh:2004hs}.
The independent results in \cite{Sorkin:2003ka} were found by using a 
more complicated approach, with
two different coordinate patches: the first patch covers the region close to the horizon while the second one 
is used in the asymptotic region  (the same approach was employed in the recent work \cite{Headrick:2009pv}). 
The equations are relaxed on both patches one after the other. 
Moreover, the patches communicate, providing boundary data to each other in the overlapping
region.
 
We have considered the problem of constructing KK black holes 
by following the approach in  \cite{Kudoh:2003ki}, \cite{Kudoh:2004hs}.
A generalization to any $D$ of the metric ansatz used there is given by
\begin{eqnarray} 
\label{ansatz-BHs}
ds^2=e^{2\bar B(r,z)}J(r,z)B_0^2(r)(dr^2+dz^2)+e^{2\bar C(r,z)}R^2(r,z)d\omega_{D-3}^2
- e^{2\bar A}A_0^2(r)dt^2~, ~~~{~~}
\end{eqnarray}
where
\begin{eqnarray} 
A_0(r)= \tanh \left((D-3)(r-r_0)/2\right),
~~B_0(r)=(1+e^{-(D-3)(r-r_0)})^{2/(D-3)},
\end{eqnarray}
and 
\begin{eqnarray} 
\label{JR}
&&J(r,z)=\frac{e^{ r}}{2\sqrt{\sinh^2 r+\cos^2 z}},
\\
\nonumber
&&R(r,z)=r+\frac{1}{2}\log(1+2 e^{-r}\sqrt{\sinh^2 r+\cos^2 z}+
e^{-2 r} \sqrt{-1+(e^{2r}+2 e^r \sqrt{\sinh^2 r+\cos^2 z })^2}~),
\end{eqnarray}
while $r_0<0$ is an input parameter corresponding to
the event horizon radius.
As  it was  extensively discussed in \cite{Kudoh:2003ki}, $r_0$ 
also specifies the size of the black hole.
Without any loss of generality, 
the compactification radius is fixed here to be $L=\pi$.

One can see from (\ref{ansatz-BHs}) that the metric functions $A,B,C$ in the generic ansatz (\ref{metric})
contain  some background terms ($e.g.$ $A(r,z)=\bar{A}(r,z)+\log A_0(r)$ etc.).
The functions $A_0(r)$, $B_0(r)$ assure that, for small
black holes,  the solution  is approximated near the horizon by the
Schwarzschild-Tangerlini metric in $D-$dimensions.
The  background terms $J(r,z)$, $R(r,z)$ are used to implement 
the topology change between event horizon and spatial infinity.
A detailed discussion of the choice (\ref{JR}) is presented in \cite{Kudoh:2003ki}.
Here we only note that $R(r,\pi/2)=0$, for $r\leq 0$ (thus also for $r=r_0$), 
and $R(r,\pi/2)>0$ elsewhere, which shows that 
$z=\pi/2$ with $r<0$ represents the symmetry axis, 
while the topology of the event horizon is $S^{D-2}$.
Also, $J(r,z)$ diverges at $r=0,z=\pi/2$, which shows that $e^{-2B}\to 0$ there.
However, this is just a coordinate singularity, as found $e.g.$ by computing the
Kretschmann scalar of the solutions.

The numerics is performed with the functions $\bar{A},\bar{B},\bar{C}$
which are finite everywhere.
The equations satisfied by these functions are derived again from the Einstein equations 
$G_t^t=0,~G_r^r+G_z^z=0$ and $G_\theta^\theta=0$.
As in the NUBS case, the remaining Einstein equations $G_z^r=0,~G_r^r-G_z^z=0$
yield two constraints, which are automatically satisfied for the set of  boundary conditions
chosen. 

For the metric ansatz (\ref{ansatz-BHs}),
the caged BH solutions are constructed subject to the following set of 
boundary conditions
\begin{eqnarray}
\nonumber
&&\bar A\big|_{r=\infty}=\bar B\big|_{r=\infty}=\bar C\big|_{r=\infty}=0,
\\
\label{bc3}
&&\partial_{r}  \bar A\big|_{r=r_0}=0,~~
\partial_{r}  \bar B \big|_{r=r_0}=1-\frac{\partial_{r} J}{2J}\big|_{r=r_0},~~
\partial_{r}  \bar C \big|_{r=r_0}=1-\frac{\partial_{r} R}{R}\big|_{r=r_0},
 \\
\nonumber
&&\partial_z  \bar A\big|_{z=0,\pi/2}=\partial_z  \bar B\big|_{z=0,\pi/2}
=\partial_z  \bar C\big|_{z=0,\pi/2}=0.
\end{eqnarray}
The regularity of the solutions requires also $ \bar B=  \bar C$ for $r\leq 0, z=\pi/2$.

The Hawking temperature and entropy of the BH solutions 
are given by\footnote{Note that the Einstein equation $G_z^r=0$  
assures that $e^{A(r,z)-B(r,z)}/\sqrt{J(r,z)}$ is constant on the horizon.}
\begin{eqnarray}
\label{temp-BH} 
&T_H=\frac{(D-3) e^{A(r_0,z)-B(r_0,z)}}{4\pi\sqrt{J(r_0,z)}} ,~~
S=  \frac{\Omega_{D-3}}{4G}\int_0^\pi dz~e^{B(r_0,z)+(D-3)C(r_0,z)} J(r_0,z)R^{D-3}(r_0,z).~~{~~~~}
\end{eqnarray}
The mass and tension of the solutions can be read in principle from the asymptotics of the metric 
functions $\bar A, \bar B$ (in practice, these quantities are computed usually 
from the event horizon quantities, by
using the Smarr relation (\ref{smarrform1}) together with the first law of thermodynamics (\ref{fl1})).
 
The numerical iteration starts with large values of $|r_0|$, $i.e.$ small black holes, which 
are well approximated by the Schwarzschild-Tangerlini solution in $D-$dimensions.
These configurations are used as initial data for the solutions with a smaller absolute value of $r_0$.
This results in a branch of caged black holes, which stops to exist for a critical value of $r_0<0$.
 
 We have constructed in this way a number of $D=6$ solutions, with relatively small values of $r_0$.
 In our approach, the numerical calculations were based on the Newton-Raphson method
and are performed with help of the program FIDISOL \cite{schoen},
which provides also an error estimate for each unknown function.
The numerical accuracy of the BHs was always lower than in the NUBS case, the errors
increasing around the
point $r=0,z=\pi/2$.
However, the results for the solutions we have constructed so far are in good agreement with those in 
 \cite{Kudoh:2004hs}, \cite{Kudoh:2003ki}.
Typical profiles of the metric functions $A,B,C$ are shown in Figure 4 (left).

%%%%%%%%%%%%%%%%%%%%%%%%%%%%%%%%%%%%%%%%%%%%%%%%%%%%%%%%%%%%%%%%%%%%%%%%%
\section{Slowly rotating, charged uniform black strings in Einstein-Maxwell theory}
%%%%%%%%%%%%%%%%%%%%%%%%%%%%%%%%%%%%%%%%%%%%%%%%%%%%%%%%%%%%%%%%%%%%%%%%%
For the parametrization considered in this paper, 
the uniform, charged black string solution
in the EM theory is given by
 \begin{eqnarray} 
 \label{charged-UBS}
 &ds^2_0=\Omega^{\frac{2}{D-3}}(r)\left( \frac{dr^2}{f(r)}+dz^2
 +r^2(
 d\theta+\sin^2 \theta d\varphi^2
 +\cos^2 \theta d\omega_{D-5}^2 
 )
 \right) -\frac{f(r)}{\Omega^2(r)}dt^2,~
 {\cal A}=V(r)dt,~~{~~~}
\end{eqnarray}
where
\begin{eqnarray} 
\Omega(r)=1+\left( \frac{r_0}{r} \right)^{D-4}\frac{U^2}{1-U^2},~~
f(r)=1-\left( \frac{r_0}{r} \right)^{D-4},
~~V(r)=\sqrt{\frac{2(D-2)}{(D-3)}}\frac{U}{\Omega(r)}f(r) ,~~{~~}
\end{eqnarray}
 $r_0>0$ being the event horizon radius and $0\leq U<1$.
%The mass to electric charge ratio of these solutions is 
%\begin{eqnarray} 
%\frac{M}{Q_e}=\frac{7U^2+9+D(D-6)(1+U^2)}{\sqrt{2(D-2)(D-3)}U}.
%\end{eqnarray} 

We consider a slowly rotating ansatz which follows from (\ref{ans-NUBS-rot}):
 \begin{eqnarray} 
 \nonumber
ds^2=ds_0^2
-2 a\sin^2 \theta j(r) dt d \varphi ,~~~
{\cal A}=V(r)dt +a h(r)\sin^2\theta d\varphi,
\end{eqnarray}
where $a$ is an infinitesimal parameter.
The functions $h(r),j(r)$ are solutions of a system of second order 
ordinary differential equations which are found from (\ref{eq-rot-NUBS}) by replacing 
there $A=-\log \Omega $, $B=\log \Omega/(D-3)$, $C=\log \Omega/(D-3)$.
 
These 
equations admit the following exact solution
\begin{eqnarray} 
\label{sol-UBS}
h(r)=-\frac{Q_e}{r^{D-4}\Omega(r)}~,
\end{eqnarray}
\begin{eqnarray}
\nonumber
 &&j(r)=\Omega(r)^{\frac{2}{D-3}}\frac{Q_e}{r^{D-4}}
\bigg(
\frac{\sqrt{2}(D-4)}{U\sqrt{(D-3)(D-2)}}
{}_2F_1\left(\frac{D-2}{D-4},\frac{3D-7}{D-3},\frac{2(D-3)}{D-4},\frac{U^2}{U^2-1}(\frac{r_0}{r})^{D-4}\right)~~^{~~~~~~~~}
\\
\nonumber
&&+\frac{(D-4)(D-1)U^2-(D-5)(D-2)}{\sqrt{2(D-2)(D-3)}U}
{}_2F_1\left(\frac{D-2}{D-4},\frac{2D-4}{D-3},\frac{2(D-3)}{D-4},\frac{U^2}{U^2-1}(\frac{r_0}{r})^{D-4}\right)  
\bigg),~~{~~}
\end{eqnarray} 
 where  ${}_2F_1$ is the hypergeometric function.
An unexpected feature here is the occurrence for $D=5$ of a logarithmic term in the expression of
$J(r)$,
\begin{eqnarray} 
&j(r)= \frac{Q_e}{\sqrt{3}U}\frac{1}{r\Omega (r)}
\left(
1-\frac{1-U^2}{r_0}r\Omega(r)
-\frac{2(1-U^2)^2r^2}{r_0^2U^2}r^2 \Omega^2(r)
+\frac{2(1-U^2)^3}{r_0^3U^4}r^3\Omega^3(r)\log \Omega(r)
\right).
~{~~~~~~}
\end{eqnarray}The angular momentum and magnetic dipole moment are read off from the asymptotic form of $j(r)$ and 
$h(r)$ (see (\ref{as-j}), (\ref{as-h}))
\begin{eqnarray} 
\nonumber
j(r)=\frac{Q_e(6+4U^2+D(D-5)(1+U^2))}{\sqrt{2(D-2)(D-3)}(D-2)U}\frac{1}{r^{D-4}}+\dots,
~~h(r)=-\frac{Q_e}{r^{D-4}}+\dots~.
\end{eqnarray}
 The expression of the gyromagnetic ratio (\ref{gyro-ratio}) derived in this way is
(here we use also (\ref{ratio}) for the mass to charge ratio):
\begin{eqnarray} 
\label{exprg}
g= D-3 -\frac{2U^2}{(D-2)(D-3)+(D-1)(D-4)U^2},
\end{eqnarray} 
which varies in the range
\begin{eqnarray} 
D-3<g<\frac{(D-2)^2(D-4)}{D(D-5)+5}.
\end{eqnarray} 
Thus, different from the corresponding case of  (-asymptotically Minkowski) 
black holes in $D$-dimensions, the gyromagnetic
ratio of the slowly rotating black strings in EM theory
is not constant even in the first order of the perturbation theory.
This can be understood as follows. After performing a KK reduction with respect to the Killing vector
$\partial/\partial z$, the static EM uniform black strings (\ref{charged-UBS}) become  black hole
solutions in a Einstein-Maxwell-dilaton theory in $D-1$ dimensions, with a value $a=1/\sqrt{2(D-2)(D-3)}$
of the dilaton coupling constant.
However, the results in \cite{Sheykhi:2008bs} show that the inclusion of a dilaton field
in the action
modifies the value of $g$ for the slowly rotating black holes in $d-$dimensions, 
which deviates from the standard result $d-2$.
This explains the relation (\ref{exprg}) we have found for charged black strings in $D-$dimensions.

 We close this part by remarking that a similar study can be performed 
  for slowly rotating
charged  solutions with all angular momenta degrees of freedom excited.
The results we have found  for a single angular momentum here (and in Section 6) are expected to hold there also.
In the first order of perturbation theory, all gyromagnetic ratio would be equal 
and given by the relation (\ref{exprg}) in the UBS limit.

 \end{document}